\documentclass[draft]{agujournal2019}
\usepackage{url} 
\usepackage{soul}

\usepackage{amsmath,amssymb}
\usepackage{xcolor,colortbl}
\usepackage{makecell}
\usepackage{mathtools}
\usepackage{accents}

\usepackage{booktabs}
\usepackage{multirow}
\usepackage{siunitx}

\usepackage{pifont}

\usepackage{comment}
\usepackage{textcomp,mathcomp}
\usepackage{algorithm}
\usepackage{algpseudocode}

\definecolor{lightgray}{cmyk}{0,0,0,0.17}

\newcommand{\harp}[1]{\accentset{\rightharpoonup}{#1}}

\journalname{Geochemistry, Geophysics, Geosystems}

\begin{document}

\title{Quantifying the influence of fault geometry via mesh morphing with applications to earthquake dynamic rupture and thermal models of subduction}

\authors{Gabrielle M. Hobson \affil{1}, Dave A. May \affil{1}, Alice-Agnes Gabriel \affil{1,2}}

\affiliation{1}{Institute of Geophysics and Planetary Physics, Scripps Institution of Oceanography, University of California San Diego, USA}
\affiliation{2}{LMU Munich, Germany}

\correspondingauthor{Gabrielle M. Hobson}{ghobson@ucsd.edu}

\textbf{This manuscript is an ArXiv preprint and has been submitted for possible publication in a peer reviewed journal. Please note that this has not been peer-reviewed before and is currently undergoing peer review for the first time. Subsequent versions of this manuscript may have slightly different content.}

\begin{keypoints}
\item We describe a method for generating ensembles of geometrically varying meshes while maintaining mesh connectivity. 
\item We apply this technique to meshes for numerical simulations of 3D earthquake dynamic rupture and 2D subduction zone thermal structure. 
\item We quantify the effects of fault dip on surface displacement in the context of rupture and of plate interface geometry on temperature.

\end{keypoints}

\begin{abstract}
Subsurface geometries are often poorly constrained, yet they exert first-order control on key geophysical processes, including subduction zone thermal structure and earthquake rupture dynamics. 
Quantifying  model sensitivity to geometric variability remains challenging due to the manual effort of mesh generation and the computational cost of exploring high-dimensional parameter spaces in high-fidelity simulations.
We present a mesh morphing approach that deforms a reference mesh into geometrically varying configurations while preserving mesh connectivity. 
This enables the automated generation of large ensembles of geometrically variable meshes with minimal user input. 
Importantly, the preserved connectivity allows for the application of data-driven, non-intrusive reduced-order models (ROMs) to perform robust sensitivity analysis and uncertainty quantification.
We demonstrate mesh morphing in two geophysical applications: (i) 3D dynamic rupture simulations with fault dip angles varying across a 40° range, and (ii) 2D thermal models of subduction zones incorporating realistic slab interface curvature and depth uncertainties informed by the Slab2 geometry dataset. 
In both cases, morphed meshes retain high quality and lead to accurate simulation results that closely match those obtained using exactly generated meshes.
For the dynamic rupture case, we further construct ROMs that efficiently predict surface displacement and velocity time series as functions of fault geometry, achieving speedups of up to  $10^9 \times$ relative to full simulations.
Our results show that mesh morphing can be a powerful and generalizable tool for incorporating geometric uncertainty into physics-based modeling. 
The method supports efficient ensemble modeling for rigorous sensitivity studies applicable across a range of problems in computational geophysics.
\end{abstract}

\section*{Plain Language Summary}

The shape of underground faults influences how earthquakes rupture, but since faults are below Earth's surface, their shape (or geometry) is not exactly known. 
Scientists building computer models of earthquakes often represent faults using a mesh, a collection of small volumes subdividing a 3D space. 
Since building a mesh representing a new fault geometry is labor-intensive and computer models are expensive to calculate, the effect of variable geometry on many earthquake-related processes has not been robustly measured. 
We present a mesh morphing method which takes a mesh representing one geometry and deforms it so that it represents a new geometry. 
This efficiently creates many geometrically-varying meshes and means that model output can be used to build reduced-order models, which are useful for measuring uncertainty in model output.  
We apply mesh morphing to models of earthquake rupture, varying the dip angle between fault and surface, and to models of subduction zone temperature, varying the subducting slab's curvature and depth.
We also use a reduced-order model to measure how changing the fault dip angle affects the surface displacement caused by a modeled earthquake.
Our results demonstrate that mesh morphing allows us to accurately and efficiently measure the effects of varying geometry.

\section{Introduction} \label{sec:intro}

Natural faults have complex, non-planar geometries with rough surfaces that are challenging to constrain observationally \cite<e.g.,>{BenZionSammis2003}. 
Descriptions of fault geometry may come from geodetic, field and laboratory observations of surface rupture \cite<e.g.,>{Brown1985,Candela2012}, locations and focal mechanisms of background seismicity and aftershocks \cite<e.g.>{Ross2019}, finite fault models of previous large earthquakes \cite<e.g.,>{ragon_et_al_2019}, and seismic reflection and tomography \cite<e.g.,>{Lee2025}, all of which include inaccuracies and uncertainties to differing extents \cite<e.g.,>{ragon_2018}.
For example, Slab2 \cite{slab2}, a global dataset of subduction zone megathrust models, assigns minimum uncertainties to each of the data types used; the assigned default uncertainty is 2.5~km for active source/seismic reflection, 10~km for relocated earthquake hypocenters, and 40~km for tomography. 
Even well-observed and extensively studied earthquakes have variable fault morphologies proposed in the literature \cite<e.g.,>{Plesch2007,hayek_et_al_2024}. 
The geometry of a plate interface and faults (even with surface rupture expressions) often remains inexactly known at depth and assumed fault geometries in models should therefore incorporate uncertainty. For example, including uncertainty in fault geometry in kinematic earthquake source inversions allows for more robust slip models, reducing the sensitivity of inferred models to geometric assumptions \cite{ragon_2018, ragon_et_al_2019}. 

Earthquake dynamic rupture is affected by complexities in fault structure, such as bends, branches, offsets, segmentation, roughness and curvature \cite{king_nabelek_1985, okubo_aki_1987, aki_1989, zhang_et_al_1991, wesnousky_2006}, including the nucleation and arrest of large earthquakes \cite{aki_1989,Stein2024}. 
Slip on a segment in a complex fault system affects the stress state of neighboring segments, which respond to loading differently depending on the segment lengths and the step size between segments \cite{segall_pollard_1980}. Elastic interactions between segments determine whether slip on one segment locally impedes or enhances slip on neighboring segments and whether multi-fault ruptures are generated \cite{bouchon_et_al_1998,wesnousky_2006,wesnousky_2008,BaiAmpuero2017}.

In subduction zones, megathrust earthquake rupture dynamics are sensitive to several geometric factors including the dip angle and curvature of the subducting slab interface \cite{wirth_et_al_2022, schellart_rawlinson_2013, bletery_et_al_2016, van_rijsingen_et_al_2018, lallemand_et_al_2018}. 
Slab geometry has been linked to both the potential for subduction zones to host large earthquakes and to tsunami generation. 
A shallow dipping slab may imply that the thermally-controlled brittle-ductile transition is pushed further from the trench, resulting in a wider seismogenic zone \cite{wirth_et_al_2022}. 
The megathrust dip angle has been found to correlate with earthquake moment magnitude, $\text{M}_\text{w}$, with $\text{M}_\text{w} \geq 8.5$ earthquakes only occurring in subduction zones with dips  $< 35^\circ$, 
and $\text{M}_\text{w} \geq 9.2$ events only occurring on megathrusts with dips $< 20^\circ$ \cite{schellart_rawlinson_2013, muldashev_sobolev_2020, wirth_et_al_2022}.

Along-dip curvature of the subducting plate interface may also control rupture extent \cite{bletery_et_al_2016, plescia_hayes_2020}, with
large earthquakes preferentially occurring on low-curvature, relatively planar slabs \cite{bletery_et_al_2016, plescia_hayes_2020,biemiller_et_al_2024}. 
Low curvature likely promotes more homogeneous shear strength across a broader region of the megathrust, allowing for wider coseismic rupture. 
In contrast, strong variations in along-strike curvature potentially limit earthquake rupture extent \cite{schellart_rawlinson_2013, bletery_et_al_2016, plescia_hayes_2020}.
For example, along-strike changes in curvature of the Chilean megathrust appear to influence the coseismic slip distributions of the 2010 $\text{M}_\text{w}$ 8.8 Maule and 2014 $\text{M}_\text{w}$ 8.2 Iquique earthquakes \cite{moreno_et_al_2012, shrivastava_et_al_2019,Herrera2024}. 

Thermal models of subduction zones are essential for estimating the limits of the seismogenic zone, the onset of metamorphic dehydration, and the distribution of slab and mantle wedge rheology. State-of-the-art models typically solve coupled heat advection-diffusion equations with prescribed kinematic slab geometries and are constrained by surface heat flow and petrological data \cite<e.g.,>{van_keken_kiefer_peacock_2002, wada_et_al_2008, currie_et_al_2004, syracuse_2010, wada_wang_2009, gerya_2022}. However, uncertainties remain due to poorly constrained material properties, modeling assumptions, uncertain slab geometry at depth, and lack of surface heat flow and petrological data coverage \cite<e.g.,>{peacock_2020}. 
Present efforts characterize different aspects of uncertainty, including: parameterizations of the maximum depth of slab-mantle decoupling \cite{syracuse_2010}; influence of 3D subduction interface geometry \cite{kneller2008effect}; shear heating \cite{kohn2018shear}; thermal number \cite{maunder_et_al_2019}; and potential rupture extent \cite{hobson_may_2025}.  
The dip angle of subducting slabs influences the thermal structure of subduction zones. 
While the temperature within the subducting slab and at the subduction interface is primarily controlled by the age of the subducting lithosphere and plate convergence rate, slab dip has a modest effect on interface temperatures and a stronger influence on the thermal structure of the mantle wedge \cite{gerya_et_al_2002, maunder_et_al_2019}. 
Shallower slab dips narrow the mantle wedge corner, potentially restricting the mantle wedge inflow of hotter material, resulting in a cooler corner region that extends farther from the trench. 
To our knowledge, studies which quantify the sensitivity of the thermal structure with respect to the dip angle do not exist.

Computational geophysics studies often involve simulations relying on meshes representing particular geometric configurations, such as fault geometries or the geometry of the plate interface in subduction zones. 
However, systematically exploring how geometry affects the simulation output is challenging due to the effort required to generate new meshes for each new geometric configuration, the high-dimensionality of the parameter space to be explored, and the high computational cost typically associated with running a single forward model. 
%

In this study, we present a \textit{mesh morphing} method that enables sensitivity analysis with respect to varying geometries in geophysical simulations. 
The approach allows for the efficient generation of ensembles of geometrically varying meshes and is applicable to a wide range of geophysical problems involving mesh-based simulations.  
We demonstrate the utility of mesh morphing in two geophysical applications  where fault geometry plays an important role: earthquake dynamic rupture modeling and subduction zone thermal structure. 
We perform a geometric variability sensitivity analysis for the earthquake dynamic rupture example. 
Our mesh morphing approach provides a general and efficient framework for exploring geometric uncertainty in computational geophysics.


\subsection{Exploring Geometric Variations Through Mesh Morphing}

\begin{figure}
     \centering
     \includegraphics[width=\textwidth]{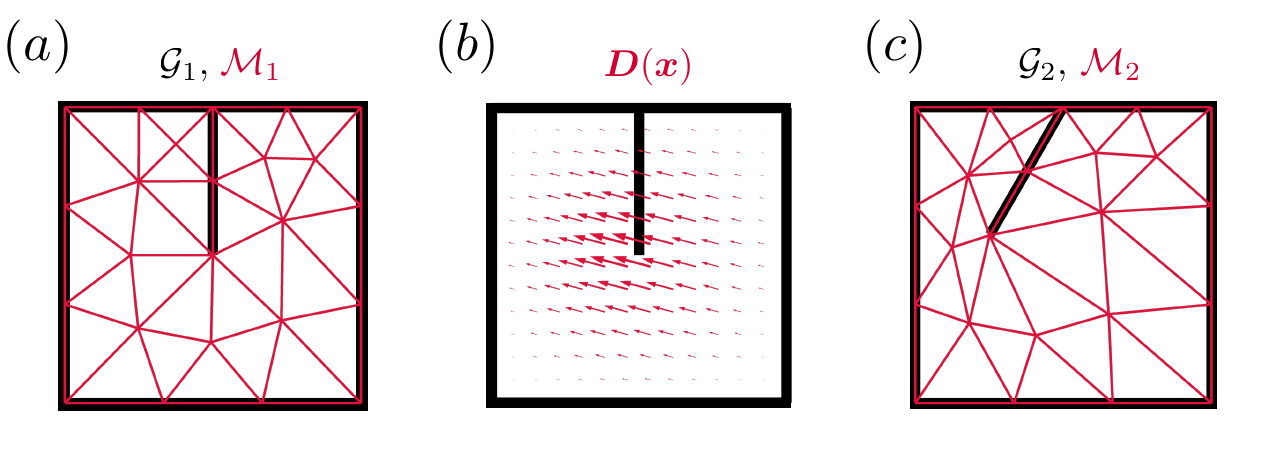}
     \caption{Mesh morphing applied to a simple 2D geometry: (a) shows the initial geometry $\mathcal G_1$ with mesh $\mathcal M_1$, while (b) shows the displacement field $\boldsymbol D$ required to morph $\mathcal M_1$ into a new configuration, shown in (c), where $\mathcal M_2$ is the morphed mesh representative of the target geometry $\mathcal G_2$.}
     \label{fig:schematic_simpler}
\end{figure}

Mesh morphing allows us to generate geometrically distinct meshes (e.g., the mesh coordinates vary) without repeating the labor-intensive meshing process, while maintaining the same mesh connectivity. 
The fundamental idea of mesh morphing is to take a mesh $\mathcal M_1$ defining a geometry $\mathcal G_1$ and then deform, or morph, the mesh into a new configuration $\mathcal M_2$ representing a new geometry $\mathcal G_2$.
The deformation of mesh entities (e.g., vertices) is achieved by prescribing displacements via an interpolant (e.g., radial basis functions, RBFs) and evaluating the interpolant at all vertices within $\mathcal M_1$ to define $\mathcal M_2$ \cite{sieger_et_al_2014}.
Under the deformation, the topology (including mesh connectivity) of $\mathcal M_1$ does not change, meaning that the mesh consists of the same mappings between vertices and cells, despite the vertices having updated spatial locations.
This process is visualized in Figure~\ref{fig:schematic_simpler}, which shows an example with initial geometry $\mathcal G_1$ and mesh $\mathcal M_1$ (Figure~\ref{fig:schematic_simpler}(a)), the displacement field that morphs $\mathcal M_1$ to $\mathcal M_2$ (Figure~\ref{fig:schematic_simpler}(b)), and the resulting morphed mesh $\mathcal M_2$ which represents the target geometry $\mathcal G_2$ (Figure~\ref{fig:schematic_simpler}(c)). 
Each geometry is described by one or more geometric parameters, and morphing is used to deform the mesh so that it conforms to a geometry with a different geometric parameter value.
Certain boundaries of a mesh can be prescribed to have zero displacement, so that those boundaries are preserved under morphing. Quantities such as the domain volume can optionally also be preserved.
In the example shown in Figure~\ref{fig:schematic_simpler}, the geometric parameter is the angle between the top boundary and the embedded line within the domain. 
The embedded line is morphed to have different angles. 
The process requires initial choices to be made that are appropriate to the given application, but each morphed mesh is then generated automatically,
removing nearly all of the manual labor associated with creating new geometries and mesh generation. 
By morphing a single mesh to represent many different geometric configurations, we can readily create an ensemble of meshes which capture the chosen geometric variations. 

To address the computational expense of the many forward model evaluations required by robust sensitivity analysis, we leverage model order reduction.
Reduced-order models are a type of surrogate model that accurately approximate the solutions of the full-order model (FOM) \cite<e.g.,>{lumley}; in the context of this study, the FOM is the geophysical simulation of interest, such as an earthquake dynamic rupture simulation or a subduction zone thermal model. 
The ROMs are highly efficient to evaluate, often many orders of magnitude faster than the FOM once constructed, permitting the use of robust global sensitivity analysis techniques requiring many model evaluations. 
Reduced-order modeling is a well-established approach in engineering and computational fluid dynamics \cite<e.g.,>{berkooz_holmes_lumley_1993, ly_tran_2001, taira_et_al_2017}, and has been applied in geophysics to approximate seismic shakemaps, surface velocity wavefields and subduction zone thermal structure \cite{rekoske_2023, rekoske_et_al_2025, hobson_may_2025}. 
The data-driven and non-intrusive reduced-order modeling methods best suited to geophysical applications are built directly from FOM solutions, where each solution corresponds to a different point in parameter space \cite<e.g.,>{lumley, sirovich}.
However, these techniques require that the same mesh topology is used for each simulation, which precludes directly building different meshes to explore geometric variations. 
By using mesh morphing to create an ensemble of meshes with identical connectivity, 
we can run the FOM on meshes with geometric variation and obtain solutions that are suitable for building a data-driven ROM. 
For one of our applications, the earthquake dynamic rupture example, we combine the advantages of mesh morphing and non-intrusive ROMs 
to perform sensitivity analysis of processes incorporating parameterizations of the geometry.

\subsection{Related Work}

The mesh morphing methodology has roots in early computer graphics, where \citeA{sederberg_parry_1986} first developed free-form deformation to deform solid geometric models. 
The computer graphics community has continued to develop mesh morphing, with applications in real-time shape editing and animation \cite{alexa_2002, botsch_kobbelt_2005}. 
The field of engineering has also advanced mesh morphing, with applications in simulation-based automatic design optimization for product development in, e.g, the automotive industry, aircraft construction, and naval architecture \cite<e.g.,>{du_et_al_2023, de_boer_2007, demo_et_al_2021}.  
A key factor in this optimization is being able to adapt an existing volumetric simulation mesh according to an updated CAD geometry. 
This is especially necessary when working with complex geometries or when many mesh variations must be generated, as required for example by stochastic global optimization \cite<e.g.,>{du_et_al_2023}. 

A range of mesh morphing techniques exist, and \citeA{staten_et_al_2012} performed a comparison of six different methods.
Methods based on generalized barycentric coordinates determine interior vertices as a linear combination of boundary vertices via generalized linear barycentric interpolation; while easy to implement and efficient, the resulting morphed meshes may not be sufficiently smooth \cite{sieger_et_al_2014}. 
Mesh smoothing based methods use updated boundary vertices as constraints and optimize based on cell quality to obtain the interior vertex locations; these methods are effective for small geometric changes, but either are not robust under larger geometric changes or become computationally complex \cite<e.g.,>{knupp_2000}. 
Mesh-based variational methods solve Laplacian or bi-Laplacian systems to compute harmonic or biharmonic deformations; these methods are numerically robust and produce good results, though they must be custom tailored to different mesh types \cite<e.g.,>{baker_2002, thoutireddy_ortiz_2004}. 
Meshless morphing techniques deform an embedding space via a space warp; these techniques are extensible to different meshes, though the results may not be as smooth as variational methods and the method requires a lattice setup. 

An approach that combines advantages of meshless and mesh-based variational methods is mesh morphing using RBFs \cite{botsch_kobbelt_2005, de_boer_2007, jakobsson_amoignon_2007, michler_2011, sieger_et_al_2014}. 
This approach prescribes displacements for vertices on curves (in 2D) or surfaces (in 3D) and uses triharmonic RBF interpolation to smoothly interpolate the displacements onto all vertices of the mesh. 
This method is straightforward to implement, is applicable to tetrahedral, hexahedral, or polyhedral meshes, and robustly produces high quality cell meshes \cite{sieger_et_al_2014}. 
Demonstrated applications of mesh morphing using RBFs include real-time freeform shape editing in computer graphics \cite{botsch_kobbelt_2005}, airfoil design \cite{de_boer_2007}, optimization of a wing in computational fluid dynamics \cite{jakobsson_amoignon_2007}, and aircraft control surface deflections \cite{michler_2011}.  
RBF mesh morphing combined with optimization frameworks can enable efficient geometric shape changes during optimization, with applications including vehicle crashworthiness design \cite{du_et_al_2023} and ship hull design optimization \cite{demo_et_al_2021}. 
Fluid-structure interaction modeling uses mesh morphing to transfer deformations computed by computational structural mechanics to a computational fluid dynamics mesh \cite{biancolini_2017}, with applications in cardiac research \cite{geronzi_et_al_2020}. 

\subsection{Outline}

The structure of this study is as follows. In Section~\ref{sec:methods} we describe the mesh morphing method for adapting an existing mesh based on geometric variations without changing the mesh connectivity. 
In Sections~\ref{sec:dr_examples} and \ref{sec:sz_int_example} we demonstrate the use of this method for meshes in 3D dynamic rupture simulations and 2D subduction zone thermal models. 
In these application sections we describe how we monitor mesh quality when geometric variations are applied, quantify the error between adapted meshes and exactly created meshes, and compare output from simulations run on morphed and exactly generated meshes. 
In Section~\ref{sec:ROM} we present model order reduction results for dynamic rupture simulations, with sensitivity analysis of the effects of geometric variability. 
In Section~\ref{sec:discussion} we outline limitations of the method, discuss the potential extension of the method to other applications, and describe some points to consider when applying the method to a new problem. 
Finally in Section~\ref{sec:conclusions} we summarize the method and the results for each of our examples. 

\section{Methods} \label{sec:methods}

\subsection{Definitions}

To facilitate the description of the mesh morphing method, below we provide several definitions which we will use throughout the remainder of this paper. 
\begin{itemize}
\item Geometry $\mathcal G$: Refers to the representation of a 2D surface or 3D volume with a specific shape, including the relative positions of, and connections between, external and internal boundaries and surfaces. 
\item Domain: The domain $\Omega$ of the geometry, which is referred to as a surface in $\mathbb R^2$, and is referred to as a volume in $\mathbb R^3$.
Frequently the domain $\Omega$ will be described in terms of non-overlapping sub-domains which we denote by $\Omega_A, \Omega_B, \dots$ with $\Omega = \underset{k=A, B, \dots}{\cup} \Omega_k$.
\item Boundary: An edge (in 2D) or surface (in 3D) of the domain; a point $\boldsymbol x_t$ lives on the boundary $\partial \Omega$ of the domain $\Omega$ if the interior of the ball with radius $r>0$ centered on $\boldsymbol x_t$ is both inside and outside the domain $\Omega$. 
\item Interface: An edge or surface $\Gamma_i$ contained within $\Omega$ where $\Gamma_i \not\subset \partial \Omega$, which is important for representing the geometry, must be respected in meshing, and may optionally form the boundary of sub-domains ($\Omega_k$) of the full domain ($\Omega$).  
\item Discretization: The process of decomposing a domain $\Omega$ or any of its sub-domains $\Omega_k$ (a volume, surface, or line) into a set of non-overlapping sub-domains called cells, dented by $\triangle$. The union of all cell sub-domains closely approximates the volume of the domain $\Omega$, or any of its sub-domains. 
\item Cell $\triangle$: The name used to refer to a single sub-domain $\triangle$ of $\Omega$ (or sub-domain $\Omega_k$). %
Cells are not typically polygons, but rather are defined by geometric primitives, e.g., in 2D these might be triangles or quadrilaterals, and in 3D tetrahedra and or hexahedra. The sub-domain $\triangle$ is defined by a finite number of vertices.
\item Vertex: Defines the spatial coordinates of a point in space $\mathbb R^n$. These points describe the geometry of each cell. 
\item Mesh $\mathcal M$: The collection of vertices and cells, as well as the connectivity that defines relationships between them, forming the discretized representation of a continuous geometry. 
\item Connectivity: Included in the mesh topology; defines relationships between mesh primitives such as vertices and cells, e.g., the cell-cell map or the cell-vertex map. 

\end{itemize}

We will describe a given geometry using the following fundamental components. 
We begin with a set of points $\{a, b, c, \ldots\} \in \mathbb{R}^n$, and we connect them using lines or curves which have a sense of direction. 
For example, $\harp{ab}$ is the curve that has point $a$ as the origin and point $b$ as the destination.
We use curve and line interchangeably, where a line is a curve with zero second derivative. 
We then consider curve loops $\mathcal{C}$, which are ordered lists of connected curves. 
Curve loops must be defined such that the destination of the last curve is the same as the origin of the first curve in the list (i.e., the loop is closed). 
For the purposes of the examples in this study, a surface $\mathcal{S}$ is defined as being the domain enclosed by a given curve loop $\mathcal C$. 
Finally, we define a volume as being the 3D domain enclosed by a set of surfaces that fully enclose the space. 

The geometry is then discretized to create a mesh.  
Curves are discretized into 1D cells, sometimes called facets, while surfaces are discretized into 2D cells (such as triangles) and volumes into 3D cells (such as tetrahedra). 
Each point $\{a, b, c, \ldots\}$ in the geometric definition has a co-located mesh vertex. 
Vertices and cells have associated information about where they lie, whether that be at a point, on a curve, on a surface, or within a volume. 
When we define a mesh, certain vertices will be considered to lie on a given curve or surface, and we consider them to still lie on that curve or surface even as it undergoes morphing and changes location. 
That is to say, the association between vertices and their fundamental object (point, curve, surface, volume) remains true even if their location changes. 

We consider mesh morphing in the context of solving an abstract parametric partial differential equation (PDE),
\begin{linenomath*}
\begin{equation}
\begin{aligned} 
\mathcal L(\phi) &= 0 \quad \text{for } \boldsymbol x \in \, \Omega, \\
\mathcal B(\phi) &= g \quad  \text{for } \boldsymbol x \in \, \partial \Omega
\label{eq:PDE}
\end{aligned}
\end{equation}
\end{linenomath*}
where $\Omega$, $\partial \Omega$ denote the domain and its boundary, $\phi$ is the unknown, $\mathcal L(\cdot), \mathcal B(\cdot)$ are differential operators and $g$ is the prescribed boundary condition. 
Denoting the parameters as $\boldsymbol m = \{m_1, m_2 \cdots \}$, we assume that
\begin{linenomath*}
\begin{align} 
\mathcal L = \mathcal L(\boldsymbol m), \,\,
\mathcal B = \mathcal B(\boldsymbol m), \,\,
g = g(\boldsymbol m), \,\,
\Omega = \Omega(\boldsymbol m), \,\,
\partial \Omega = \partial \Omega(\boldsymbol m).
\end{align}
\end{linenomath*}
That is, we regard the geometry as being parameterized, and geometry parameters are included in $\boldsymbol m$ in addition to coefficients in the PDE such as material or physical constants. 
For clarity we let $\boldsymbol m = (\boldsymbol p, \boldsymbol q)$ where $\boldsymbol q$ denote parameters related to the geometry and $\boldsymbol p$ be all other non-geometric parameters. Thus the abstract parametric PDE in Equation~\eqref{eq:PDE} is written as
\begin{linenomath*}
\begin{equation}
\begin{aligned} 
\mathcal L(\phi, \boldsymbol p) = 0 \quad \text{for } \boldsymbol x \in \, \Omega(\boldsymbol q), \\
\mathcal B(\phi, \boldsymbol p) = g(\boldsymbol p) \quad  \text{for } \boldsymbol x \in \, \partial \Omega(\boldsymbol q).
\label{eq:p_pde_f}
\end{aligned}
\end{equation}
\end{linenomath*}

\begin{table}
\begin{center}
\caption{Notation used throughout this paper.}
\label{table:notation}
\begin{tabular}{  l l }
Notation & Definition \\ \hline

$n$ & Dimension of parameter space: number of geometric parameters. \\

$\boldsymbol q$ & Vector of parameters defining geometric variation.  \\

$\boldsymbol Q$ & Parameter space, the cartesian product of the ranges of each parameter in $\boldsymbol q$, where $\boldsymbol Q \in \mathbb{R}^n$. \\

$\boldsymbol q_\text{new} $  &  Vector of samples in the parameter space $\boldsymbol Q$. \\

$\boldsymbol q_\text{ref} $ & The parameter vector that generates the reference geometry and mesh. \\

$\boldsymbol q_\text{eval} $ & Vector of samples in $\boldsymbol Q$ that describe the geometries to which the reference mesh is morphed. \\

$\boldsymbol D(x,y)$ & The displacement field  between points on the reference geometry and geometries generated by $\boldsymbol q_\text{new} $. \\

$\boldsymbol X_\Gamma$ & Points on the boundary $\Gamma$. \\ 

$\Omega$ & The domain of the reference geometry. \\

$M$ & The reference mesh. \\

$\boldsymbol X_M$ & Vertices in the discretized reference mesh. \\

$\Gamma_a$ & The union of all boundaries and interfaces that will be morphed. \\

$\Gamma_k$ & The union of all boundaries and interfaces displaced only in the $k$ direction, where $k = x, y, z$. \\

$\Gamma_0$ & The union of all boundaries and interfaces that have zero displacement. \\

\hline
\end{tabular}
\end{center}
\end{table}

\subsection{The Mesh Morphing Method}

The objective of mesh morphing is to take a mesh $\mathcal M_1$ defining a geometry $\mathcal G_1$ and then deform (i.e., morph) the mesh into a new configuration $\mathcal M_2$ representing a new geometry $\mathcal G_2$. 
Displacements are prescribed for certain points and used to build an interpolant (e.g., an RBF interpolant). 
Vertices in the mesh are deformed by applying displacements obtained by evaluating the interpolant at all vertices of the mesh \cite{sieger_et_al_2014}.
The mesh vertices are displaced to have new positions, but the mesh connectivity does not change during deformation. 

\begin{figure}
     \centering
     \includegraphics[width=\textwidth]{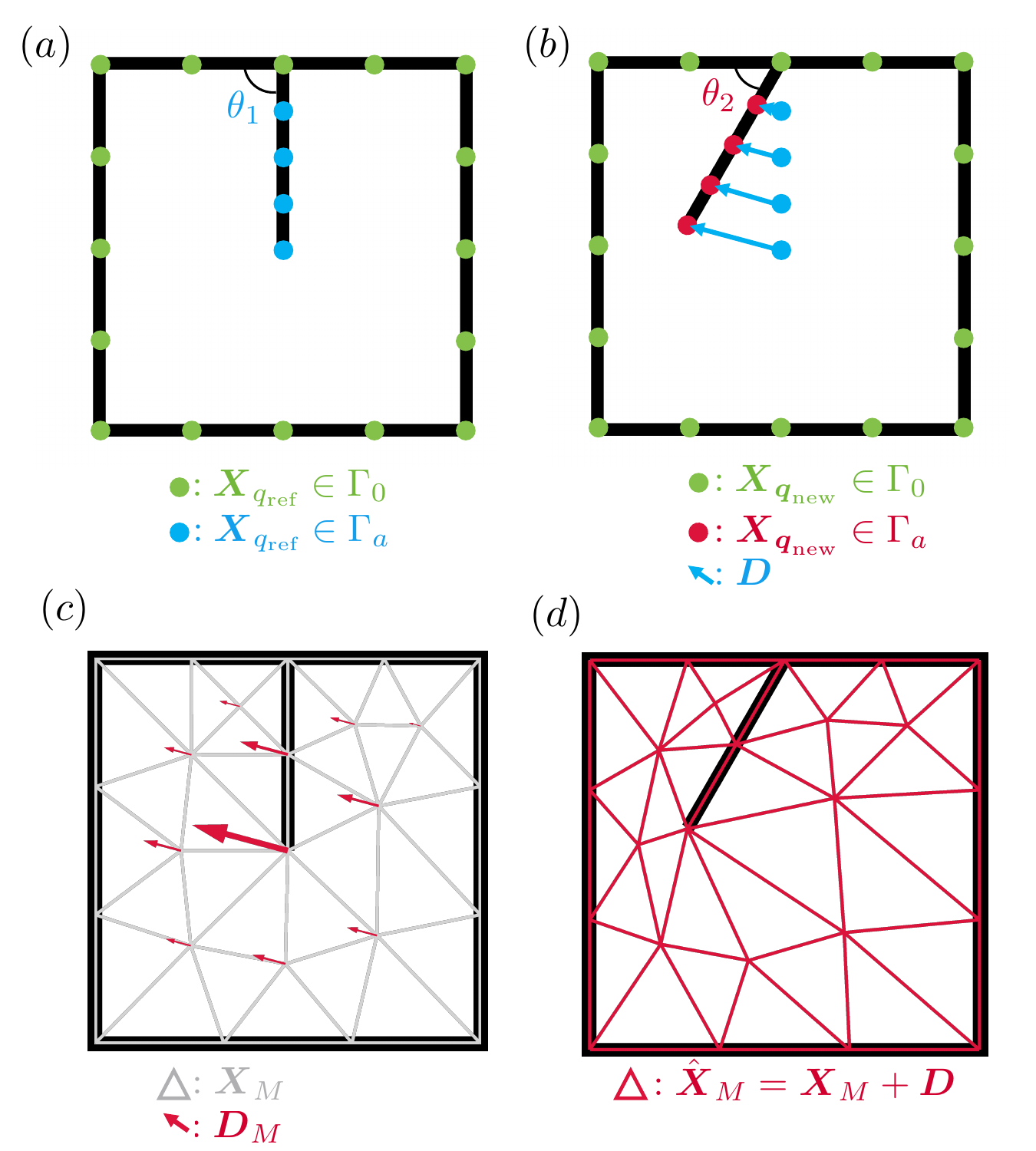}
     \caption{Illustration of the mesh morphing method. (a) The reference geometry, in black, as well as points $\boldsymbol X_{\boldsymbol q_\text{ref}}$ which lie on $\Gamma_0$, in green, and those which lie on $\Gamma_a$, in blue. (b) The new geometry, in black, as well as points $\boldsymbol X_{\boldsymbol q_\text{new}}$ which lie on $\Gamma_0$, in green, and those which lie on $\Gamma_a$, in blue. (c) the reference mesh in gray, with vertices $\boldsymbol X_M$, and the displacement applied to each vertex, $\boldsymbol D_M$. (d) The morphed mesh, $\Hat{\boldsymbol X}_M$. }
     \label{fig:schematic_mm_method}
\end{figure}

For example, consider a volume containing an embedded plane at a given angle $\theta$ (e.g., a fault with given dip) which we would like to vary. 
This example is visualized in Figure~\ref{fig:schematic_mm_method}. 
The angle $\theta$ is the geometric parameter (e.g., $\boldsymbol q = (\theta)$), and the objective is to morph a reference mesh $\mathcal M_1$ with vertices $\boldsymbol X_M$ defined for some angle $\theta_1$ into a new mesh $\mathcal M_2$  with vertices $\Hat{\boldsymbol X}_M$ which has an embedded plane at a different angle $\theta_2$.
To begin, points $\boldsymbol X_{\boldsymbol q_\text{ref}}$ are defined, which lie on the boundaries and interface (the embedded plane) of the reference geometry (Figure~\ref{fig:schematic_mm_method}(a)). 
Similarly, points $\boldsymbol X_{\boldsymbol q_\text{new}}$ are defined, which lie on the boundaries and interface of a target geometry (Figure~\ref{fig:schematic_mm_method}(b)); this is done for several values of $\theta$ to obtain several sets of $\boldsymbol X_{\boldsymbol q_\text{new}}$ representative of different geometric configurations. 
The locations of $\boldsymbol X_{\boldsymbol q_\text{ref}}$ and $\boldsymbol X_{\boldsymbol q_\text{new}}$ are chosen to be equidistantly spaced along each of the boundaries or interface.
Some points in $\boldsymbol X_{\boldsymbol q_\text{ref}}$ lie on the boundary $\Gamma_0$ and will have zero displacement during morphing (colored in green in Figure~\ref{fig:schematic_mm_method}(a)), while others lie on the embedded plane, $\Gamma_a$ and will have non-zero displacement (colored in blue in Figure~\ref{fig:schematic_mm_method}(a)). 
The displacement $\boldsymbol D$ between $\boldsymbol X_{\boldsymbol q_\text{ref}}$ and each set of $\boldsymbol X_{\boldsymbol q_\text{new}}$ points is computed. 
The $\theta$ values used to define the $\boldsymbol X_{\boldsymbol q_\text{new}}$ points and the resulting displacements are used to build an RBF interpolant $F_\text{RBF}$.

To apply the morphing to $\boldsymbol X_M$, the interpolant $F_\text{RBF}$ is evaluated at the desired value of $\boldsymbol q_\text{eval} = \theta_2$ to obtain displacements $\boldsymbol D_j$.  
That displacement is then used to create another RBF interpolant, $G_\text{RBF}$, which takes $\boldsymbol X_{\boldsymbol q_\text{ref}}$ and $\boldsymbol D_j$ as input. 
This second interpolant defines the displacement field and can be evaluated at any point in $\Omega$ to obtain the corresponding displacement. 
Finally, $G_\text{RBF}$ is evaluated at each point in $\boldsymbol X_M$ to obtain the displacements $\boldsymbol D_M$, shown in Figure~\ref{fig:schematic_mm_method}(c). 
The displacements are added to each point in $\boldsymbol X_M$ to obtain $\Hat{\boldsymbol X}_M$, shown in Figure~\ref{fig:schematic_mm_method}(d). 

The example above illustrates several subtleties. First, some mesh vertices are required to be morphed (e.g., those on the embedded fault plane), whilst other mesh vertices should not be morphed (e.g., those on the boundary).
Second, some geometric features may be required to be conserved under morphing -- e.g., the deformation field must obey a constraint. 
In the example above the length of the fault is constrained to remain constant, i.e., the fault movement must be defined via a rigid body rotation.
Additionally, the volume of the domain remains constant since the boundaries are prescribed to have zero displacement. 
For each application, the user must define the geometric parameters, choose which boundaries of the geometry will be morphed (i.e., deformed), generate a reference mesh as a starting point for morphing, and calculate displacements between reference and target boundaries.
 
Each example presented in this paper follows the formalism described here, using Algorithm~\ref{alg:mmm} to build the first RBF interpolant and Algorithm~\ref{alg:mm_eval} to evaluate it for a given parameter vector $\boldsymbol q_j$ and apply the morph to the reference mesh. 
For each example, the geometric parameter vector $\boldsymbol q$ is defined, and the parameter space $\boldsymbol{Q}$ is defined as the Cartesian product of the ranges of each parameter in $\boldsymbol q$. 
A choice of reference value for the geometric parameter, $\boldsymbol q_\text{ref}$, can then be used to generate a reference geometry, which is discretized into a reference mesh $\boldsymbol X_M$. 
Each of the boundaries of the geometry, whether curves for a 2D geometry or surfaces for a 3D geometry, are assigned to one of the following sets: 
$\Gamma_0$, the union of all boundaries that will have zero displacement; 
$\Gamma_a$, the union of all boundaries that will be morphed when $\boldsymbol q$ varies; 
and finally $\Gamma_k$, which consists of boundaries that will be displaced in only one direction, where $k = x, y$ or $z$ directions.
Additionally, the parameter space $\boldsymbol{Q}$ is sampled to obtain a vector of parameter values $\boldsymbol q_\text{new}$ that define different geometric configurations.

Each of these ($\boldsymbol q_\text{ref}$, $\boldsymbol q_\text{new}$, $\Gamma_a$, $\Gamma_0$, $\Gamma_k$, and $\boldsymbol X_M$) is passed as input into Algorithm~\ref{alg:mmm}. 
Algorithm~\ref{alg:mmm} requires defining $N$ points $\boldsymbol X_{\boldsymbol q_\text{ref}} \in \Gamma(\boldsymbol q_\text{ref})$, where $\Gamma = \Gamma_a \cup \Gamma_0 \cup \Gamma_k$.
Equivalent points $\boldsymbol X_{\boldsymbol q_\text{new}} \in \Gamma(\boldsymbol q_i)$ are also defined for each parameter value in $\boldsymbol q_\text{new}$ using the same procedure as used to define $\boldsymbol X_{\boldsymbol q_\text{ref}}$ (as chosen by the user).
These points are not required to correspond to reference mesh vertices and can be arbitrarily placed on the relevant geometric boundaries; there are typically fewer points than mesh vertices.
The displacement $\boldsymbol D$ between $\boldsymbol X_{\boldsymbol q_\text{ref}}$ and each $\boldsymbol X_{\boldsymbol q_\text{new}}$ is computed using Algorithm~\ref{alg:compute_D}. 
Finally, Algorithm~\ref{alg:mmm} returns an RBF interpolant $F_\text{RBF}$ defined for $\boldsymbol q_\text{new}, \boldsymbol D $. 

Algorithm~\ref{alg:mm_eval} is then used to obtain a morphed mesh for a given parameter vector $\boldsymbol q_j$. 
This is done by evaluating $F_\text{RBF}$ at a given value $\boldsymbol q_j$ to obtain the corresponding displacement field $\boldsymbol D_j$. 
This displacement field is then used to define another RBF interpolant $G_\text{RBF}$ using $\boldsymbol X_{\boldsymbol q_\text{ref}}$ and $ \boldsymbol D_j$.
By evaluating $G_\text{RBF}$ at each mesh vertex, the corresponding displacement is obtained and can be added to the vertex coordinate, resulting in the morphed mesh. 
The mesh quality and validity is verified before the morphed mesh is returned. 

RBF interpolation is a type of scattered data interpolant which can be applied to data of any dimension, and is constructed in terms of the Euclidean distance between input scattered data \cite{fasshauer_2007}. Given a function $f(\boldsymbol x)$, the RBF interpolant is a linear combination of kernel functions, $\Theta_j(\boldsymbol x) = \Theta \left( \| \boldsymbol x - \boldsymbol x_j \|_2  \right)$, which are centered at scattered data points $\boldsymbol x_j$ and weighted by $w_j$:
\begin{linenomath*}
\begin{equation} 
f(\boldsymbol x) \approx f_{RBF}\left(\boldsymbol x \right) = \sum_{j=1}^{m} w_j \Theta_j (\| \boldsymbol x - \boldsymbol x_j \|_2). 
\label{eq:RBF}
\end{equation}
\end{linenomath*}
In the mesh morphing steps in this study we use the linear kernel, $\Theta_j(\boldsymbol x) = - \| \boldsymbol x - \boldsymbol x_j \|_2$. 
To ensure that the RBF interpolant is unique,
the interpolant is augmented with a complete polynomial basis \cite{fasshauer_2007}.

As described here, it is possible to morph the mesh to be representative of a geometry with a given geometric parameter value, while keeping certain boundaries fixed in their original locations.
However, it is often important that the morphed mesh satisfy certain additional requirements, as suited to the geophysical application. 
For example, the user may require that the mesh volume be preserved, or that certain lines remain straight or planes remain planar after morphing. 
We use Algorithm~\ref{alg:successive_disp} to apply additional morphing steps to ensure that the morphed mesh respects specific constraints. 
To use Algorithm~\ref{alg:successive_disp}, points on boundaries of the previously morphed mesh are considered $\boldsymbol X_{\boldsymbol q_\text{ref}}$, while target points are defined by the user as $\boldsymbol X_{\boldsymbol q_\text{new}}$.
Then an interpolant is constructed using $\boldsymbol X_{\boldsymbol q_\text{ref}}$ and the displacement between $\boldsymbol X_{\boldsymbol q_\text{ref}}$ and $\boldsymbol X_{\boldsymbol q_\text{new}}$. 
The interpolant is evaluated at each mesh vertex to obtain a further morphed mesh that respects the specified constraint. 

\begin{algorithm}
\caption{Building the mesh morphing RBF interpolant.}\label{alg:mmm}
\begin{algorithmic}
\Require $\boldsymbol q_\text{ref}$, $\boldsymbol q_\text{new}$, $\Gamma_a$, $\Gamma_0$, $\Gamma_k$, $\boldsymbol X_M$.
\State Define $N$ points $\boldsymbol X_{\boldsymbol q_\text{ref}} \in \Gamma(\boldsymbol q_\text{ref})$, where $\Gamma = \Gamma_a \cup \Gamma_0 \cup \Gamma_k$. 
\State $\boldsymbol D = \emptyset$. \Comment{$\boldsymbol D$ is the displacement field array.}
\For{$i = 1, \ldots, m$}: \Comment{$m$ is the number of vectors in $\boldsymbol q_\text{new}$}
	\State $\boldsymbol q_i = \boldsymbol q_\text{new}^i$.  \Comment{Select the $i^\text{th}$ column of $\boldsymbol q_\text{new}$.}
	\State Define $N$ points $\boldsymbol X_{\boldsymbol q_\text{new}} \in \Gamma(\boldsymbol q_i)$. 
	\State $\boldsymbol D_i = \text{ComputeDisplacement}\left(  \boldsymbol X_{\boldsymbol q_\text{ref}}, \boldsymbol X_{\boldsymbol q_\text{new}}, \Gamma_a, \Gamma_0, \Gamma_k  \right) $.  
	\State $\boldsymbol D \leftarrow \boldsymbol D \cup \boldsymbol D_i $. \Comment{Augment $\boldsymbol D$ with the displacement corresponding to $\boldsymbol q_i$.}
\EndFor
\State $F_\text{RBF} =  F_\text{RBF} \left( \boldsymbol q_\text{new}, \boldsymbol D \right)$.  \Comment{Define the RBF interpolant as in Equation~\eqref{eq:RBF}.}
\State \Return $F_\text{RBF}$, $\boldsymbol X_{\boldsymbol q_\text{ref}}$.
\end{algorithmic}
\end{algorithm}

\begin{algorithm}
\caption{Morphing the mesh to represent a configuration defined by $\boldsymbol q_j$.}\label{alg:mm_eval}
\begin{algorithmic}
\Require  $\boldsymbol q_j$, $F_\text{RBF}$, $\boldsymbol X_{\boldsymbol q_\text{ref}}$, $\boldsymbol X_M$.
\State  $\boldsymbol D_j = F_\text{RBF} \left( \boldsymbol q_j \right)$. 
\State $G_\text{RBF} =  G_\text{RBF} \left( \boldsymbol X_{\boldsymbol q_\text{ref}}, \boldsymbol D_j \right) $.
\State $\boldsymbol D_M = G_\text{RBF} \left( \boldsymbol X_M \right)$.
\State $\Hat{\boldsymbol X}_M = \boldsymbol X_M + \boldsymbol D_M$.
\State \texttt{valid} $\gets$ CheckValidMesh($\Hat{\boldsymbol X}_M$).  
\If{\texttt{valid} = \texttt{True}}: 
	\State \Return $\Hat{\boldsymbol X}_M$.
\Else: 
	\State Raise error. 
\EndIf
\end{algorithmic}
\end{algorithm}

\begin{algorithm}
\caption{ComputeDisplacement. A function that calculates the displacement $\boldsymbol D$ between $\boldsymbol X_{\boldsymbol q_\text{ref}}$ and $\boldsymbol X_{\boldsymbol q_\text{new}}$ based on whether they lie on $\Gamma_a, \ \Gamma_0$, or $\Gamma_k$. }\label{alg:compute_D}
\begin{algorithmic}
\Require $\boldsymbol X_{\boldsymbol q_\text{ref}}$ , $\boldsymbol X_{\boldsymbol q_\text{new}}$, $\Gamma_a$, $\Gamma_0$, $\Gamma_k$.
\State $\boldsymbol D = \emptyset$
\For{$i = 1, \ldots, N$}:
\State $\boldsymbol x_\text{r} = \boldsymbol X_{\boldsymbol q_\text{ref}, i}$ \Comment{Extract $i^{\text{th}}$ point in $\boldsymbol X_{\boldsymbol q_\text{ref}}$}
\State $\boldsymbol x_\text{t} = \boldsymbol X_{\boldsymbol q_\text{new}, i}$ \Comment{Extract $i^{\text{th}}$ point in $\boldsymbol X_{\boldsymbol q_\text{new}}$}
\State $\boldsymbol d = \boldsymbol x_\text{t} - \boldsymbol x_\text{r}$ \Comment{Compute displacement between $i^{\text{th}}$ points in $\boldsymbol X_{\boldsymbol q_\text{ref}}, \boldsymbol X_{\boldsymbol q_\text{new}}$}.
\If{$\boldsymbol x_\text{r},  \boldsymbol x_\text{t} \in \Gamma_a$}:
\State $\boldsymbol D \gets \boldsymbol D \cup  \boldsymbol d$

\ElsIf{$\boldsymbol x_\text{r},  \boldsymbol x_\text{t} \in \Gamma_0$}:
\State $\boldsymbol D \gets \boldsymbol D \cup  \boldsymbol 0$

\ElsIf{$\boldsymbol x_\text{r},  \boldsymbol x_\text{t} \in \Gamma_k$}:
\State $\boldsymbol d_\text{uni} = \left[0, 0, 0 \right]$
\State $\boldsymbol d_{\text{uni},k} = \boldsymbol d_k$ \Comment{Replace $k^\text{th}$ element of $\boldsymbol d_\text{uni}$ with $k^\text{th}$ element of $\boldsymbol d$.}
\State $\boldsymbol D \gets \boldsymbol D \cup  \boldsymbol d_\text{uni}$
\Else:
\State Raise error. 
\EndIf

\EndFor
\State \Return $\boldsymbol D$.
\end{algorithmic}
\end{algorithm}

\begin{algorithm}
\caption{Applying an additional morph so that a constraint is respected.}\label{alg:successive_disp}
\begin{algorithmic}
\Require $\boldsymbol X_{\boldsymbol q_\text{ref}}$ , $\boldsymbol X_{\boldsymbol q_\text{new}}$, $\Gamma_a$, $\Gamma_0$, $\Gamma_k$, $\boldsymbol X_M$.
\State $\boldsymbol D \coloneqq  \text{ComputeDisplacement}\left(  \boldsymbol X_{\boldsymbol q_\text{ref}}, \boldsymbol X_{\boldsymbol q_\text{new}}, \Gamma_a, \Gamma_0, \Gamma_k \right) $.
\State $H_\text{RBF} = H_\text{RBF}( \boldsymbol X_{\boldsymbol q_\text{ref}}, \boldsymbol D)$. 
\State $\boldsymbol D_M = H_\text{RBF} \left( \boldsymbol X_M \right)$.
\State $\Hat{\boldsymbol X}_M = \boldsymbol X_M + \boldsymbol D_M$.
\If{\texttt{valid} = \texttt{True}}: 
	\State \Return $\Hat{\boldsymbol X}_M$.
\Else: 
	\State Raise error. 
\EndIf
\end{algorithmic}
\end{algorithm}

\subsection{Evaluating Mesh Quality} \label{sec:mesh_qual_defs}

In this study we focus on unstructured triangular (2D) and tetrahedral (3D) meshes, and we require that meshes are valid and of acceptable quality. 
A valid mesh is defined as a mesh with no ``inverted'' cells; an inverted cell has (i) edges that have crossed each other, resulting in a self-intersecting geometry, and or (ii) has an cell edge with zero length.
For a mesh to be valid, each cell must be such that the coordinate transformation from physical (global) coordinates to ``local'' (cell-based) coordinates must be invertible everywhere in the cell. 
Given a valid mesh, we measure the mesh quality using several metrics. 
The aspect ratio (AR) of a cell is defined as the ratio of the largest edge length to the inner-sphere radius of the cell; it has an ideal value of 1, and the higher the AR value, the more elongated and lower quality the cell. 
The minimum angle (MA) metric is the minimum angle between edges (for triangular cells) or faces (for tetrahedral cells); it has an ideal value of $60^{\circ}$, and lower values indicate lower quality cells. 
Finally, the scaled Jacobian (SJ) metric is the ratio of the Jacobian to the maximum of the products of edge lengths adjacent to each vertex; it has an ideal value of 1, with lower values indicating lower cell quality. 

\section{Dynamic Rupture Simulation Application: Varying Dip of a Planar Normal Fault} \label{sec:dr_examples}

In this section we demonstrate the application of the mesh morphing approach to 3D dynamic rupture simulations performed using SeisSol, a freely available open-source software for simulating dynamic rupture and seismic wave propagation \cite{seissol}. 
SeisSol enables realistic simulation of seismic wave propagation generated by a dynamic rupture earthquake source, which is governed by a frictional constitutive law on geometrically complex faults \cite<e.g.,>{ramos_et_al_2022}. 
SeisSol is based on the Arbitrary high-order accurate DERivative Discontinuous Galerkin method (ADER-DG), has been verified against a range of community benchmarks \cite{kaser_dumbser_2006, pelties_gabriel_ampuero_2014} and is optimized for HPC infrastructure \cite<e.g.,>{heinecke_et_al_2014,uphoff_et_al_2017}. We use polynomial basis functions of degree 3, ensuring $\mathcal{O}4$ accuracy of the seismic wavefield in space and time.

\begin{figure}
     \centering
     \includegraphics[width=\textwidth]{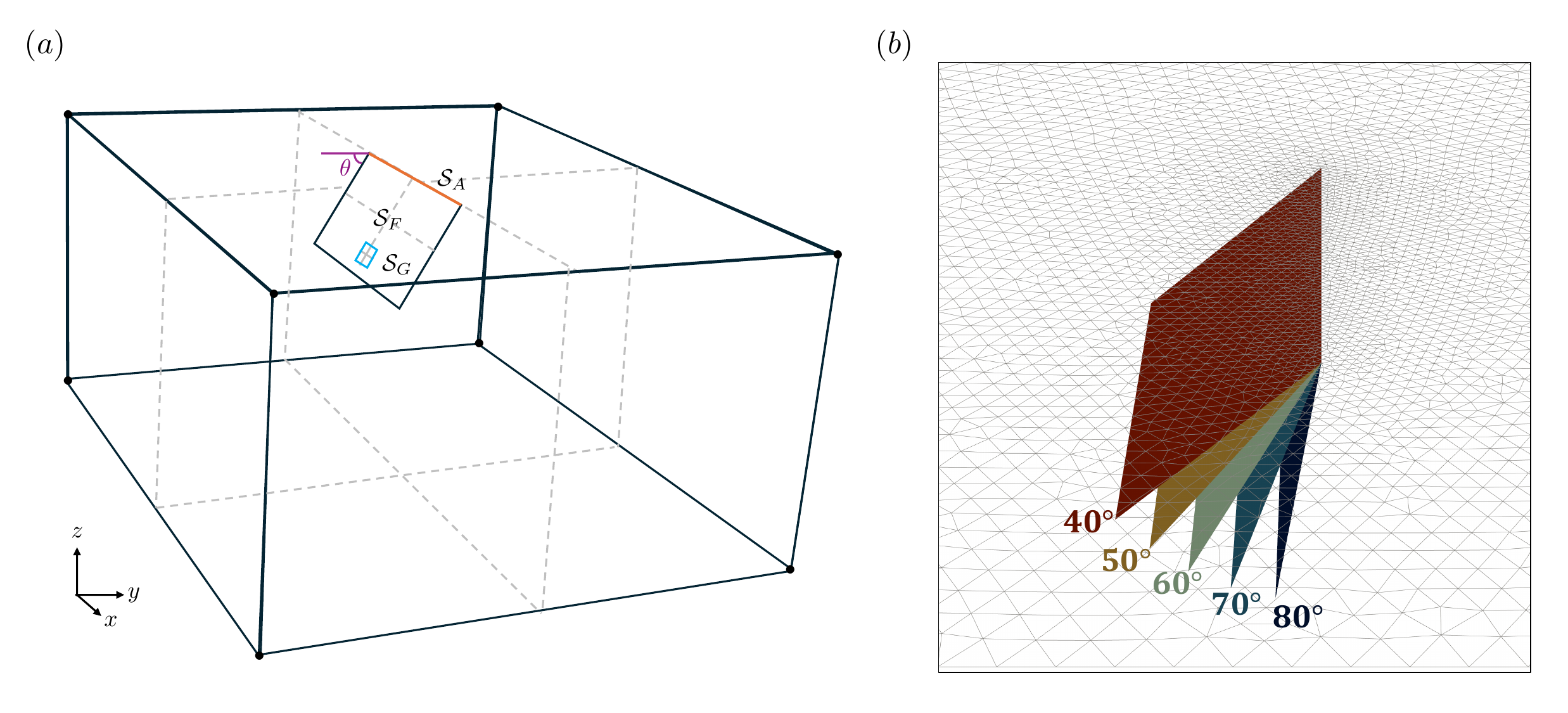}
     \caption{Subpanel (a): the TPV13 geometry. The fault trace is highlighted in orange, the boundary of the nucleation patch is in blue, and the fault dip is in purple. The union of all outer surfaces is denoted $\mathcal{S}_A$, while the surface defining the embedded fault plane is $\mathcal{S}_F$, and the nucleation patch is $\mathcal{S}_G$, where $\mathcal{S}_G \subset \mathcal{S}_F$. 
     Subpanel (b): Fault planes for different $\theta$ values, with the surface mesh shown as a light gray wireframe.}
     \label{fig:combined_DR_schematics}
\end{figure}

We present mesh morphing results for the SCEC/USGS TPV 13-3D benchmark exercise \cite{harris_et_al_2018}; we will later extend this example to include reduced-order modeling results in Section~\ref{sec:ROM}.
TPV13 describes spontaneous earthquake rupture on a 2D planar normal fault at 60$^\circ$ dip, with initial stress conditions that lead to supershear rupture, and governed by linear slip-weakening friction \cite{ida_1972,palmer_et_al_1973,andrews_1976}. The model includes off-fault plasticity governed by a non-associative Drucker-Prager visco-plastic rheology \cite{andrews_2005,wollherr_et_al_2018}. 
The TPV13 geometry is shown in Figure~\ref{fig:combined_DR_schematics}(a) and we report the parameters we used in Table~S3. 
We define $\mathcal{S}_A$ to be the union of all outer surfaces of the geometry, while the surface defining the embedded fault plane is $\mathcal{S}_F$, and the nucleation patch is $\mathcal{S}_G$, where $\mathcal{S}_G \subset \mathcal{S}_F$. 
The fault plane has a dip $\theta$, which is our geometric parameter, thus $\boldsymbol q = ( \theta )$. 
Fault planes for selected values of $\theta$ is shown in Figure~\ref{fig:combined_DR_schematics}(b).
Our aim is to morph the entire 3D mesh such that the embedded planar fault has different dip angles. 
We select $\boldsymbol q_\text{ref} = 60^\circ$, and $\boldsymbol Q = [40^\circ, 80^\circ]$. 
Let $\Gamma_a = \{ \mathcal{S}_F \}$, while $\Gamma_0 = \{ \mathcal{S}_A \} $ and $\Gamma_k = \{ \emptyset \}$. 
Since the union of all outer surfaces of the geometry, $\mathcal{S}_A$, is prescribed to have zero displacement, the volume of the domain will be preserved in this example. 

\begin{figure}
     \centering
     \includegraphics[width=\textwidth]{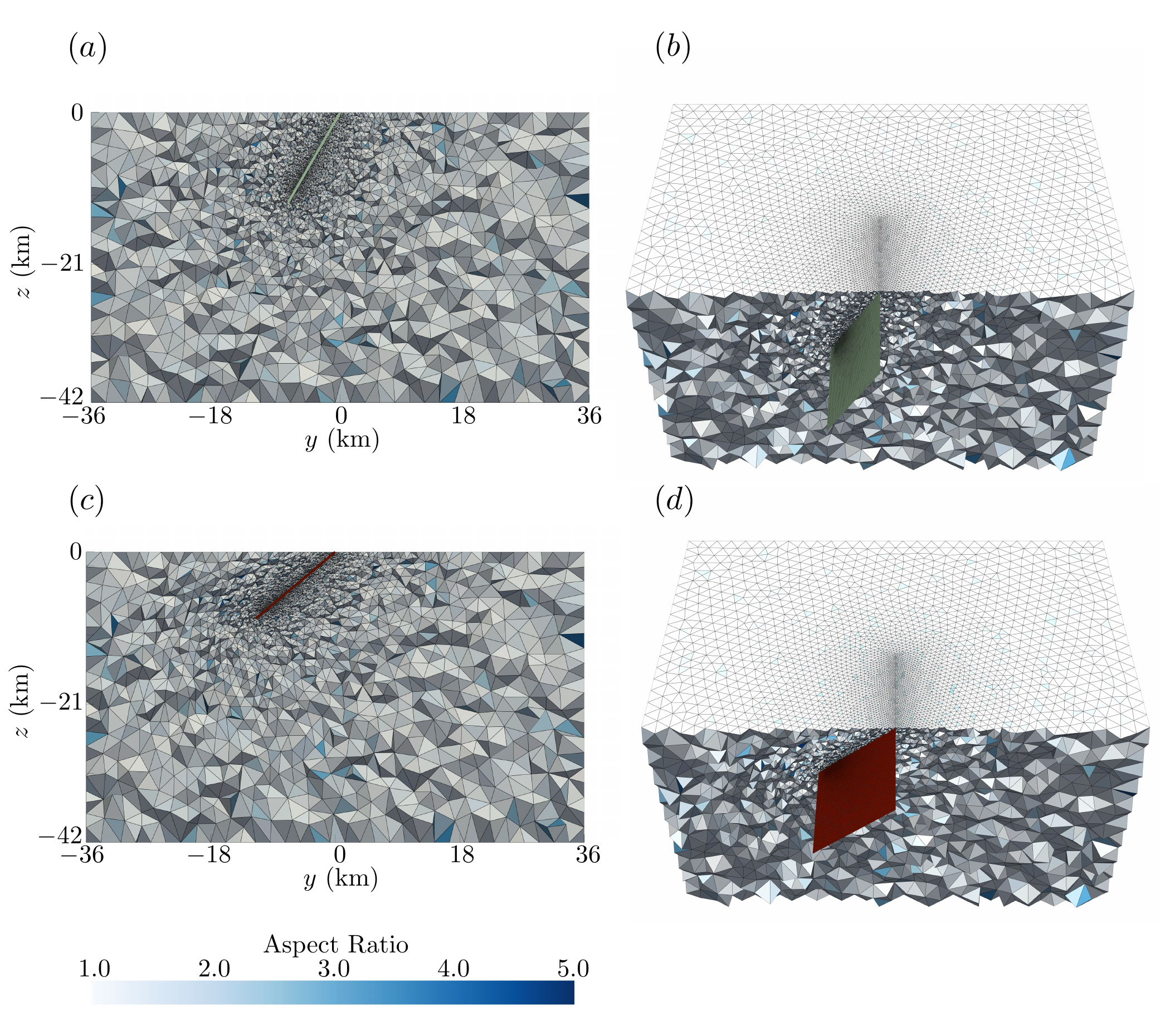}
     \caption{The reference mesh ($\theta = 60^\circ$) shown from a side on view (a) and angled view (b). The morphed mesh with $\theta = 40^\circ$ is shown in (c) and (d). Faults are colored as in Figure~\ref{fig:combined_DR_schematics}. Cells are colored by mesh quality based on the Aspect Ratio metric.}
     \label{fig:crinkle_reference}
\end{figure}

All our SeisSol simulations are performed on unstructured tetrahedral meshes, which employ adaptive coarsening away from the fault. 
We use the open-source software GMSH to generate a reference mesh $M$ with vertices $\boldsymbol X_M$, locally refined to have a mesh size of 100~m in the nucleation patch, 250~m on the fault, and 2.5~km far from the fault \cite{gmsh}. 
Finally, we select 18 $\boldsymbol q_\text{new}$ values, uniformly sampled in $\boldsymbol Q$. 
With these components defined and passed to Algorithm~\ref{alg:mmm}, we call Algorithm~\ref{alg:mm_eval} to obtain morphed meshes with varying fault dips and outer boundaries that have not been displaced. 
Figure~\ref{fig:crinkle_reference} shows the reference mesh (a-b), and a morphed mesh with $\theta = 40^\circ$ (c-d). 

The morphing process can result in the embedded fault becoming non-planar, and even slight non-planarity of the fault (in particular, at the nucleation patch) affects the dynamic rupture. 
We use Algorithm~\ref{alg:successive_disp} to apply an additional constraint after morphing the mesh to ensure that the morphed fault is nearly exactly planar. 
Algorithm~\ref{alg:successive_disp} is called, with $\boldsymbol X_{q_\text{ref}}$ defined as the points in $\Hat{\boldsymbol X}_M$ that lie in $\mathcal{S}_F$. 
A plane is defined that goes through each corner of the fault, and each point on the fault is projected onto that plane, giving the points $\boldsymbol X_{q_\text{new}}$. 
As before, $\Gamma_a = \{ \mathcal{S}_F \}$, while $\Gamma_0 = \{ \mathcal{S}_A \} $ and $\Gamma_k  = \{ \emptyset \}$. 
Applying this correction results in morphed meshes with faults that have points within 1 m of a plane fitted through them. 
It also results in dynamic rupture output that more closely matches output from simulations run on an exactly meshed fault at a given dip. 

Another consequence of the morphing is that the area of the nucleation patch does not exactly match that of an exactly generated mesh, which can affect rupture dynamics \cite<e.g.,>{galis_et_al_2014}.
Each exactly generated mesh will have a nucleation patch with an area of 9 km$^2$, while the morphed meshes can have slightly larger nucleation patches, by up to 0.03 km$^2$.
Given the close match of rupture dynamics with exact meshes (Sec. \ref{sec:DR_mesh_qual_accuracy}), we consider this an acceptable amount of variation, but this effect should be kept in mind for future studies, and handled with additional corrections if required.  

The initial stress is prescribed as a Cartesian stress tensor, hence the initial stress resolved on the fault does change as the dip changes.
As a result, we find that this inhibits rupture at dip angles beyond $3^{\circ}-5^{\circ}$ of the reference dip angle of $60^{\circ}$. 
If all parameters are kept the same as the benchmark description, we find that earthquake rupture only nucleates on faults with $\theta \in [55^\circ, 63^\circ]$. 
To examine dynamic rupture output in a wider $\theta$ interval, we decrease the static friction coefficient inside the nucleation patch from the benchmark value of 0.54 to 0.48. 
This results in earthquake rupture nucleating on faults with $\theta \in [50^\circ, 70^\circ]$, illustrating the trade-offs of fault geometry, prestress, fault strength and nucleation size. 

\subsection{Mesh Quality and Accuracy} \label{sec:DR_mesh_qual_accuracy}

As described in Section~\ref{sec:mesh_qual_defs}, we measure the mesh quality of the morphed dynamic rupture meshes via several metrics, and compare to the reference mesh. 
Figure~\ref{fig:crinkle_reference}(c-d) shows the mesh quality for a morphed mesh with $\theta =  40^\circ$.
The mesh quality for morphed meshes as $\theta$ varies is reported for all tetrahedra in the volume in Table~\ref{table:mesh_qual_volume_theta} and all triangles on the fault plane in Table~S2. 
Overall, we find that there is little variation in mesh quality relative to the reference mesh as $\theta$ varies. 
For example, maximum AR varies between 5 and 6; for SeisSol, AR $< 20-40$ is typically considered acceptable \cite{seissol}. 
Therefore we conclude that meshes can be morphed in a wide range of fault dips, $\theta \in [40^\circ, 80^\circ]$, while maintaining acceptable mesh quality standards. 

We next assess the accuracy of the morphed meshes, meaning the distance between the faults in morphed and exactly generated meshes. We find that for $\theta \in [40^\circ, 80^\circ]$, the maximum distance between exactly meshed and morphed faults is typically 140 - 150 m. 
This is similar to the mesh resolution near the nucleation patch (100~m) and less than the mesh resolution on the fault away from the nucleation patch (250~m).
Though  higher accuracy could be obtained by including more values in $\boldsymbol q_\text{new}$, we consider this to be an acceptable accuracy for the fault location. 

\begin{table}
\begin{center}
\caption{Mesh quality metrics for the full mesh volume as $\theta$ varies.}
\label{table:mesh_qual_volume_theta}
\begin{tabular}{  l l l l l l l }
\hline
~ & \multicolumn{2}{l}{Aspect Ratio}& \multicolumn{2}{l}{Scaled Jacobian} & \multicolumn{2}{l}{Minimum Angle ($^{\circ}$)} \\
$\theta$ & Avg & Max & Avg & Min & Avg & Min \\
\hline
40$^\circ$ & 1.6 & 6.35 & 0.6  & 0.11 & 47.99 & 9.06 \\
50$^\circ$ &  1.55 & 5.19 & 0.62  & 0.13 & 49.06 & 11.21 \\
60$^\circ$ (ref) &  1.5 &  5.08 & 0.63  & 0.15 & 49.41 & 10.74 \\
70$^\circ$ & 1.56 & 5.36 & 0.62  & 0.15 & 49.14 & 9.61 \\
80$^\circ$ & 1.56 & 5.36 & 0.62  & 0.15 & 49.14 & 9.61 \\
  \hline
\end{tabular}
\end{center}
\end{table}

\subsection{Efficiency and Computational Resources} \label{sec:dr_eff_comp_res}

The computational resources required for morphing are low - all calculations are performed on a personal laptop (MacBook Pro 2020 with a 2 GHz Quad-Core Intel Core i5 processor and 16 GB memory).
The reference mesh for this example has 116,657 vertices and 744,505 cells. 
Performing the steps in Algorithm~\ref{alg:mmm}, including building the first RBF interpolant, takes 4.6 s. 
Morphing the mesh by performing the steps in Algorithm~\ref{alg:mm_eval} takes 17~s
and applying the correction to obtain a planar fault takes a further 80~s. 
The correction is time-consuming because it involves building an RBF interpolant that takes every vertex on the fault as data - more than twice as many points as used to build the RBF in Algorithm~\ref{alg:mm_eval}. 

Creating the reference and exactly generated comparison meshes takes 50-55 s per mesh using GMSH version \verb|4.10.1| \cite{gmsh}.
The main motivation behind using mesh morphing is to automatically, without manual user input, create ensembles of geometrically varying meshes with the same connectivity, not to create meshes more quickly than by using GMSH. 
We note the efficiency of the mesh morphing to illustrate that many samples can be drawn in parameter space in a reasonable amount of time. 

\subsection{Verification of Simulation Output} \label{sec:verification_sim_output}

\begin{figure}
     \centering
     \includegraphics[height=0.8\textheight]{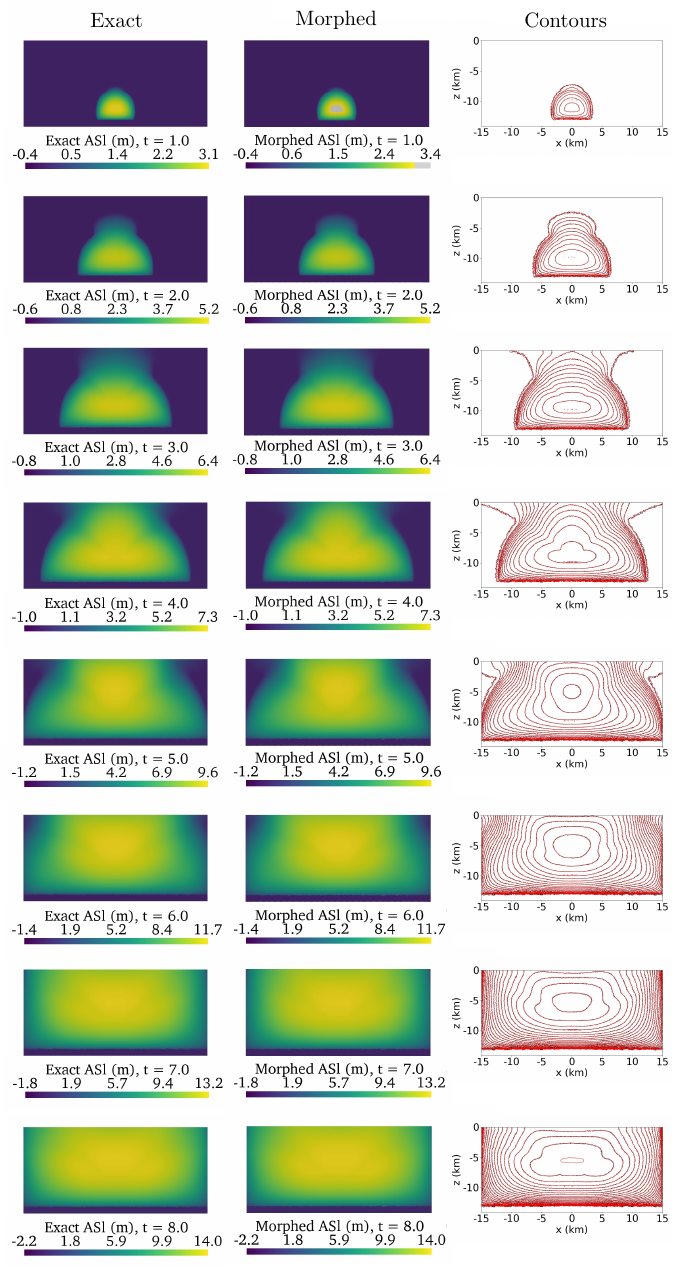}
     \caption{Accumulated slip (ASl) on the fault for $\theta = 70^\circ$. The left column shows output from exactly generated meshes, while the center column shows output from morphed meshes. The right column shows contour plots with both exact and morphed output interpolated to the same mesh, with exact mesh output contours in black and morphed mesh output contours in red.}
     \label{fig:subplots_ASl_dip_70}
\end{figure}

Beyond assessing the mesh quality and accuracy, we also require that simulations on morphed meshes produce output that agrees well with output obtained by running the same simulation on an exactly generated mesh.
We compare on-fault and off-fault simulation output for a range of different simulation output quantities. 
A comparison of on-fault accumulated slip is shown in Figure~\ref{fig:subplots_ASl_dip_70}, while peak slip rate on the fault is shown in Figure~S3.
These results are for $\theta=70^\circ$, with similar or better agreement observed for other $\theta$ values. 
We find that the morphed meshes give simulation output which agrees well with that obtained from simulations calculated on exactly generated meshes; there is slight over-prediction of accumulated slip of 0.3 m/s at $t=1$ s, and good agreement at other time steps, as shown in Figure~\ref{fig:subplots_ASl_dip_70}. 

\begin{figure}
     \centering
     \includegraphics[width=\textwidth]{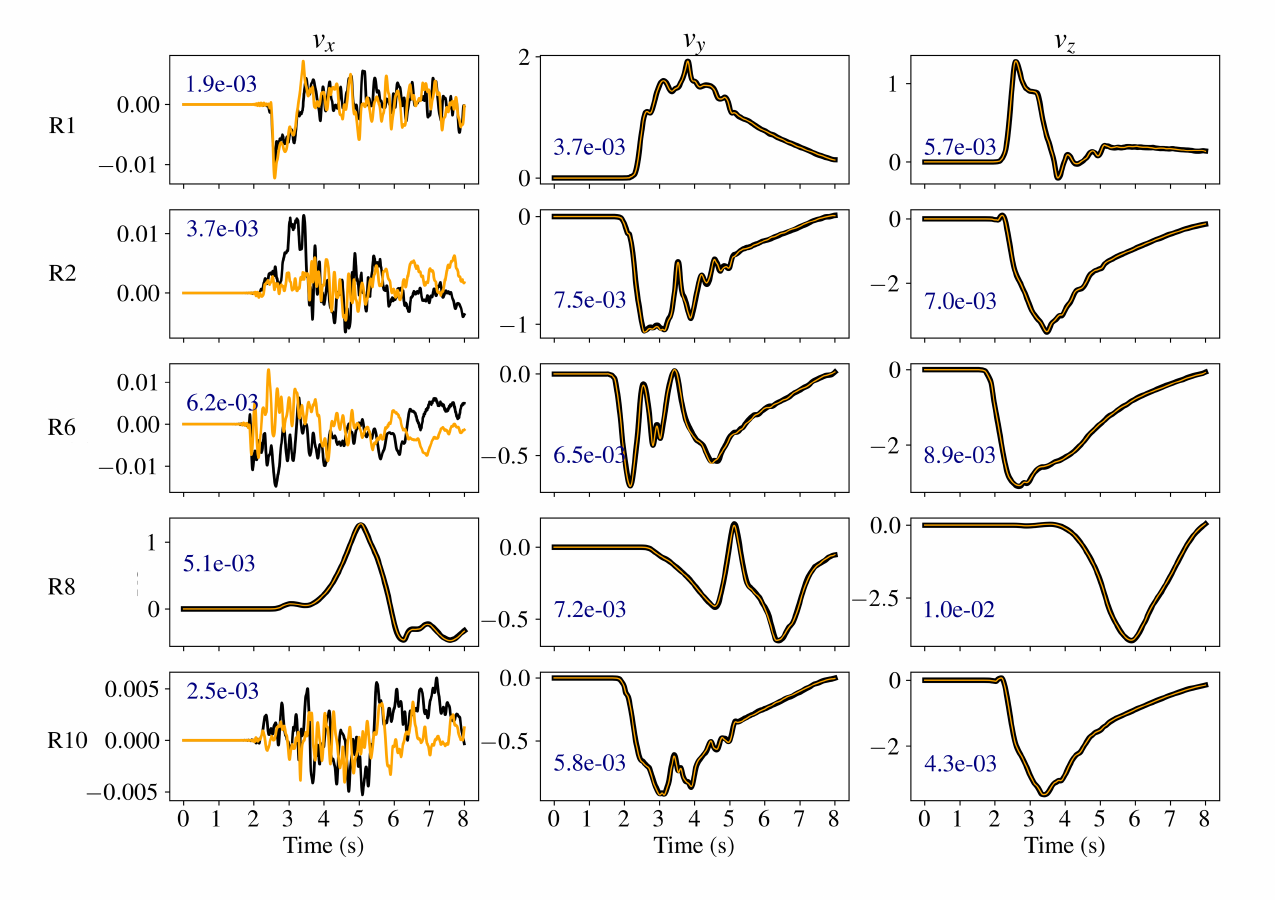}
     \caption{Measured components of velocity at 5 of the receivers for $\theta = 50^\circ$. Output from the exactly generated mesh is in black, while output from the morphed mesh is in orange.}
     \label{fig:receiver_dip_50}
\end{figure}

We also compare off-fault simulation outputs for exact vs. morphed meshes.
Off-fault receivers placed on the surface and at depth record the time history of the stress tensor and particle velocities during the simulation (see Table~S1 and Figure~S2). 
Velocities recorded by 5 receivers are compared in Figure~\ref{fig:receiver_dip_50} for $\theta = 50^\circ$, with similar agreement shown for $\theta = 70^\circ$ in Figure~S4. 
We find that the morphed and exactly generated meshes produce similar velocity recordings at all 12 receivers, with strong agreement for $v_y$ and $v_z$. 
Though there are some differences for the $v_x$ recordings (excepting receivers R7 and R8), the magnitude of velocity is much smaller. 
The RMS error between morphed and exact mesh velocity recordings is typically on the order of $10^{-3}$ m/s. 
We also see strong agreement between the uplift and subsidence profiles along a line perpendicular to the fault (line shown in Figure~S2 and profiles compared in Figure~S5). 
We conclude that simulations calculated on the morphed meshes produce output that agrees sufficiently well with the output from exactly generated meshes. 

\section{Subduction Zone Slab Interface Geometry} \label{sec:sz_int_example}

A well-established modeling approach to study the thermal structure in a subduction zone is the kinematic-dynamic approach, 
in which the kinematic behavior of the subducting slab is prescribed, and viscous flow in the mantle wedge is solved for dynamically \cite<e.g.,>{van_keken_kiefer_peacock_2002, currie_et_al_2004, wada_et_al_2008, community_benchmark_van_keken_2008}. 
Typically, when this approach is used, the slab interface geometry is treated as an input and does not vary during the simulation. 
To date, to assess the effect of the slab interface geometry on the subduction thermal structure, one must run models on different meshes with different slab interface geometries.

In this section we use mesh morphing to apply geometric variability to a thermal model mesh. 
Mesh connectivity is maintained, so a ROM could be constructed using model output, though we leave the ROM construction and full sensitivity analysis to a future study.
Here, we focus on two examples representing different aspects of slab interface variability.
In the first, we explore variation in slab interface curvature representative of the global variation across subduction zones. 
In the second, we apply geometric variation within the depth uncertainty range estimated by the Slab2 subduction interface dataset \cite{slab2}, to assess the impact of depth uncertainty on the thermal structure. 
These examples demonstrate that geometric variability can be applied while maintaining robust mesh quality and respecting constraints required by the thermal models. 

\subsection{Kinematic-Dynamic Thermal Model of Subduction in 2D}

We model temperature in a subduction zone using a standard approach commonly used to understand the present day thermal structure of subduction zones  \cite<e.g.,>{van_keken_kiefer_peacock_2002, wada_et_al_2008, currie_et_al_2004, syracuse_2010, wada_wang_2009}. 
The thermal model used in this section 
is a 2D $(x, y)$, steady-state, kinematically-driven, regional-scale model of subduction.
The equations for the conservation of mass, momentum and energy are solved to obtain velocity, pressure, and temperature solutions, given a temperature- and strain-rate dependent shear viscosity $\eta(\boldsymbol u, T)$ and a shear heating source along the slab interface.
The model domain is discretized using an unstructured triangular mesh, locally refined near the slab interface, created using the same GMSH version as above, \verb|4.10.1| \cite{gmsh}.
The coupled governing equations are solved using the Finite Element Method via FEniCS version \verb|2019.1.0| \cite{fenics}.
This model was presented in \citeA{hobson_may_2025}, which contains the full description of the modeling approach and the accompanying software. 

\subsection{Description of Geometry} \label{sec:sz_geom}

\begin{figure}
     \centering
     \includegraphics[width=\textwidth]{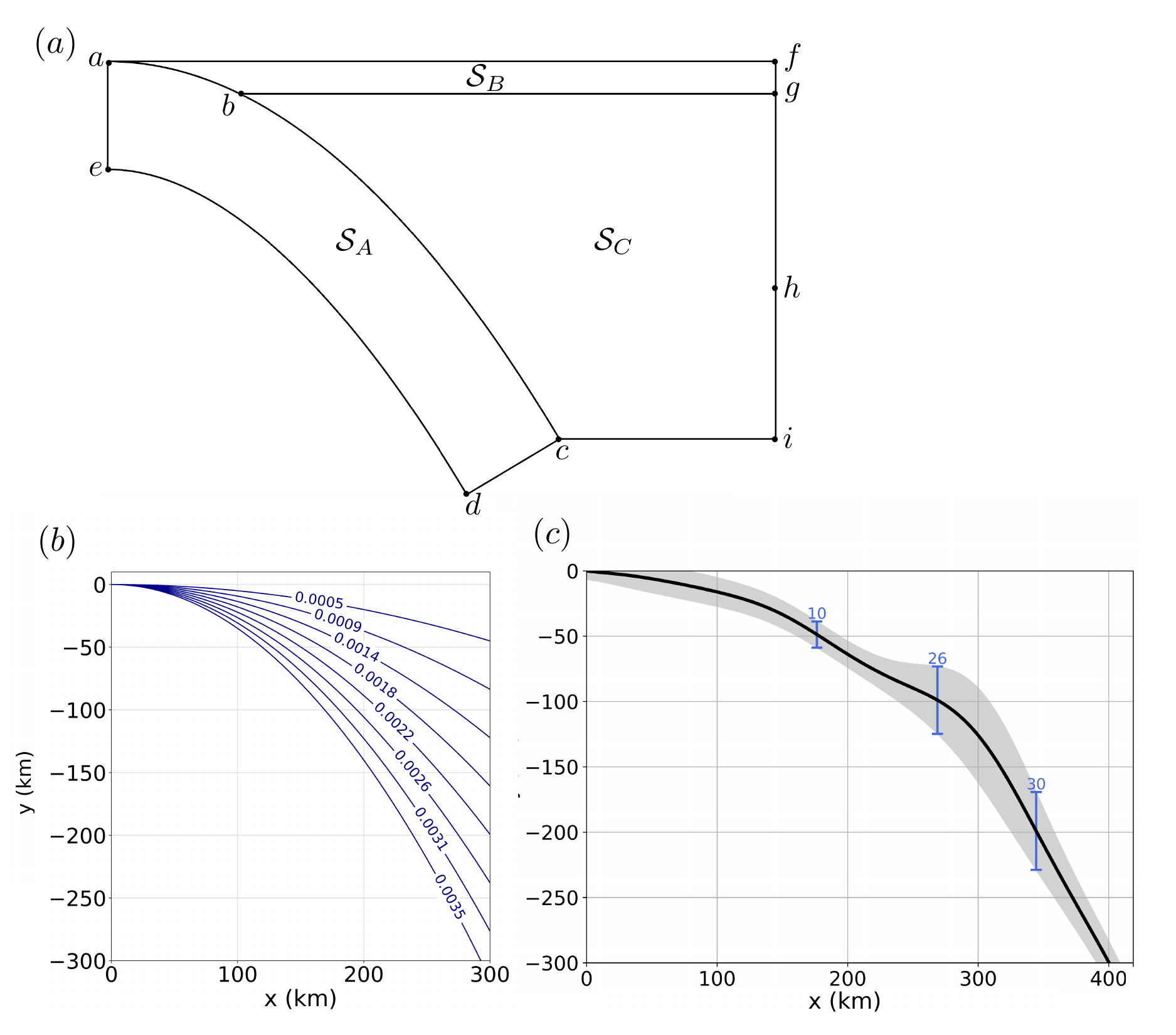}
     \caption{(a) The subdomains used in 2D subduction models, where $\mathcal{S}_A$, $\mathcal{S}_B$, and $\mathcal{S}_C$ correspond to the slab, overlying plate, and mantle wedge respectively.
     The model domain is given by $\mathcal{S}_D = \mathcal{S}_A \cup \mathcal{S}_B \cup \mathcal{S}_C$. 
     (b) The slab interfaces obtained using Equation~\eqref{eqn:polyn_slab_int} for $\alpha \in \left[ 5 \times 10^{-4},  3.5 \times 10^{-3} \right]$. 
     (c) The 2D profile of Slab2 depth data corresponding to slice B - B' of the Cascadia subduction zone (see Figure~S1), with uncertainty envelopes shown in gray. Blue error-bars highlight the uncertainty at 50, 100, and 200 km depth, with blue numbers showing the standard deviation at that depth.}
     \label{fig:combined_thermal_schematic}
\end{figure}

Both of the examples in this section use the same reference geometry, which is shown in Figure~\ref{fig:combined_thermal_schematic}(a) and is defined as follows. 
Let the points $\{a, b, c, d, e, f, g, h, i\}$ be connected by directional curves. 
Let the slab subdomain be the surface $\mathcal{S}_A$ defined by the curve loop $\mathcal{C}_A = \{ \harp{ab}, \harp{bc}, \harp{cd}, \harp{de}, \harp{ea}  \}$. 
Similarly, let the overlying plate subdomain $\mathcal{S}_B$ be defined by the curve loop $\mathcal{C}_B = \{ \harp{af}, \harp{fg}, \harp{gb}, \harp{ba}\}$.
Finally, let the mantle wedge subdomain $\mathcal{S}_C$ be defined by the curve loop $\mathcal{C}_C = \{ \harp{bg}, \harp{gh}, \harp{hi}, \harp{ic}, \harp{cb} \}$.
The entire model domain is $\mathcal{S}_D = \mathcal{S}_A \cup \mathcal{S}_B \cup \mathcal{S}_C$. 

\subsection{Global Slab Interface Variability} \label{sec:alpha_example}

This example demonstrates morphing of the slab interface geometry in a range representative of the global variation. 
The slab interface geometry is described using a parabolic expression,
\begin{linenomath*}
\begin{equation} 
y = - \alpha x^2,  \label{eqn:polyn_slab_int}
\end{equation}
\end{linenomath*}
where $y$ is the vertical coordinate of the slab interface (increasingly negative with depth) at a distance $x$ from the trench, and $(x,y) = (0,0)$ at the trench. 
This expression was shown by \citeA{england_may_2021} to represent most profiles taken from Slab2 geometries \cite{slab2}, with an RMS misfit that is less than the depth uncertainty of Slab2.
The expression in Equation~\eqref{eqn:polyn_slab_int} generates a slab interface that is concave down, with constant curvature set by the parameter $\alpha$. 
\citeA{england_may_2021} examined 154 profiles, and of these, 138 were reported to have curvature such that $5 \times 10^{-4} < \alpha < 3.5 \times 10^{-3}$. 
Figure~\ref{fig:combined_thermal_schematic} (b) shows the slab interfaces generated by several values of $\alpha$. 

In this example, we take $\alpha$ to be our geometric parameter of interest, thus $\boldsymbol q = ( \alpha )$. 
In order to be able to generate morphed meshes in the full range $\left[ 5 \times 10^{-4},  3.5 \times 10^{-3} \right]$ 
we select a slightly wider range, $\boldsymbol Q = \left[ 4.5 \times 10^{-4},  4 \times 10^{-3} \right]$.
We select a reference value of $\boldsymbol q_\text{ref} = 1.75 \times 10^{-3}$. 
While this value is not in the middle of the parameter interval, the profile it generates is approximately in the middle of the spatial geometric variation of the profiles. 
We uniformly sample in $\boldsymbol Q$ to obtain $m = 20$ vectors $\boldsymbol q_\text{new}$, where $\boldsymbol q_\text{new} \neq \boldsymbol q_\text{ref}$.  
We use Equation~\eqref{eqn:polyn_slab_int} to define slab interface curves for the $\boldsymbol q_\text{ref}$ and $\boldsymbol q_\text{new}$ values, and $\boldsymbol X_{\boldsymbol q_\text{ref}}$ and $\boldsymbol X_{\boldsymbol q_\text{new}}$ consist of points along these curves. 
The slab interface curve associated with $\boldsymbol q_\text{ref}$ is used to generate the reference mesh $M$ with vertices $\boldsymbol X_M$ using the same process as \citeA{hobson_may_2025}, with the modification that the mesh subdomain for the overlying plate is only 2 km. 

We begin with the main morphing step, which morphs the slab interface (curve $\harp{ab}~\cup~\harp{bc}$) to have a different $\alpha$ value. 
Let $\Gamma_a = \{ \harp{ab}, \harp{bc}, \harp{cd}, \harp{de}, \harp{ea} \}$, $\Gamma_0 = \{ \harp{af}, \harp{fg} \}$, and $\Gamma_k = \{ \harp{ci}, \harp{gh}, \harp{hi} \}$, where $\harp{ci}$ is morphed only in the $x$ direction and $\harp{gh}, \harp{hi}$ are morphed only in the $y$ direction. 
These components are passed to Algorithm~\ref{alg:mmm} and Algorithm~\ref{alg:mm_eval}, which returns the vertices of the morphed mesh, $\Hat{\boldsymbol X}_M$. 
We note that since the outer boundaries of the domain are morphed, the volume of the domain is not preserved in this example. 

\begin{figure}
     \centering
     \includegraphics[width=0.8\textwidth]{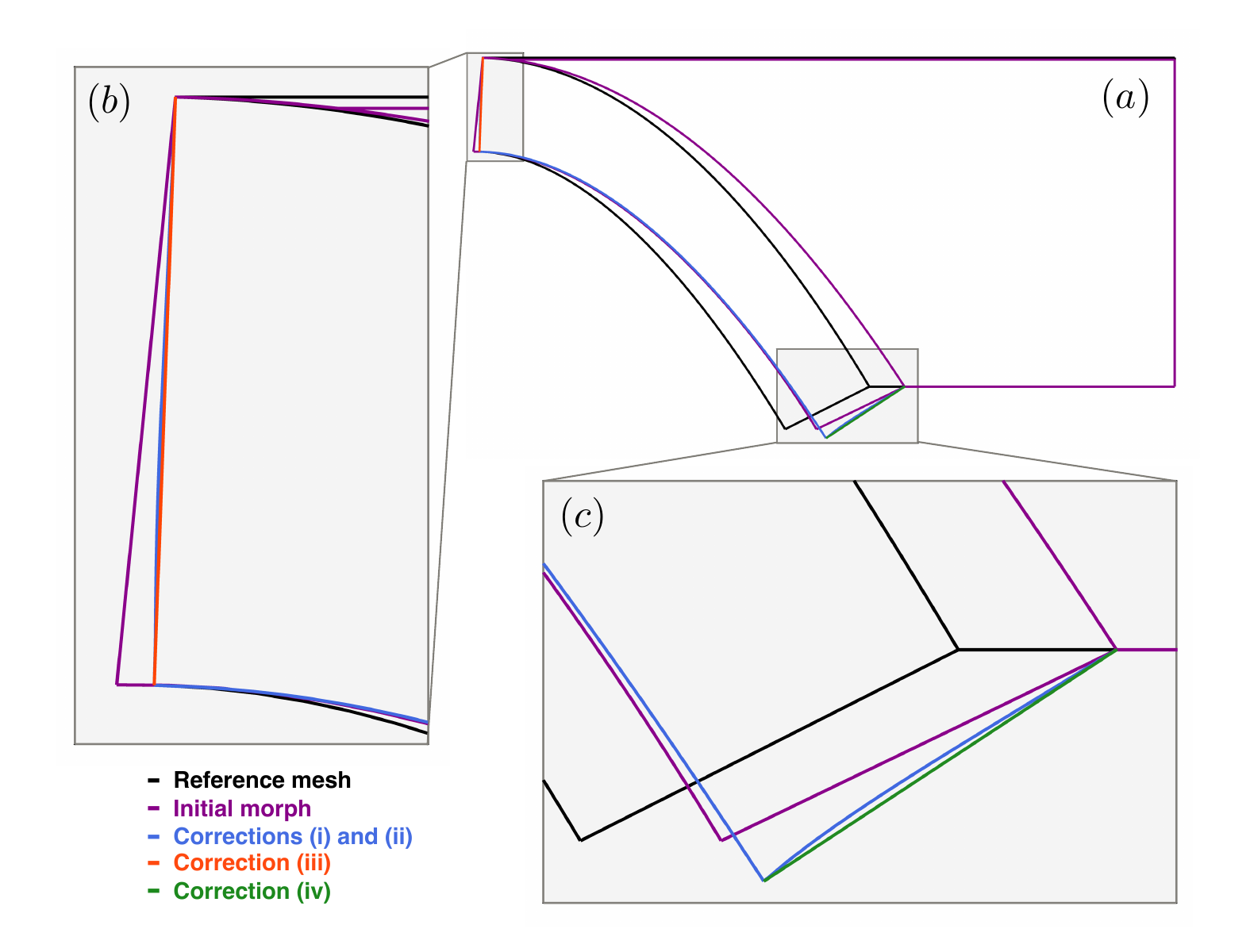}
     \caption{Boundaries of the mesh after each successive morph is applied, with the full domain in panel (a) and insets shown in panels (b) and (c). Boundaries are colored according to whether they are moved when a particular correction is applied.}
     \label{fig:successive_morphs}
\end{figure}

Upon inspection, we see that this initial morphing step produces a valid mesh with the desired $\alpha$ value, but does not satisfy several other characteristics which are required for the subduction thermal model application. 
Specifically, (i), the slab must have a specific width $w_\text{slab}$ (i.e. the distance between $\harp{ab} \, \cup \, \harp{bc}$ and $\harp{ed}$ must be $w_\text{slab}$), 
(ii), the boundaries $\harp{ae}$ and $\harp{cd}$ must be normal to the slab interface $\harp{ab} \cup \harp{bc}$, 
(iii), boundary $\harp{ae}$ must be a straight line, 
and (iv), boundary $\harp{cd}$ must be a straight line. 
These conditions must be met in order to have good velocity and pressure solutions when solving the equations for conservation of mass and momentum in $\mathcal{S}_A$.
We apply three successive morphs so that each of these requirements are satisfied.
The mesh boundaries after each of the successive morphs is shown in Figure~\ref{fig:successive_morphs}. 

To morph such that conditions (i) and (ii) are satisfied, let $\Gamma_a = \{ \harp{cd}, \harp{de}, \harp{ea} \}$ and $\Gamma_0 = \{ \harp{af}, \harp{fg}, \harp{gh}, \harp{hi}, \harp{ic}, \harp{cb}, \harp{ba}, \harp{bg} \}$;
in short, the boundaries of the slab subdomain $\mathcal{S}_A$, excepting the interface, $\harp{ab}  \cup  \harp{bc}$, will be morphed, while all other boundaries in the geometry will be held fixed. 
Let $\boldsymbol X_{\boldsymbol q_\text{ref}}$ be points evenly spaced (in terms of distance along-curve) along $\harp{ed}$. 
For a set of similarly evenly spaced points along $\harp{ab} \, \cup \,  \harp{bc}$, compute the positions of a set of points $\boldsymbol X_{\boldsymbol q_\text{new}}$, each of which is a distance $w_\text{slab}$ normal to $\harp{ab} \cup \harp{bc}$. 
These components get passed to a call to Algorithm~\ref{alg:successive_disp}; the returned morphed mesh has a slab of uniform width $w_\text{slab}$, and $\harp{ae}$ and $\harp{cd}$ are normal to $\harp{ab} \, \cup \, \harp{bc}$ (see Figure~\ref{fig:successive_morphs}, blue curves in (b-c)). 

To satisfy condition (iii), let $\Gamma_a = \{ \harp{ea} \}$, while $\Gamma_0 = \{ \harp{ab}, \harp{bc}, \harp{dc}, \harp{de}, \harp{af}, \harp{fg}, \harp{gh}, \harp{hi}, \harp{ic}, \harp{bg} \}$ (all curves that are not $\harp{ea}$), and $\Gamma_k = \{ \emptyset \}$. 
Let $\boldsymbol X_{\boldsymbol q_\text{ref}}$ be points evenly spaced along $\harp{ea}$.
The target points $\boldsymbol X_{\boldsymbol q_\text{new}}$ are points evenly spaced along a straight line drawn between points $e$ and $a$. 
The point $e$ is already at a position $w_\text{slab}$ normal to the interface, due to the previous constraint. 
Again, these components are passed to Algorithm~\ref{alg:successive_disp}, which returns a morphed mesh with a curve $\harp{ea}$ which is straight and normal to $\harp{ab} \, \cup \, \harp{bc}$ (see Figure~\ref{fig:successive_morphs}(b), orange curve).
To satisfy condition (iv), we then apply the same process but with $\Gamma_a = \{ \harp{cd} \}$ instead of $\Gamma_a = \{ \harp{ea} \}$ (see Figure~\ref{fig:successive_morphs}(c), green curve).
The result is a morphed mesh which satisfies constraints (i)-(iv), and is suitable for use in the subduction zone thermal model.

\subsubsection{Mesh Quality}

\begin{figure}
     \centering
     \includegraphics[width=0.8\textwidth]{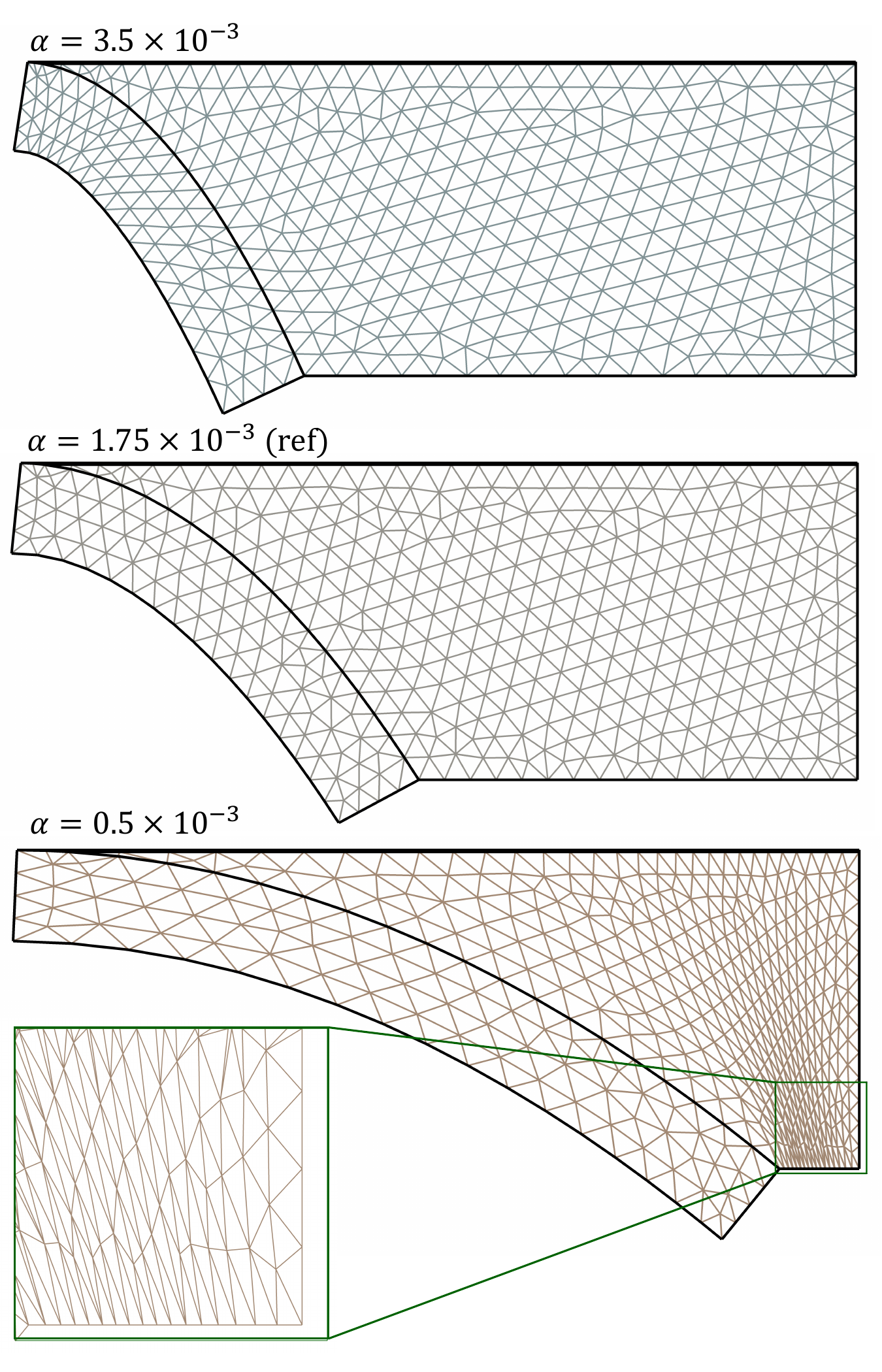}
     \caption{The mesh morphing method applied to a coarse subduction zone reference mesh for different values of $\alpha$.
     Middle panel: Reference mesh using $\alpha = 1.75 \times 10^{-3} = \alpha_\text{ref}$ with cell diameter of $\sim$30 km. 
     Top panel: $\alpha = 3.5 \times 10^{-3}> \alpha_\text{ref}$. 
     Bottom panel: $\alpha = 0.5 \times 10^{-3} < \alpha_\text{ref}$.      
     In the bottom panel, the inset highlights that high-aspect-ratio cells are generated by morphing when using a low value of $\alpha$.}
     \label{fig:demo_mesh_coarse}
\end{figure}

We show an example of the mesh morphing process applied to a coarse mesh in Figure~\ref{fig:demo_mesh_coarse}, to demonstrate how mesh morphing affects the shape of triangular mesh cells. We see that higher $\alpha$ values result in modest stretching and compression of cells relative to the reference mesh, while smaller $\alpha$ values result in high-aspect-ratio cells in the highly compressed lower right corner of the mesh. 

\begin{figure}
     \centering
     \includegraphics[width=\textwidth]{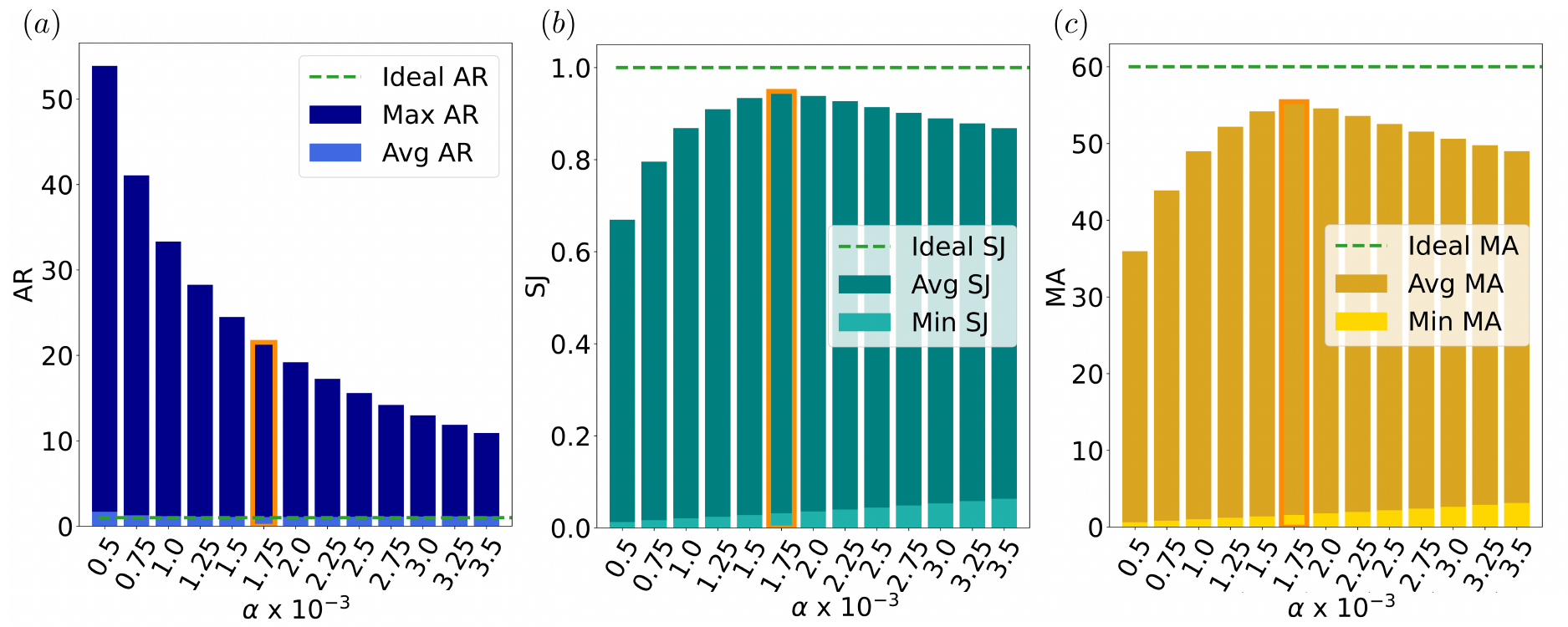}
     \caption{Bar plots of the mesh quality as $\alpha$ varies, measured by (a) the Aspect Ratio (AR), (b) the Scaled Jacobian (SJ), and (c) the Minimum Angle (MA) metrics. The ideal values for each metric is shown by the dotted green line, and the mesh quality for the reference (not morphed) mesh is outlined in orange.}
     \label{fig:mesh_qual_barplots_alpha}
\end{figure}

We then apply mesh morphing to a reference mesh with cell sizes appropriate for making the thermal model calculations; the mesh is locally refined near the slab interface with cell diameter of 0.5 km, and coarsens to a 8 km cell diameter far from the interface.
We generate morphed meshes for 13 uniformly sampled values in $\alpha \in \left[ 5 \times 10^{-4},  3.5 \times 10^{-3} \right]$, and compute the mesh quality. 
Figure~\ref{fig:mesh_qual_barplots_alpha} shows the average and minimum or maximum (whichever represents the lowest mesh quality) values for the morphed and reference meshes. 
Similar trends are seen for the three metrics, with the morphed meshes having similar mesh quality to the reference mesh. The mesh quality is lower for lower values of $\alpha$ because these meshes have shallower dip and so the reference mesh is compressed, forcing triangular cells to have higher aspect ratios, lower scaled Jacobians, and lower minimum angles. For higher $\alpha$  values, triangular cells in the mesh are being stretched in a less exaggerated way than their compression in the low-$\alpha$ cases, so the mesh quality is similar or even marginally better than the reference mesh. There is also a notable difference between the average mesh quality and the lowest mesh quality values. For example, the maximum AR for $\alpha = 0.5 \times 10^{-3}$ is 53, but the average is 1.6, implying that most of the mesh is of high quality with a few lower-quality cells included. Figure~S6 shows the region of lower mesh quality in the lower mantle wedge subdomain. A higher-$\alpha$ example is shown in Figure~S7, which shows the overall good mesh quality with only slight lowering of mesh quality where the mesh in the lower left part of the subducting slab domain is most affected. Overall we consider the mesh quality of the morphed meshes to be sufficiently similar to the reference mesh, and sufficient to run our model. 

\subsubsection{Efficiency and Computational Resources} \label{sec:slab_eff_com_res}

As before, all morphing calculations are performed on a personal laptop (MacBook Pro 2020 with a 2 GHz Quad-Core Intel Core i5 processor and 16 GB memory). 
The reference mesh for this example has 81,376 vertices and 164,251 cells. 
It takes $< 1$ second to build the displacement RBF interpolant using the target slab interface points, and 6 - 7 seconds to perform each morph for this example. 
By comparison, creating the reference and comparison meshes takes 26 - 44 seconds per mesh using GMSH version \verb|4.10.1| \cite{gmsh}. 
As mentioned in Section~\ref{sec:dr_eff_comp_res}, the aim in using mesh morphing is not to create meshes more quickly than by using GMSH, but rather to create ensembles of geometrically varying meshes with the same connectivity. 
We make the comparison between timings for mesh morphing and exact mesh generation to illustrate the computational feasibility of drawing many samples in parameter space. 

\subsubsection{Accuracy}

\begin{figure}
     \centering
     \includegraphics[width=\textwidth]{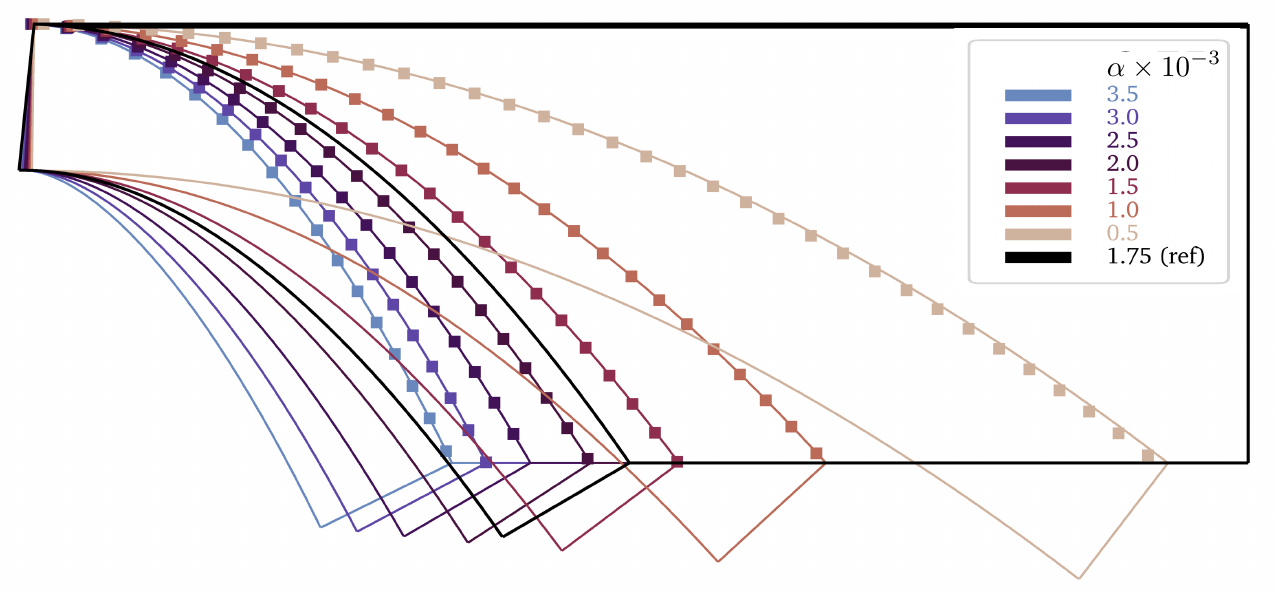}
     \caption{Boundaries of the morphed meshes as $\alpha$ varies (solid lines), with the reference mesh boundaries shown in black and the exact slab interface points as scatter plots. }
     \label{fig:variability_alpha_wireframe}
\end{figure}

\begin{figure}
     \centering
     \includegraphics[width=\textwidth]{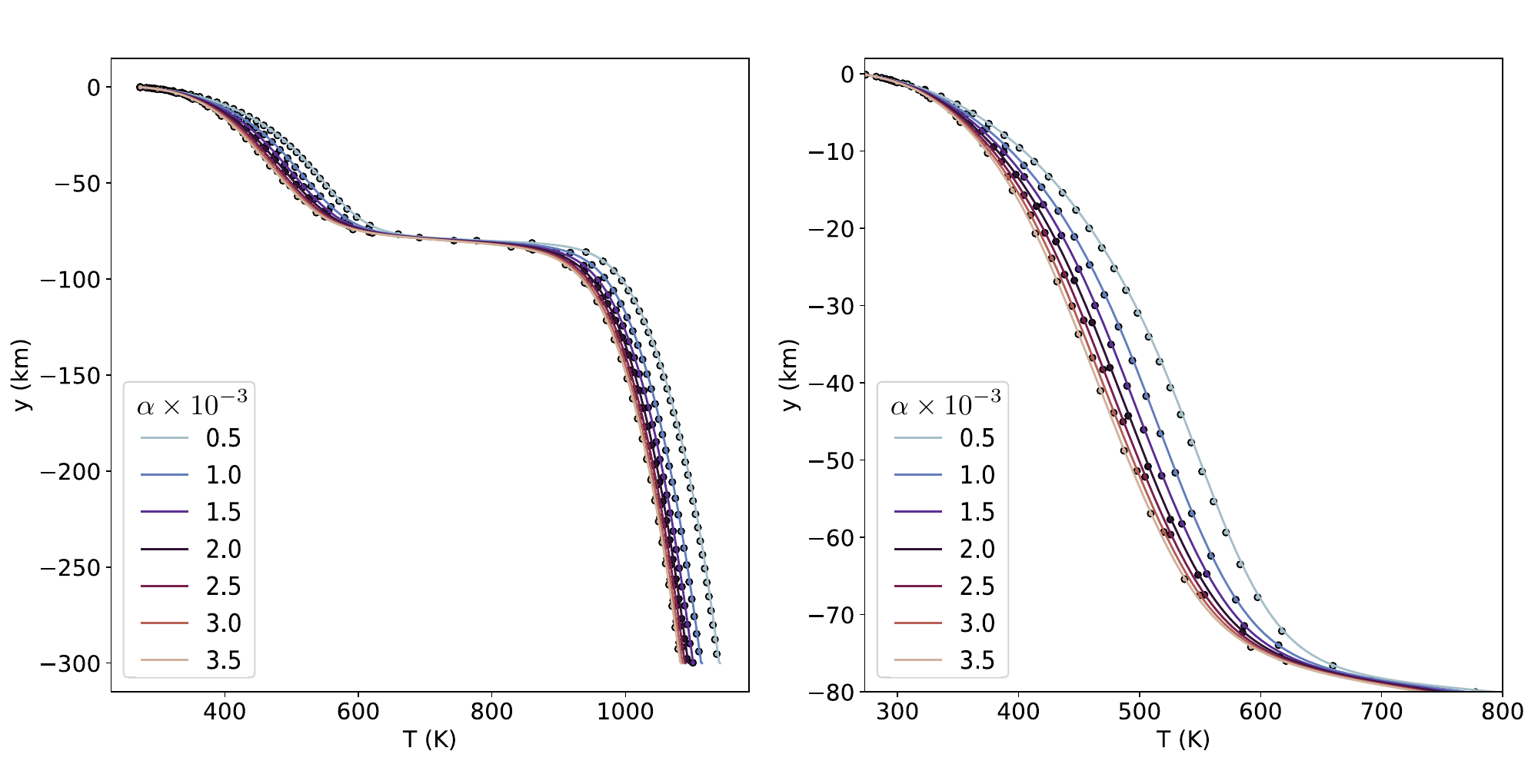}
     \caption{Temperature vs. depth curves as $\alpha$ varies, with the morphed mesh results shown in solid lines and values from the exactly generated meshes shown as scatter plots.}
     \label{fig:P_T_comparison_alpha}
\end{figure}

For the purposes of comparison, we generate meshes for the 13 $\alpha_\text{eval}$ values which are uniformly sampled in $\alpha \in \left[ 5 \times 10^{-4},  3.5 \times 10^{-3} \right]$, 
and compute the difference in where the slab interface lies. 
The maximum distance between vertices lying on the slab interface in the morphed meshes vs. the exact meshes is between 1 and 8 km for $\alpha_\text{eval}$ values in the range $[1.25 \times 10^{-3}, 0.5 \times 10^{-3}]$, and is less than 1 km for all other $\alpha_\text{eval}$ values. 
Figure~\ref{fig:variability_alpha_wireframe} shows good agreement between the morphed mesh boundaries (solid curves) and the exact slab interfaces (square points).
Ideally, the morphed slab interface should lie directly on the exactly meshed interface. 
The slight difference of at most 8 km can be seen for $\alpha = 0.5 \times 10^{-3}$, where the morphed slab interface does not correspond exactly to the exact slab interface points in the deepest region of the interface. 
If fewer $\boldsymbol q_\text{new}$ points were used during mesh morphing, this disagreement would be higher. 
We also run the thermal model on both the morphed and exact meshes and find good agreement in temperature vs. depth for morphed and exact meshes, as shown in Figure~\ref{fig:P_T_comparison_alpha}. 

\subsubsection{Temperature Variability} \label{sec:alpha_temp_var}

The slab interface temperatures in Figure~\ref{fig:P_T_comparison_alpha} are from a calculation with no shear heating; a viscosity flow law with composite diffusion and dislocation creep; a plate convergence rate of $u_s = 5.0$ cm/yr; and an age of the incoming lithosphere of $a_s = 20 $ Myr (see \citeA{hobson_may_2025} for other parameter values). 
We find that the temperature varies by up to 85 K along the slab interface as $\alpha$ varies, with all other model input parameters held fixed. 

\subsection{Variability From Slab2 Uncertainty}

In this example, we estimate the variability in temperature due to geometric variation of the subduction interface that falls within reported Slab2 depth uncertainties. 
Slab2 calculates subduction zone geometries by combining information from different data sources, including earthquake hypocenters, active source/seismic reflection, receiver functions, and tomography \cite{slab2}. 
Each data type has an associated uncertainty, and \citeA{slab2} compute and report the estimated vertical uncertainty at each location (for more detail, see Supporting Information Section~1 and the Supporting Information for \citeA{slab2}). 
Maps of Slab2 data for Cascadia and the corresponding depth uncertainty are shown in Figure~S1(a-b). 
We take a profile through the highest uncertainty part of the margin, which is shown in Figure~\ref{fig:combined_thermal_schematic}(c), and use the slab interface depth and associated uncertainty in the results in this section. 

\subsubsection{Applying Mesh Morphing}

The aim of this example is to morph the slab interface up or down by a given fraction $\beta$ of the Slab2 depth uncertainty.
So for a slab interface profile with coordinates $(x_B, y_B)$ and uncertainty values $(x_B, U_B)$, the target interface coordinates will be $(x_B, y_B + \beta U_B)$
Therefore in this example, $\boldsymbol q = ( \beta )$, and we set $\boldsymbol Q = [-1.1, 1.1]$ so that we can select $\boldsymbol q_\text{eval}$ values in the range $[-1.0, 1.0]$. 
We define $\boldsymbol q_\text{ref} = \beta_\text{ref} = 0$, and define $\boldsymbol q_\text{new}$ by selecting $m = 9$ values uniformly sampled in $\boldsymbol Q$. 
This example uses the same geometry as described in Section~\ref{sec:sz_geom}, but with a 35 km thick overlying plate and a 40 km thick slab (instead of 50 km; the slightly thinner slab is necessary to generate meshes for highly curved slab interfaces using GMSH, such as the exactly generated mesh corresponding to the $\beta = 1.0$ curve in Figure~\ref{fig:variability_beta_wireframe}). 
Let $\Gamma_o = \{ \harp{fg}, \harp{ea} \}$, while $\Gamma_a = \{ \harp{ab}, \harp{bc} \}$ as expected, and $\Gamma_k = \{ \harp{gh}, \harp{hi} \}$ where $\harp{gh}, \harp{hi}$ are displaced only in the $y$ direction. 
Note that here we are specifying only the curves which are being used to constrain the morph, and other curves could be considered to fall into these sets: 
for example, $\harp{ci}$ is displaced only in the $y$ direction, though its displacement is entirely guided by the vertical displacement of $\harp{bc}$ and $\harp{hi}$. 
Similarly, though $\harp{af}$ is not specified to have zero displacement, its displacement is linearly interpolated using values from $\harp{ea}$ and $\harp{fg}$ which have zero displacement, and therefore it too has zero displacement. 
As in Section~\ref{sec:alpha_example}, the slab interface curve for $\boldsymbol q_\text{ref} = \boldsymbol 0$ is used to generate the reference mesh $M$ with vertices $\boldsymbol X_M$.  
These components are passed into Algorithm~\ref{alg:mmm} and Algorithm~\ref{alg:mm_eval}, which returns a morphed mesh whose interface is $\beta \times $ Slab2 uncertainty away from the reference mesh interface.  
Successive morphs are then applied using Algorithm~\ref{alg:successive_disp} exactly as described in Section~\ref{sec:alpha_example}, so that the final morphed mesh satisfies conditions (i)-(iv). 

\subsubsection{Mesh Quality}

Mesh morphing is applied for 9 $\boldsymbol q_\text{eval}$ values, uniformly sampled in $[-1.0, 1.0]$. 
As before, we measure the mesh quality as $\beta$ varies and find that morphed mesh quality is generally similar to the reference mesh for most $\beta$ values; the exception is $\beta = 1.0$ which has a distinctly higher maximum Aspect Ratio of 16. 
This is due to the compression of already high-Aspect Ratio cells in the upper left corner of the mesh when the slab interface is made shallower.  
Figure~S8 shows the average and minimum or maximum (whichever represents the lowest mesh quality) values for the morphed and reference meshes, as $\beta$ varies.
The mesh quality is well within acceptable limits for this example. 
The full morphed mesh for $\beta = 1.0$ is shown in Figure~S9, which highlights the two regions with the lowest mesh quality.
The increasing curvature of the slab interface as $\beta$ increases means that cells near the highly curved interface are morphed more strongly than in other regions, and cell quality is lower as a result. 
Similarly, cells near the base of the slab are compressed as the curvature increases, decreasing cell quality. 

\subsubsection{Accuracy}

\begin{figure}
     \centering
     \includegraphics[width=\textwidth]{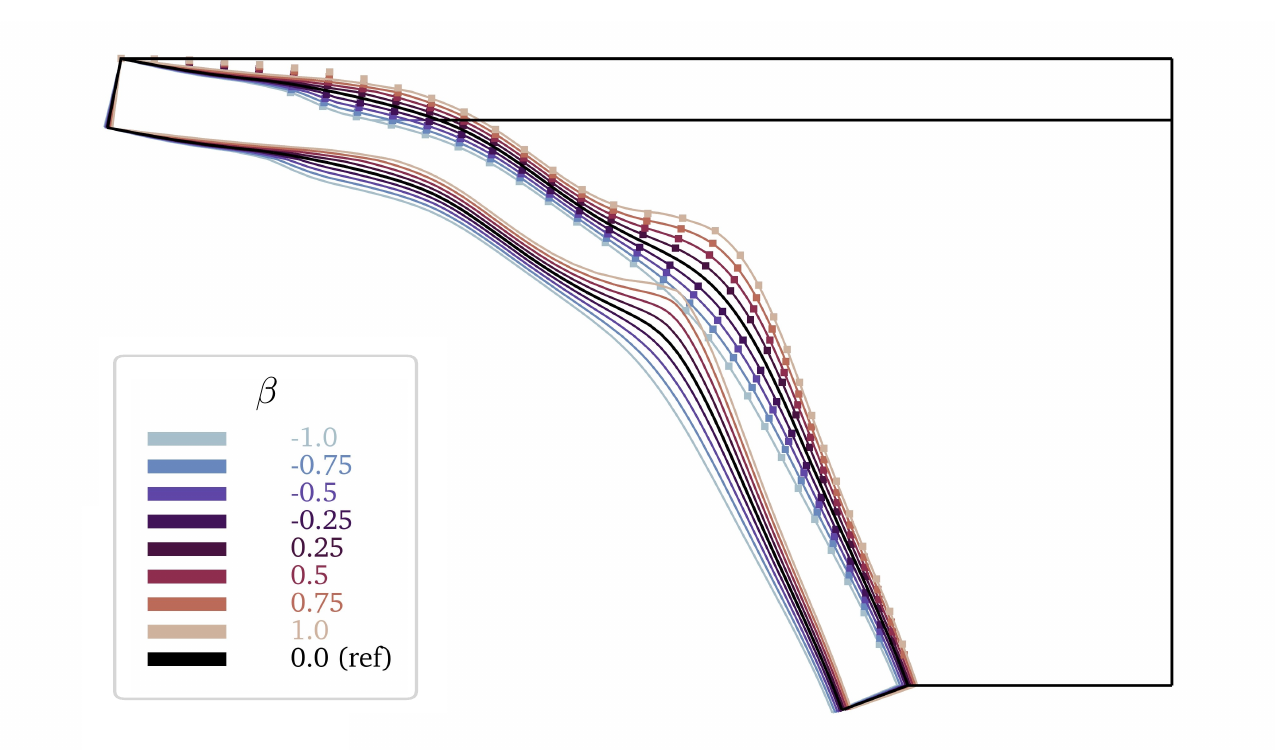}
     \caption{Boundaries of the morphed meshes as $\beta$ varies (solid lines), with the reference mesh boundaries shown in black and the exact slab interface points as scatter plots. }
     \label{fig:variability_beta_wireframe}
\end{figure}

We generate exact meshes for all of the $\beta_\text{eval}$ values and compute the difference in where the slab interface lies as compared to the morphed meshes. 
The maximum distance between vertices lying on the slab interface in the morphed meshes vs. the exact meshes is between 5 and 5.5 km for each $\beta$ value. 
This difference is found in the upper 10-20 km of the mesh, where the morphing slab interface is very close to the surface of the domain which has zero displacement; at deeper depths, the agreement between morphed and exact mesh slab interfaces is much closer. 
This is shown in Figure~\ref{fig:variability_beta_wireframe}, which visualizes the morphed mesh boundaries (solid curves) in comparison to the exact slab interfaces (square points).

\subsubsection{Temperature Variability}

The slab interface temperatures are calculated with parameters as in Section~\ref{sec:alpha_temp_var}, but with $a_s = 8 $ Myr to reflect Cascadia's relatively young age of the incoming lithosphere. 
We compute the variability in temperature at each point along the slab interface as $\beta$ varies in the range $[-1.0, 1.0]$ and find variability of as much as 40 K.
Between 0 and 150 km depths, the mean variability is 22 K.  
Below 150 km depth, the variability is much lower, with a mean variability of 3 K. 
We conclude that the uncertainty in the depth of the Slab2 interface for profile B-B' in Cascadia translates to at most a 40 K uncertainty in temperature on the slab interface. 

\section{Reduced-Order Models for Dynamic Rupture Output with Geometric Variability} \label{sec:ROM}

A central requirement of the mesh morphing approach presented in this paper is the explicit specification of a geometric parameterization, such as using the dip angle in Sec.~\ref{sec:dr_examples} $\theta$ and slab curvature $\alpha$ in Sec.~\ref{sec:alpha_example}.
Such geometric parameterizations can be naturally incorporated into surrogate and ROM frameworks, which approximate the map between parameters $\boldsymbol m = (\boldsymbol p, \boldsymbol q)$ and simulation output (see Equations~\eqref{eq:p_pde_f}).
Once constructed, surrogates and ROMs enable extremely rapid evaluation of approximate simulation output and thus are suitable for uncertainty quantification.
In the subsequent sections, we present two ROMs constructed using simulation data from the TPV13-3D model with varying fault dip angles. The first ROM predicts time series of velocity recorded at off-fault receivers, and the second ROM approximates vertical surface displacements.
Both ROMs employ the interpolated Proper Orthogonal Decomposition (iPOD) approach \cite{ly_tran_2001, bui-thanh_2003, my-ha_2007, walton}, which has recently been applied in several geophysics applications \cite{rekoske_2023,rekoske_et_al_2025,hobson_may_2025}.

\subsection{Reduced-Order Model for Velocity Time History at Receivers} \label{sec:rom_receivers}

As described in Section~\ref{sec:verification_sim_output}, the TPV13-3D model contains off-fault receivers which record time-series of ground velocities. 
We build a unique ROM for each receiver, where the data matrix contains the time history of a given velocity component recorded at that receiver in the row direction, and different dip values of the fault in the column direction.
The data matrix contains 21 FOM solutions, sampled in parameter space at every $1^{\circ}$ dip over the interval $[50^\circ, 70^\circ]$, i.e. $\theta = [50^\circ, 51^\circ, \ldots, 69^\circ, 70^\circ]$.
POD coefficients are interpolated using an RBF interpolant as in Equation~\eqref{eq:RBF} with a quintic kernel, 
$\Theta_j(\boldsymbol d) = - \| \boldsymbol d - \boldsymbol d_j \|_2^5$. 

We perform a leave-one-out cross validation (LOOCV), meaning that for each sampled point in parameter space, the corresponding FOM solution is dropped from the data matrix and a ROM built without that solution; the ROM is then used to approximate the solution at that point in parameter space, and the error between the FOM and ROM solutions is calculated.
The LOOCV error is high at the edges of the parameter interval, as might be expected since the RBF interpolation is one-sided in these regions, but lower within the interval. 
The LOOCV $L_\infty$ error for $v_x$ across all receivers is, in the worst cases, $0.09$ m/s for $\theta=50^\circ$ and $0.3$ m/s for $\theta=70$; for parameter values $>2^\circ$ away from the edges of the parameter interval, $\theta \in [52^\circ, 68^\circ]$, the error is low, at most $0.05$ m/s and typically $< 0.01$ m/s. 
For $v_y$ and $v_z$, the LOOCV $L_\infty$ error is also higher at the edges of parameter space but it is low elsewhere with at most $<0.2$ m/s for $\theta \in [52^\circ, 68^\circ]$ and typically $<0.02$ m/s. 

\begin{figure}
     \centering
     \includegraphics[width=\textwidth]{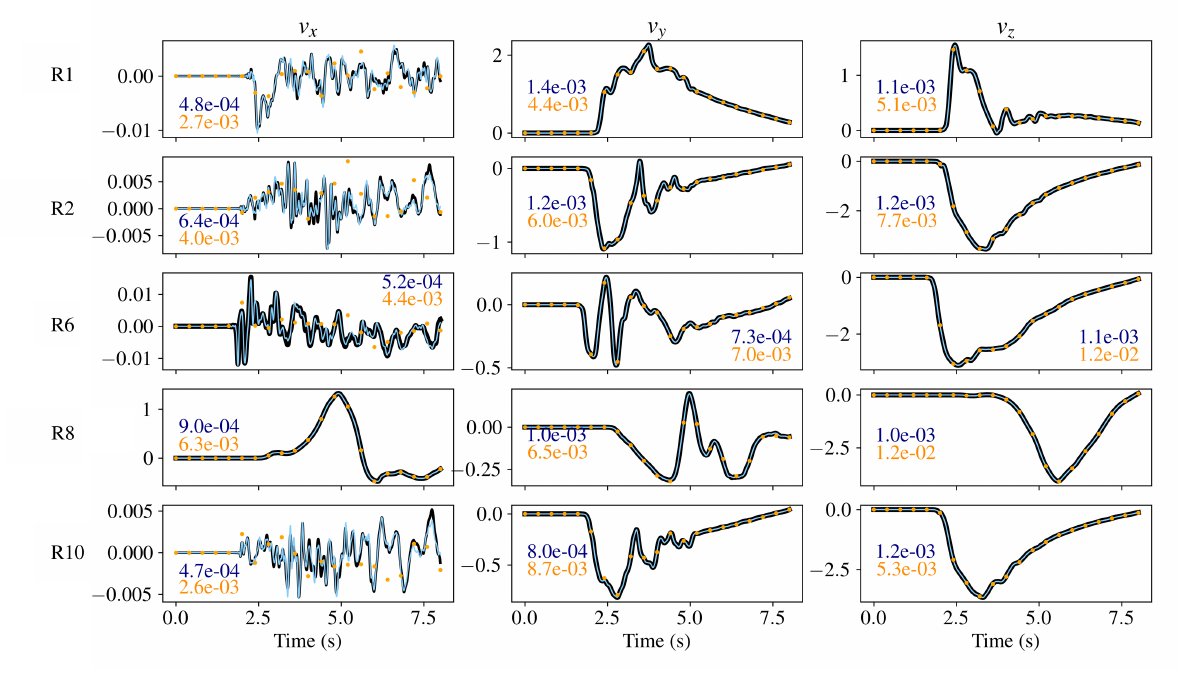}
     \caption{Time series of ground velocity at 5 of the TPV13-3D off-fault receivers for $\theta = 54^\circ$, with the simulated velocity on a morphed mesh shown in black and the ROM approximation shown in blue. Scatter points in orange show the solution for a simulation on an exact mesh with $\theta = 54^\circ$. The RMSE values between the morphed mesh snapshot and the ROM evaluation are shown in dark blue, while the RMSE between the ROM approximation and the simulation on an exact mesh is shown in orange.}
     \label{fig:rom_receiver_comparison}
\end{figure}

In Figure~\ref{fig:rom_receiver_comparison}, we show a comparison between the velocity from a dynamic rupture simulation with fault dip $\theta = 54^\circ$ and a ROM which was constructed using a data matrix where the column corresponding to $\theta = 54^\circ$ had been dropped. 
By approximating the velocity at $\theta = 54^\circ$, we obtain an estimate of the ROM accuracy. 
We find good agreement between simulated and ROM-approximated velocities for the $v_y$ (where $y$ is the horizontal, fault-perpendicular direction) and $v_z$ (where $z$ is the vertical direction) components. 
While the $v_x$ component (where $x$ is the horizontal, fault-parallel direction) is well approximated for receivers R7 and R8, which are far from the rupture nucleation, the $v_x$ component is noisy for other receivers and less well approximated, though the RMSE values are low since the velocity magnitude is low relative to $v_y$ and $v_z$. 

The cost of the ROM can be divided into two categories: the offline costs, for steps involved in building the ROM which only need to be performed once, and the online cost, for steps required each time the ROM is evaluated. 
The offline cost includes running the SeisSol simulations to obtain solutions for the data matrix. 
We run SeisSol simulations using one MPI rank and 64 OpenMP threads on a machine with two AMD EPYC 7552 48-core processors and 256 GB of system memory. 
The total walltime required for running the 21 simulations was \ $\sim31,500$ s, or $\mathcal{O}(10^4)$ s.
Computing the SVD of the data matrix required a walltime of less than $10^{-3}$ s, while building the RBF interpolant for the POD coefficients required $10^{-4}$ s. 
The online cost (walltime) required to evaluate the ROM is $5 \times 10^{-5}$ s. 
The speed of the ROM, once built, means that a ROM evaluation can be obtained $10^9 \times$ faster than a FOM evaluation. 

\subsection{Reduced-Order Model for Surface Displacement}

We build a second unique ROM for vertical displacement $u_z$ on the surface of the model domain at the final time step, $t=8$ s.
As in Section~\ref{sec:rom_receivers}, the ROM data matrix contains 21 FOM solutions, corresponding to $\theta$ values sampled at every 1$^\circ$ dip over the interval $[50^\circ, 70^\circ]$, i.e. $\theta = [50^\circ, 51^\circ, 52^\circ, \ldots, 69^\circ, 70^\circ]$. 
We note that the ordering of the FOM solution data is affected by the local time-stepping (LTS) employed by SeisSol \cite{Dumbser_2007,Breuer_2016}. 
Since SeisSol computes a time step for each cell based on the cell size and the local wave speed, and then groups cells into clusters, morphed meshes sometimes have cells that fall into different LTS clusters than those in the reference mesh.
Therefore, the output files written by SeisSol have different orderings of the output, despite the cell connectivity remaining the same. 
A permutation of the output vector is applied by associating cell centers with a dense background grid; this imposes a uniform ordering before output vectors are added to the ROM data matrix. 
After taking the SVD of the data matrix, we again use a quintic kernel when performing the RBF interpolation of the POD coefficients. 
We calculate the LOOCV $L_{\infty}$ error and again find that the error is higher at the edges of the parameter space: $0.03$ m for when $\theta = 50^\circ$ is dropped, and $0.06$ m when $\theta = 70^\circ$ is dropped. 
Within the parameter interval, the error is at most $0.01$ m and typically $<0.005$ m. 

\begin{figure}
     \centering
     \includegraphics[width=\textwidth]{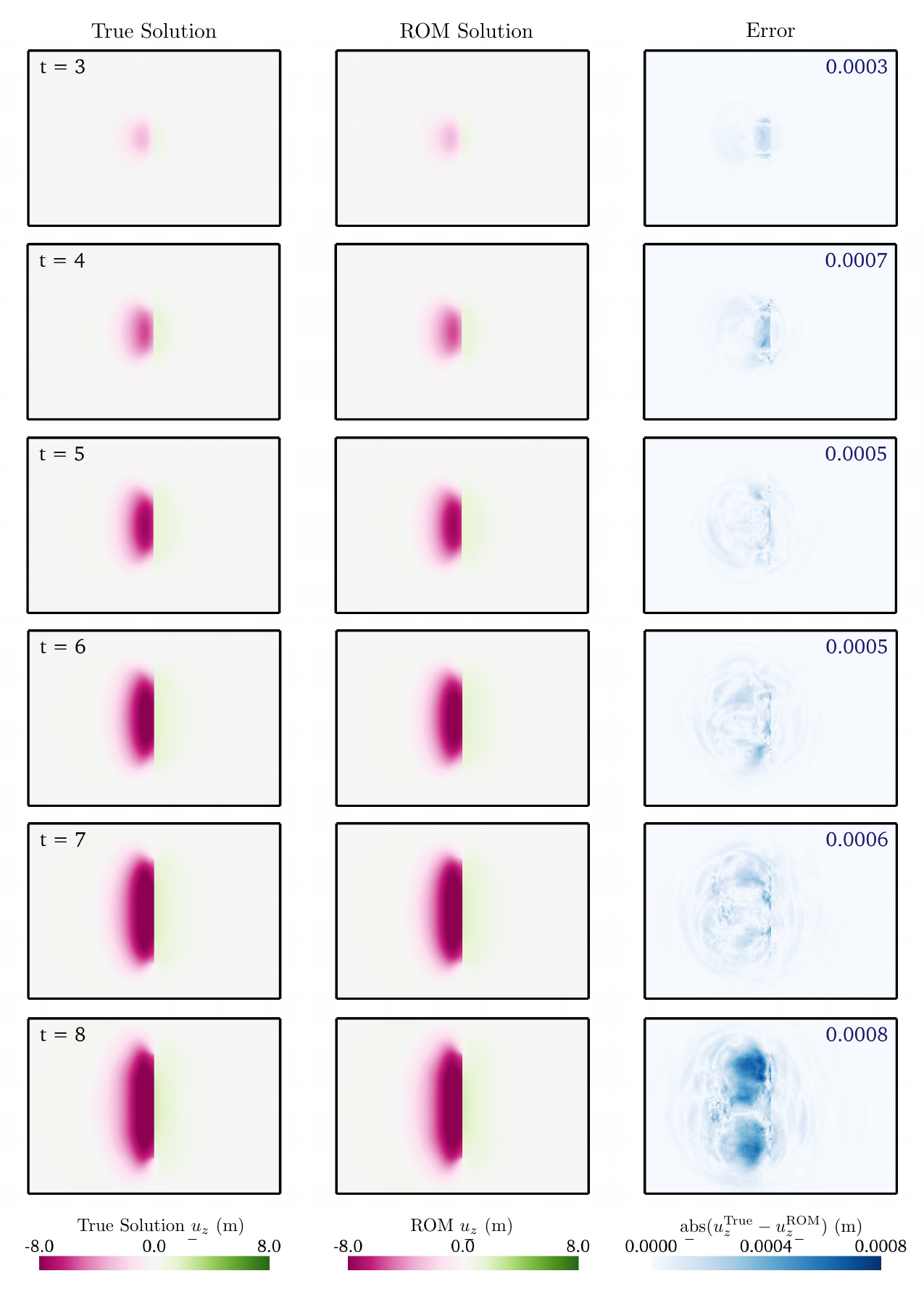}
     \caption{Full order model (FOM) vs. reduced order model (ROM) approximations of the vertical ground velocity component $u_z$ for a fault dip of $\theta = 52.5^\circ$. The maximum error is reported in blue text.}
     \label{fig:rom_surface_comp_u3}
\end{figure}

We also run the FOM on meshes morphed to have $\theta$ values sampled at every 1$^\circ$ in the range $[50.5^\circ, 69.5^\circ]$, i.e. $\theta = [50.5, 51.5, \ldots, 68.5, 69.5]$. 
These parameter values are not included in the ROM's data matrix, and this allows us to make a direct comparison between modeled $u_z$ and the ROM approximation at the same $\theta$ values. 
We find that the $L_\infty$ error between FOM and ROM solutions is at most 0.007 m, for $\theta = 69.5^\circ$, and for all other testing parameter values, the maximum error is $< 0.003$ m. 
The comparison between FOM and ROM solutions evaluated at $\theta = 52.5^\circ$ is shown in Figure~\ref{fig:rom_surface_comp_u3}. 
To perform this comparison, a unique ROM is constructed for each simulation time step between $t = 3$ s to $t=8$ s. 
Each ROM approximates the FOM solution to high accuracy, with a maximum error of $8 \times 10^{-4}$ m between FOM and ROM solutions for $\theta = 52.5^\circ$.

\begin{figure}
     \centering
     \includegraphics[width=\textwidth]{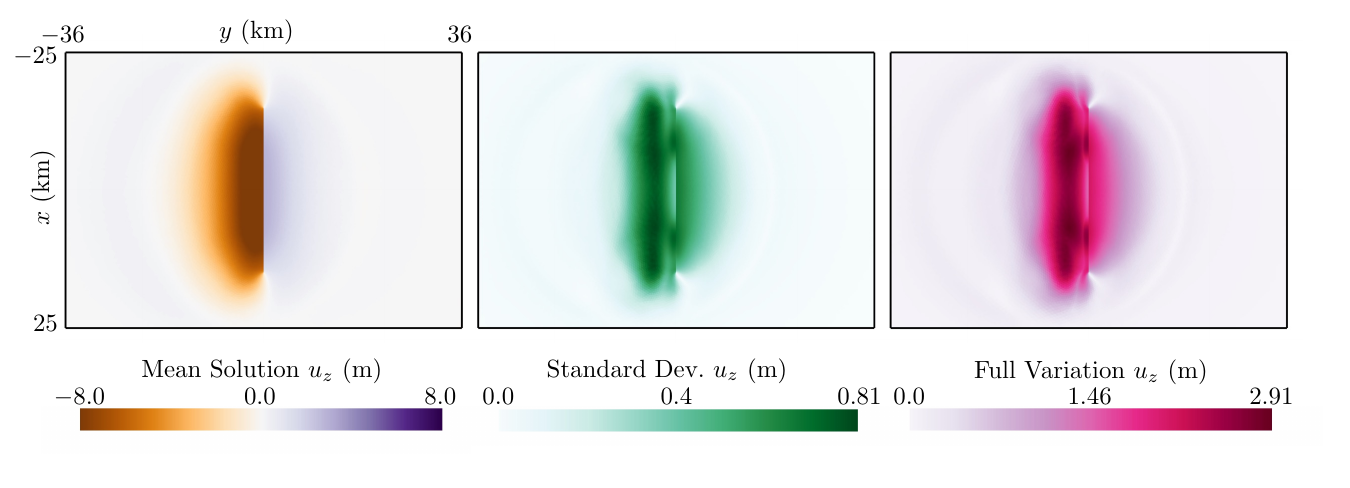}
     \caption{Mean field, standard deviation, and maximum variability as dip varies, for vertical ground velocity $u_z$ at simulation time $t=8$ seconds.}
     \label{fig:rom_surface_variability}
\end{figure}

The ROM can also be used to quantify the variability in $u_z$ as $\theta$ varies. 
We evaluate the ROM at $10^4$ samples in parameter space and compute the mean field, standard deviation, and maximum variability in $u_z$ as $\theta$ varies, shown in Figure~\ref{fig:rom_surface_variability}.
The fault dips towards the negative side of the $y$ axis, and there is the most variability on this side of the fault. 
From this analysis we see that the vertical ground velocity $u_z$ varies by as much as 2.9 m, though there is also $> 1$ m variability on the positive-$y$ axis side of the fault trace.  

The cost of building the ROM is similar to that in Section~\ref{sec:rom_receivers}, since the same simulations are used to obtain solutions for the data matrix, with a total walltime for 21 simulations of $\sim31500$ s, or $\mathcal{O}(10^4)$ s.
However, the data matrix for this example is larger since each column contains the solution on the full surface of the domain, rather than just the velocity time history at one receiver. 
This means computing the SVD requires a larger walltime, here 0.04 s. 
Building the RBF interpolant for the POD coefficients still requires a walltime of $10^{-4}$ s. 
The online cost (walltime) of evaluating the ROM in this case is $5 \times 10^{-4}$ s.
This is a $10^8 \times$ speedup for the ROM approximation over the FOM evaluation.

\begin{table}
\begin{center}
\caption{LOOCV errors for ROMs built for all components of velocity and displacement at time $t=8$ s. Maximum absolute and relative errors are reported for the edges of parameter space, $\theta \in \theta_\text{edge}$ (within 2$^\circ$ of the bounds of $\boldsymbol Q$), and for the interior, $\theta \in \theta_\text{int} =[53^\circ, 57^\circ]$.}
\label{table:LOO_error_all_components}
\begin{tabular}{  l l l l l l l }
\hline
~ & $u_x$ (m) & $u_y$ (m) & $u_z$ (m) & $v_x$ (m/s) & $v_y$ (m/s) & $v_z$ (m/s) \\
\hline
Max abs. error, $\theta \in \theta_\text{edge}$ & 3.3$\times 10^{-2}$ & 7.1$\times 10^{-2}$ & 6.7$\times 10^{-2}$ & 1.6$\times 10^{-1}$ & 5.3$\times 10^{-1}$ & 2.2$\times 10^{-1}$ \\
Max abs. error, $\theta \in \theta_\text{int}$ & 3.9$\times 10^{-3}$ &  7.5$\times 10^{-3}$ & 5.9$\times 10^{-3}$ & 2.3$\times 10^{-2}$ & 7.0$\times 10^{-2}$ & 3.0$\times 10^{-2}$ \\
Max rel. error, $\ \theta \in \theta_\text{edge}$ & 1.6$\times 10^{-2}$ & 1.1$\times 10^{-2}$ & 6.7$\times 10^{-3}$ & 1.7$\times 10^{-1}$ & 4.4$\times 10^{-1}$ & 1.3 $\times 10^{-1}$ \\
Max rel. error, $\ \theta \in \theta_\text{int}$ & 1.9$\times 10^{-3}$ & 1.1$\times 10^{-3}$ & 6.0$\times 10^{-4}$ & 2.4$\times 10^{-2}$ & 5.9$\times 10^{-2}$ & 1.8$\times 10^{-2}$ \\
Max value of component & 2.01 & 6.75 & 10.04 & 0.95 & 1.19 & 1.70 \\

\hline 
\end{tabular}
\end{center}
\end{table}

\section{Discussion} \label{sec:discussion}

\subsection{Balancing Efficiency and Accuracy Against Reduced Meshing Time}

Though there is an initial time cost required for setting up the mesh morphing process for a given application (further discussed in Section~\ref{sec:challenges}), once the process is in place, a morphed mesh can be generated in 10-100 s as shown in Section~\ref{sec:dr_eff_comp_res} and Section~\ref{sec:slab_eff_com_res} without manual user interaction. 
However, the advantage of being able to efficiently generate geometrically varying meshes must be balanced against potential increases in numerical simulation time and decreases in accuracy. 
Morphed meshes can contain high aspect ratio cells, which can decrease the accuracy and efficiency of the numerical simulation.
In the case of the thermal model examples shown in this study, high aspect ratio cells will affect the accuracy of the simulation output because they have a long length scale in one direction, which makes it harder to accurately resolve a field over that cell. 
In the dynamic rupture simulation example, the aspect ratio of mesh cells affects the time step used in the simulation. 
Higher aspect ratio cells will require a smaller time step, and increase the time to solution. 
Therefore, there may be a tradeoff between time to solution and the reduction in meshing time afforded by mesh morphing.

\subsection{Outlook for Other Types of Meshes}

In this study, we limit our scope to triangular and tetrahedral meshes and do not consider meshes of other types, such as hexahedral meshes. 
However, previous studies have presented examples of mesh morphing applied to hexahedral meshes.
\citeA{de_boer_2007} demonstrated the application of RBF-based mesh morphing to both unstructured triangular and unstructured hexahedral meshes, the latter example including a mesh surrounding a moving airfoil.
In their comparison of different mesh morphing methods, \citeA{staten_et_al_2012} apply their methods to both tetrahedral and hexahedral meshes of four different test geometries. 
They observe that hexahedral meshes maintain better cell quality under morphing than tetrahedral meshes. 
They note that this is especially the case for structured hexahedral meshes, and suggest that this is due to hexahedral cell having a more flexible cell topology than tetrahedral cells. 
\citeA{sieger_et_al_2014} use the same test geometries as \citeA{staten_et_al_2012}, and also show that their implementation of RBF-based mesh morphing can be successfully applied to unstructured and structured hexahedral meshes as well as tetrahedral meshes. 
While these examples indicate that there may be potential for application of mesh morphing to geophysical simulations utilizing hexahedral meshes, we do not explore this in the scope of this study. 
As with tetrahedral meshes, the suitability of mesh morphing for hexahedral meshes will depend greatly on the reference mesh quality and its level of topological complexity, as well as the geometric variations to be applied. 

\subsection{Limitations} \label{sec:limitations}

\subsubsection{Failure Under Extreme Morphing}

\begin{figure}
     \centering
     \includegraphics[width=\textwidth]{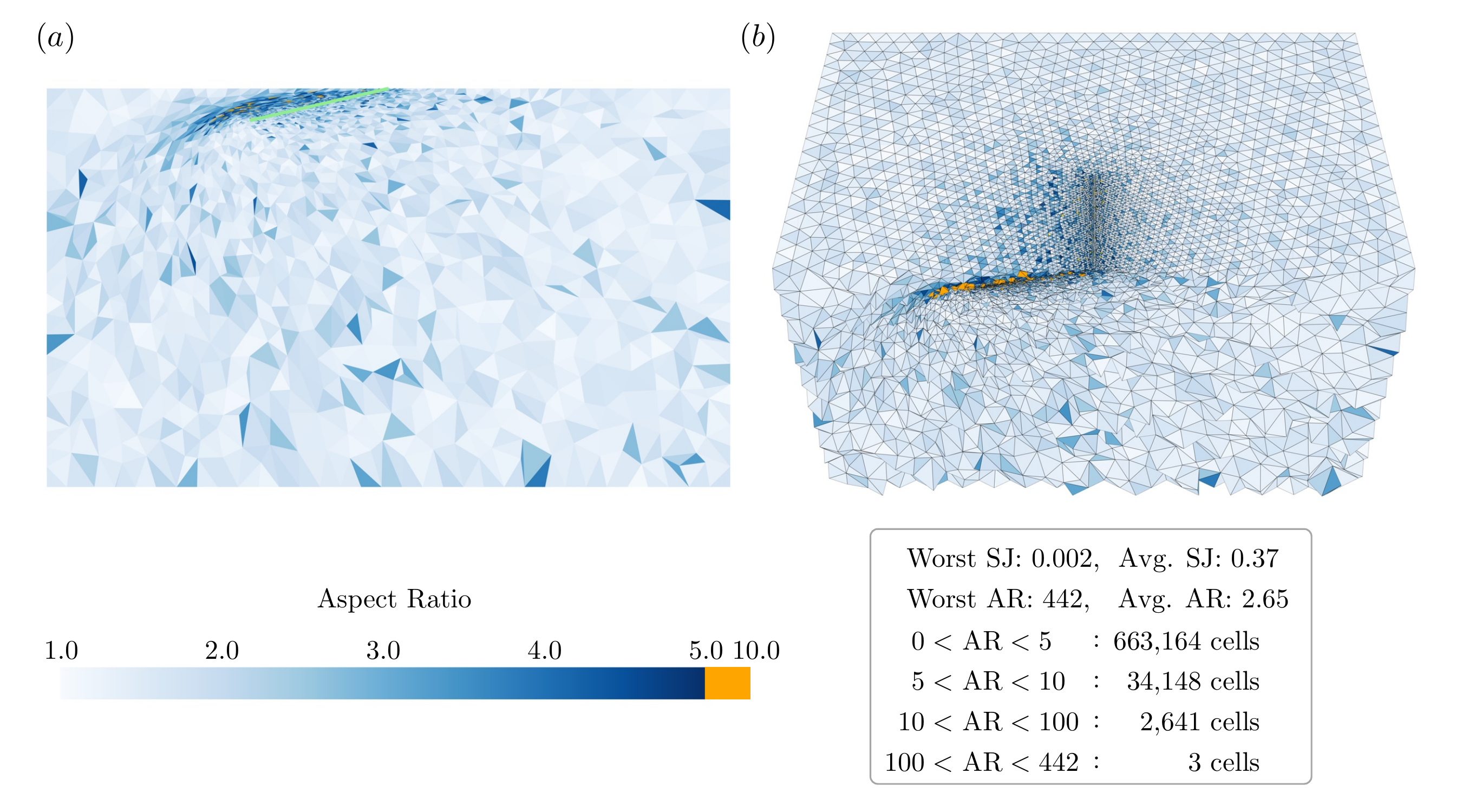}
     \caption{Mesh quality for a morphed mesh with $\theta = 13^\circ$, shown from (a) a side-on view with the fault in light green, and (b) an angled view with a summary of the mesh quality information.}
     \label{fig:extreme_morphing}
\end{figure}

It is possible to use the mesh morphing method presented here to apply substantial displacement to a reference mesh and obtain a valid morphed mesh (no inverted cells) that has very poor mesh quality (cells with high aspect ratio, and or low minimum interior angles). 
Measuring mesh quality is a key step, and each user must determine how much geometric variability can be applied while maintaining mesh quality suitable for their particular application. 
For example, we find that it is possible to morph the TPV13 mesh beyond the $\theta \in [40^\circ, 80^\circ]$ range presented in Section~\ref{sec:DR_mesh_qual_accuracy}, and still obtain a valid mesh. 
Selecting $\boldsymbol q_\text{new}$ values in the range $\left[1^\circ, 90^\circ \right]$, we can morph the reference mesh (which has $\theta = 60^\circ$) to $\boldsymbol q_\text{eval}$ values in the range $\left[13^\circ, 89^\circ \right]$ and obtain valid meshes. 
The outer boundaries of the domain are respected, and there are no inverted cells.
However, the mesh quality degrades significantly for low fault dip angles. 
The maximum aspect ratio (AR) values for $\boldsymbol q_\text{eval}$ values in [20, 60] are at most 21, which is still reasonable. 
For $\boldsymbol q_\text{eval}$ values in [13, 20], the max AR values increase rapidly as the dip decreases, with max AR values up to 442. 
Interestingly, the mean AR value does increase as $\boldsymbol q_\text{eval}$ decreases, but not as dramatically as the max AR value, indicating that there are relatively few ill-shaped tetrahedra in the overall mesh. 
Indeed, we see that the poor-quality tetrahedra are concentrated in the region just above the fault, and just downdip of the fault, where the mesh is being most compressed. 
This is visualized in Figure~\ref{fig:extreme_morphing}, which shows the effect on mesh quality of a morph to $\theta = 13^\circ$, highlighting the region above and just downdip of the fault where mesh quality is degraded, as well as the relatively few ill-shaped tetrahedral cells with higher AR values. We note that a morph from $60^\circ$ to $89^\circ$ does not decrease the mesh quality nearly as much as a morph from $60^\circ$ to $30^\circ$, despite being a nearly equivalent change in dip. This is because the low dip angles cause all of the tetrahedra above the fault to become compressed; when the dip angle increases, there is space on the other side of the domain to accommodate the movement of tetrahedra without undue compression.  

We note that cases where the target points $\boldsymbol X_{\boldsymbol q_\text{new}}$ are selected to be outside of the model domain can result in the creation of an invalid mesh (e.g., one containing inverted cells) if there is a different constraint that conflicts with their placement.
For example, in the thermal model example shown in Figure~\ref{fig:combined_thermal_schematic}(a), if $\boldsymbol X_{\boldsymbol q_\text{new}}$ is set such that point $c$ would be to the right of the original point $i$, and any of points $f, g, h$ or $i$ are constrained to not move to the right, these constraints will conflict and cause a mesh with inverted cells. 
For the same example, if the base of the mantle wedge (curve $\harp{ci}$) is not wide enough in the reference mesh, significant morphing can cause inverted cells. 
The user must carefully choose their constraints and the extent of the geometric variability they wish to apply for their given application.

\subsubsection{Challenges in Designing and Guiding Mesh Morphing} \label{sec:challenges}

There are several challenges to consider when implementing mesh morphing for a new application. 
The first challenge is in describing geometric variability using a few parameters. 
In some applications, the geometric parameters that should be used may be intuitive and natural; for example, faults are typically described in terms of strike and dip.
But for other applications, the parameterized description of geometry may be less evident. 
Another important consideration is that the mesh morphing process is not fully automated start-to-finish, but requires some initial experimentation when applied to a new application. 
The first step in this experimentation is to carefully select which boundaries in the domain are set to morph in one or several directions and which are set to have zero displacement. 
Another non-evident part of the process is choosing how to place $\boldsymbol X_{\boldsymbol q_\text{ref}}$ and $\boldsymbol X_{\boldsymbol q_\text{new}}$ points on the boundaries of the domain. 
This must be done consistently across different geometric configurations, and may be done with code tailored to an application, or by coarsely discretizing the boundaries of the domain using a tool such as GMSH (i.e., creating a 1D mesh for the boundary curves of a 2D domain, or a 2D mesh for the boundary surfaces of a 3D domain). 
Another important choice is the configuration of the reference mesh; ideally, a reference mesh should be selected that minimizes the maximum displacement required to morph between configurations across parameter space
(this does not necessarily mean the configuration corresponding to the middle of the geometric parameter range, as seen in Section~\ref{sec:alpha_example}). 
Reference meshes at one end of parameter space may not be able to be morphed all the way to configurations corresponding to the other end of parameter space. 
It is also necessary to determine acceptable bounds for the parameter space by monitoring mesh quality as morphing is applied. 
Different applications will have varying levels of acceptable mesh quality and will permit different amounts of morphing.  
This initial experimentation phase is also a good time to check whether the morphed meshes satisfy all the necessary conditions, or whether additional morphs should be performed.
For example, the volume of a domain can be preserved if the outer boundaries are prescribed to have zero displacement. 
However, if the outer boundaries are morphed, the volume will change; the user must determine if this is acceptable for the application, or whether additional corrections can be applied to preserve volume.
After these initial experiments are performed and a parameter space with acceptable mesh quality is selected, the mesh morphing can be applied in an automated way and very little user interaction is required; however, these steps are necessary to build a robust mesh morphing process for a new application. 

\subsection{Outlook for complex fault systems}

The demonstrated uses of mesh morphing in this study are limited to settings with a single fault.
A natural extension is to consider a fault network with multiple branching, intersecting faults, and or non-intersecting faults. 
Settings where this is of interest include adjacent fault segments with stepovers, branching and intersecting faults, and megathrusts with splay faults \cite<e.g.,>{Lauer_Saffer_2012,Gabriel2024}.
Varying the dip angle, trace, curvature, or roughness of a fault may be of interest. 
We expect that morphing neighboring faults which are unconnected will be tractable, though careful monitoring of mesh quality at fault edges and in the regions between faults will be required, especially if the distance between faults is being decreased.  
Morphing branching or intersecting faults in 3D is more challenging because the displacement of one fault will necessarily affect the fault it intersects; since the mesh connectivity remains identical under morphing, cells on either side of the intersection must remain connected, so the intersection may get ``dragged'' or warped under morphing. 
This means that even if only one fault has displacements prescribed, the other fault will likely be locally displaced near the intersection.
It may be possible to mitigate this effect using successive morphs as described in this study. 
In this case, the resulting morphed faults may each generally represent the desired new geometric configuration, with some local disagreement near the intersection.  
Therefore, while the method does not preclude application to multi-fault systems, intersections between faults should be handled carefully and may require additional development or local corrections to obtain close agreement near intersections. 
Further, the topology of a fault network cannot be changed through morphing: intersecting faults must remain intersecting, non-intersecting faults cannot intersect, branching faults must remain branched, etc. 

\section{Conclusions} \label{sec:conclusions}

Mesh morphing enables the efficient generation of ensembles of geometrically varying meshes while preserving mesh connectivity, offering a useful tool for quantifying geometric sensitivity in geophysical models.
This systematic approach reduces the manual effort typically associated with mesh generation and 
facilitates the construction of data-driven, non-intrusive reduced-order models (ROMs), which require consistent mesh connectivity. 
ROMs can be orders of magnitude faster to evaluate than full-order models, and can therefore be used for robust sensitivity analysis. 

We demonstrate mesh morphing in two geophysically-relevant applications: 
(i) 3D earthquake dynamic rupture simulations with varying fault dip angles, and (ii) 2D subduction zone thermal modeling with parameterized slab curvature and uncertainty-informed interface depths.
In all cases, morphed meshes maintain high quality across a wide range of geometric configurations. 
Simulation results run on morphed meshes closely match those obtained from exactly generated meshes.  
We also demonstrate the successive application of morphing steps, to obtain morphed meshes satisfying specific conditions, such as maintaining fault planarity or consistent slab thickness during morphing. 

For the earthquake rupture application, we further construct ROMs to analyze sensitivity of surface displacement and receiver velocity time histories to fault dip angle. 
The ROMs reproduce full-model behavior with high fidelity up to $10^9 \times$ faster, enabling rapid exploration of geometric parameter space and advancing our understanding of subduction processes and earthquake rupture processes. 
To our knowledge, this is the first application of model order reduction to earthquake dynamic  rupture modeling.

This study seeks to lay a foundation for further applications of mesh morphing in computational geophysics problems involving uncertain geometries. 
We provide a careful description of a practical framework alongside several relevant examples with different characteristics to illustrate the flexibility of the method.
We describe the limitations of the method and provide guidance about important points to consider when applying the method to new problems.
By lowering the barrier to incorporating realistic geometric variability, mesh morphing can help bridge the gap between idealized models and natural fault systems.

\section{Acknowledgments}

The authors acknowledge financial support from the National Science Foundation (GMH through MTMOD, grant No. EAR-2121568; DAM and A-AG through MTMOD, grant Nos. EAR-2121568 and QUAKEWORX, OAC-2311208). 
The authors thank T.~W. Becker for providing the computational resources required for running the forward models in this study.
A-AG acknowledges additional support from Horizon Europe (ChEESE-2P grant No. 101093038, DT-GEO grant No. 101058129, and Geo-INQUIRE grant No. 101058518), the National Science Foundation (CRESCENT, grant no. EAR-2225286,  CSA-LCCF, grant no. OAC-2139536), the National Aeronautics and Space Administration (grant no. 80NSSC20K0495) and from the Statewide California Earthquake Center (SCEC, grant no. 25341).

\section{Open Research}

The code and data files required to perform the mesh morphing presented in this study are available in a Zenodo repository \cite{zenodo_repo}. 
SeisSol is available open-source and the results shown here were run using \verb|v1.1.3-217-gc278571f| \cite{seissol}. 
The thermal models were run using the code that accompanies \citeA{hobson_may_2025}, which is also available open-source.  


\appendix

\makeatletter
\def\@seccntformat#1{\@ifundefined{#1@cntformat}%
   {\csname the#1\endcsname\space}
   {\csname #1@cntformat\endcsname}}
\newcommand\section@cntformat{} 
\makeatother
\renewcommand{\thesection}{S}
\counterwithin{equation}{section}
\counterwithin{figure}{section}
\counterwithin{table}{section}

\newpage

\section{Supporting Information}

\subsection*{S1 Uncertainty in Slab2 data} \label{sec:slab2_uncertainty}

Each of the data types used to calculate the Slab2 interfaces, which include earthquake hypocenters, active source/seismic reflection, receiver functions, and tomography, is assigned a default minimum uncertainty. 
For example, active source/seismic reflection is assigned a default uncertainty of 2.5 km, while tomography is assigned an uncertainty of 40 km.  
To compute the depth at a given location, the relevant data (within 2.5$^{\circ}$ of the node) is collected. 
They assume that the depth of a data point can be described by a Probability Density Function (PDF) which is centered at the preferred depth and has a standard deviation equal to the assigned uncertainty for that data type. 
The PDFs for all points of a certain data type are summed together, and then the PDF for each type of data is summed to obtain the PDF describing the slab center depth at that location. 
A further step uses information about the age of the oceanic lithosphere at the trench to infer the depth of the slab interface. 
The Slab2 dataset provides files with slab interface depths and the corresponding vertical uncertainties (the standard deviations of the comprehensive summed PDFs).
Slab2 only considers vertical uncertainty and does not consider horizontal uncertainty \cite{slab2}. 
Many of the types of data they use have either small or negligible horizontal uncertainties, which are less than the resolution of the slab models. 
The teleseismic earthquake locations dataset has higher uncertainties, but these vary spatially across subduction zones, and a bias correction process was included in the catalog used in Slab2. 
Uncertainty estimates for each margin are included in the Slab2 data products, and details of how the uncertainty estimates were made are found in the Supporting Information for \citeA{slab2}.

\begin{table}
\centering
\caption{Receiver locations.}
\label{table:receivers}
\begin{tabular}{  l l l l }
\hline
Receiver & x (m) & y (m) & z (m) \\
R1 &  0 &   1.0   &      0  \\
R2  &  0   &  -1.0  &      0  \\
R3  &  0  &   2.0  &      0   \\
R4       &     0  &  -2.0   &      0   \\
R5       &     0   &   3.0  &      0   \\
R6       &     0  &  -3.0   &      0   \\
R7       & -12.0   &   3.0  &      0   \\
R8       & -12.0   &  -3.0   &      0   \\
R9       &     0   &   -0.673205 &   -0.3   \\
R10       &     0   &  -1.17320 &   -0.3   \\
R11      &     0   &    0.326795 &   -0.3   \\
R12       &     0   &    0.826795 &   -0.3  \\
\hline
\end{tabular}
\end{table}

\begin{table}
\begin{center}
\caption{Mesh quality metrics on the fault plane as $\theta$ varies.}
\label{table:mesh_qual_fault_theta}
\begin{tabular}{  l l l l l l l }
\hline
~ & \multicolumn{2}{l}{Aspect Ratio}& \multicolumn{2}{l}{Scaled Jacobian} & \multicolumn{2}{l}{Minimum Angle ($^{\circ}$)} \\
$\theta$ & Avg & Max & Avg & Min & Avg & Min \\
\hline
40$^\circ$ & 1.03 & 1.4 & 0.98  & 0.77 & 57.87 & 41.84 \\
50$^\circ$ & 1.03 & 1.4 & 0.98  & 0.77 & 57.92 & 41.94\\
60$^\circ$ (ref) & 1.03 & 1.4 & 0.98  & 0.77 & 57.94 & 41.97 \\
70$^\circ$ & 1.03 & 1.4 &  0.98 & 0.77 & 57.92 & 41.94 \\
80$^\circ$ & 1.03 & 1.4 & 0.98  & 0.77 & 57.92 & 41.94 \\
  \hline
\end{tabular}
\end{center}
\end{table}

\begin{table}
\centering
\caption{Parameter values used in our simulations. All values match the benchmark description except the bulk cohesion, which is $1.0\times 10^6$ Pa instead of the benchmark value of $5.0\times 10^6$ Pa, and the static friction coefficient inside the nucleation zone, which is 0.48 instead of the benchmark value of 0.54.}
\label{table:tpv13_params}
\begin{tabular}{  l c l l }
\hline
\textbf{Plasticity} & & & \\ 
Density & $\rho$ & 2700 & kg/m$^3$ \\
Shear-wave velocity & $V_s$ & 3300 & m/s \\
Pressure-wave velocity & $V_p$ & 5716 & m/s \\
Bulk cohesion & $c$ & $1.0\times 10^6$ & Pa \\
Bulk friction & $\phi$ & 0.85 & - \\
Plastic relaxation & $T_v$ & 0.03 & - \\
\hline
\textbf{Friction} & & & \\ 
Dynamic friction coefficient & $\mu_d$ & 0.1 & - \\
Static friction coeff., outside nucleation zone & $\mu_s$ & 0.7 & - \\
Static friction coeff., inside nucleation zone & $\mu_s$ & 0.48 & - \\
Frictional cohesion & $c_0$ & -200,000 & Pa \\
Critical distance & $d_c$ & 0.5 & m \\ 
\hline
\textbf{Initial stress, above 11951.15 m depth} & & & \\ 
Max compressive principal stress & $\sigma_1$ & 26460 $\times H$ & Pa/m \\
Min compressive principal stress & $\sigma_3$ & 15624.3 $\times H$ & Pa/m \\
Intermediate principal stress & $\sigma_2$ & $(\sigma_1 + \sigma_3)/2 $ & Pa/m \\
Fluid pressure & $P_f$ & 1000 kg/$\text{m}^3 \times 9.8 \ \text{m/s}^2 \times H$ \\
\hline
\textbf{Initial stress, below 11951.15 m depth} & & & \\ 
All principal stresses & $\sigma_1, \sigma_2, \sigma_3$ & $2700 \ \text{kg/m}^3 \times 9.8 \ \text{m/s}^2 \times H$ \\
\hline
\end{tabular}
\end{table}

\begin{figure}
     \centering
     \includegraphics[width=\textwidth]{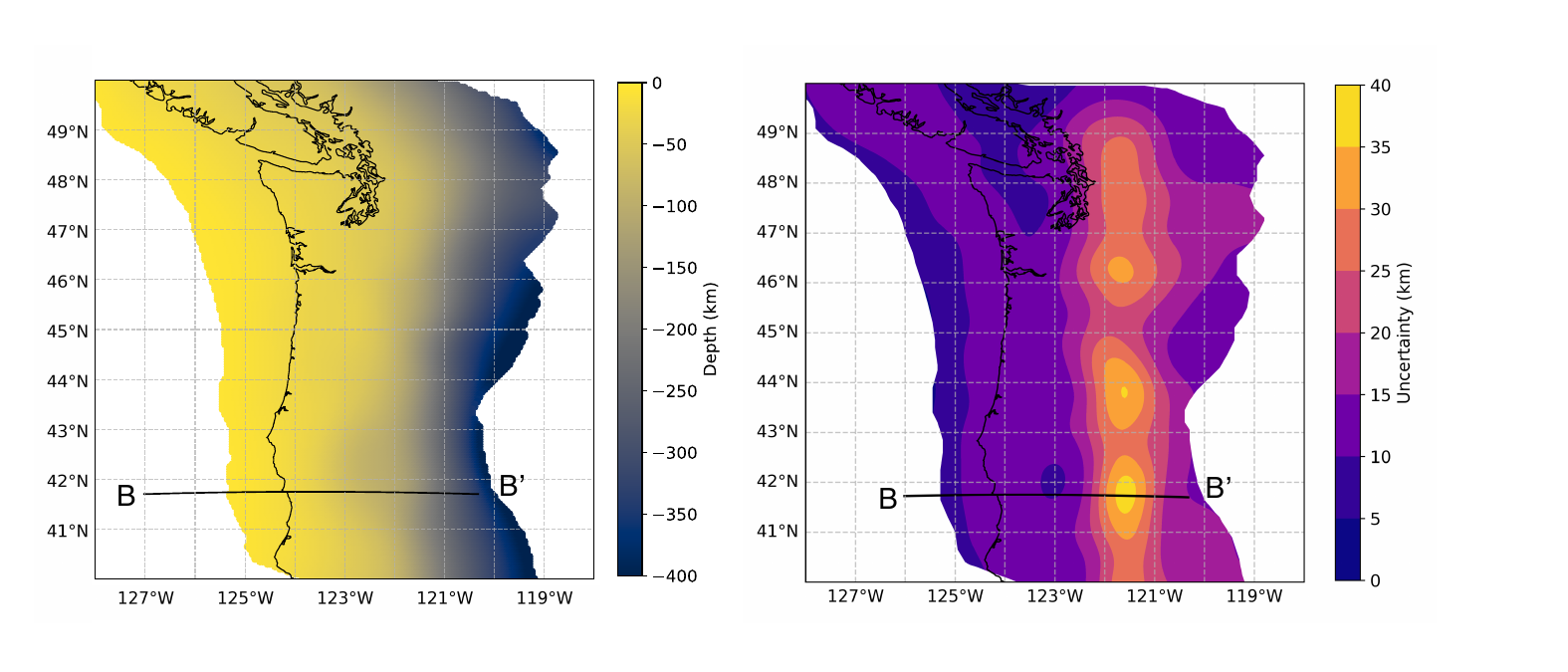}
     \caption{Slab2 data for the Cascadia subduction zone, with (a) depth of the subduction interface, (b) depth uncertainty as calculated by Slab2.}
     \label{fig:slab2_uncertainty_maps_only}
\end{figure}

\begin{figure}
     \centering
     \includegraphics[width=\textwidth]{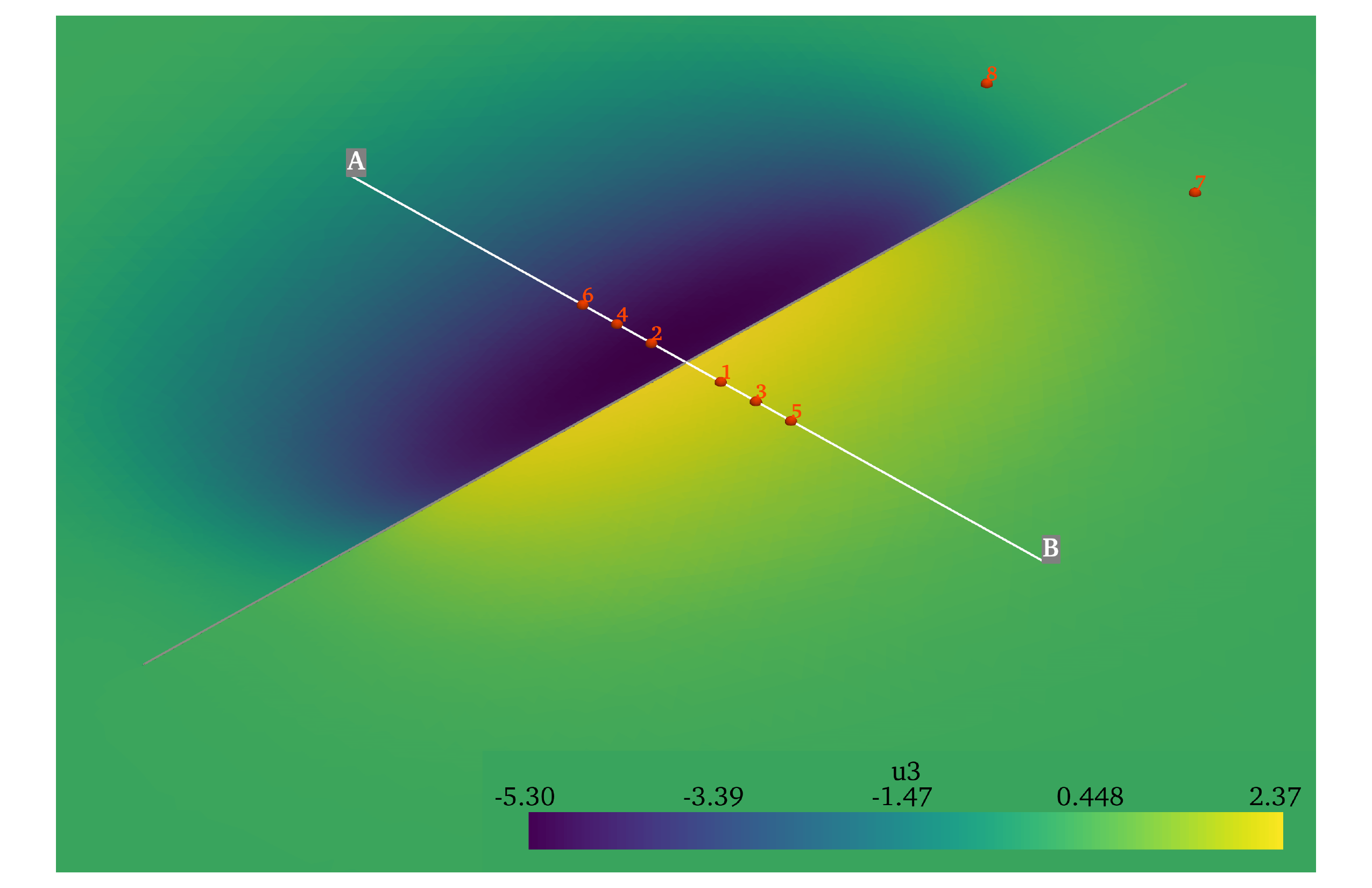}
     \caption{Surface displacement for $\theta = 70$, with receiver locations numbered and plotted in orange, the fault in gray, and the line A-B along which uplift and subsidence is measured in white.}
     \label{fig:surface_with_receivers}
\end{figure}

\begin{figure}
     \centering
     \includegraphics[height=0.8\textheight]{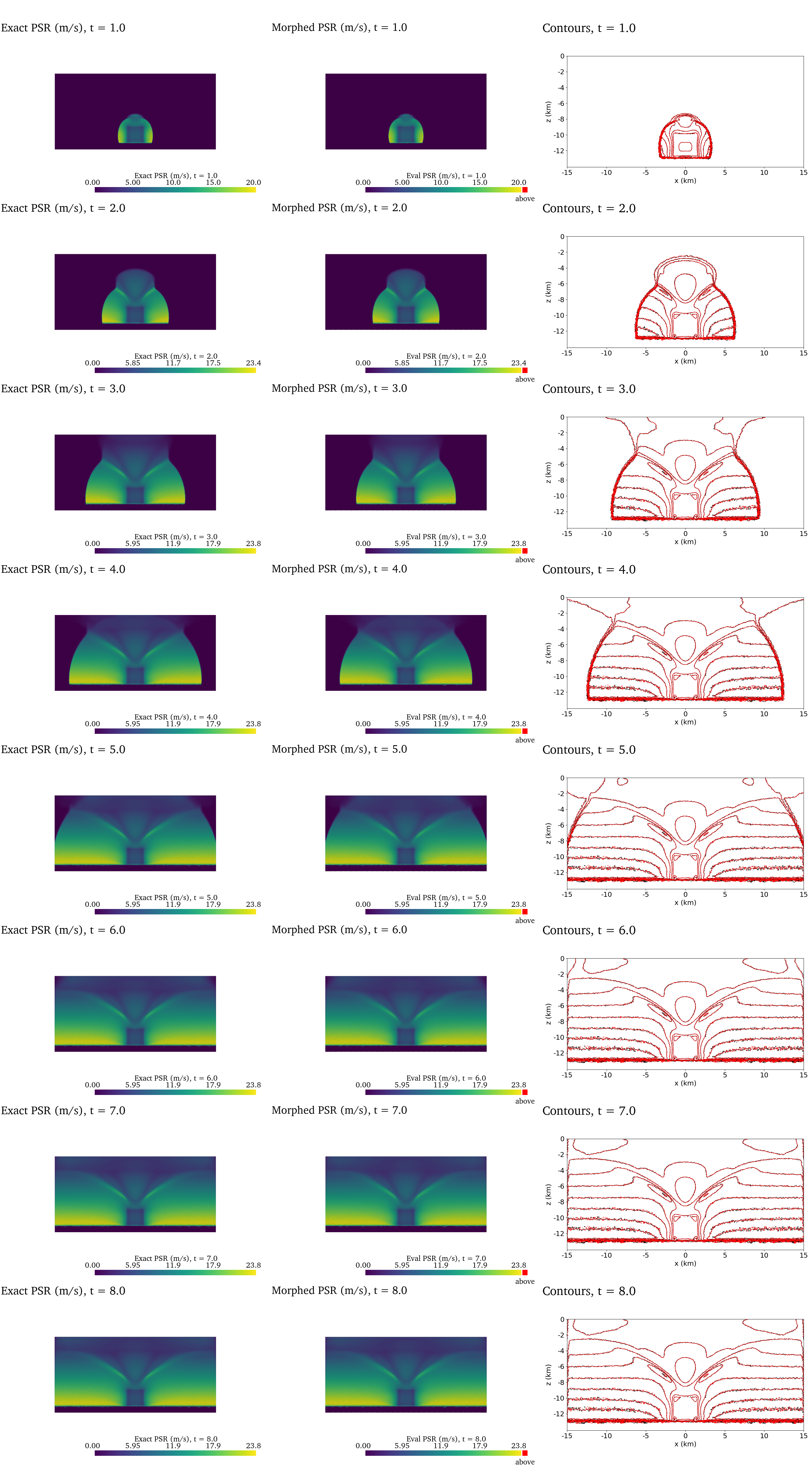}
     \caption{Peak slip rate on the fault for $\theta = 70$. The left column shows output from exactly generated meshes, while the center column shows output from morphed meshes. The right column shows contour plots with both exact and morphed output interpolated to the same mesh, with exact mesh output contours in black and morphed mesh output contours in red.}
     \label{fig:subplots_PSR_dip_70}
\end{figure}

\begin{figure}
     \centering
     \includegraphics[width=\textwidth]{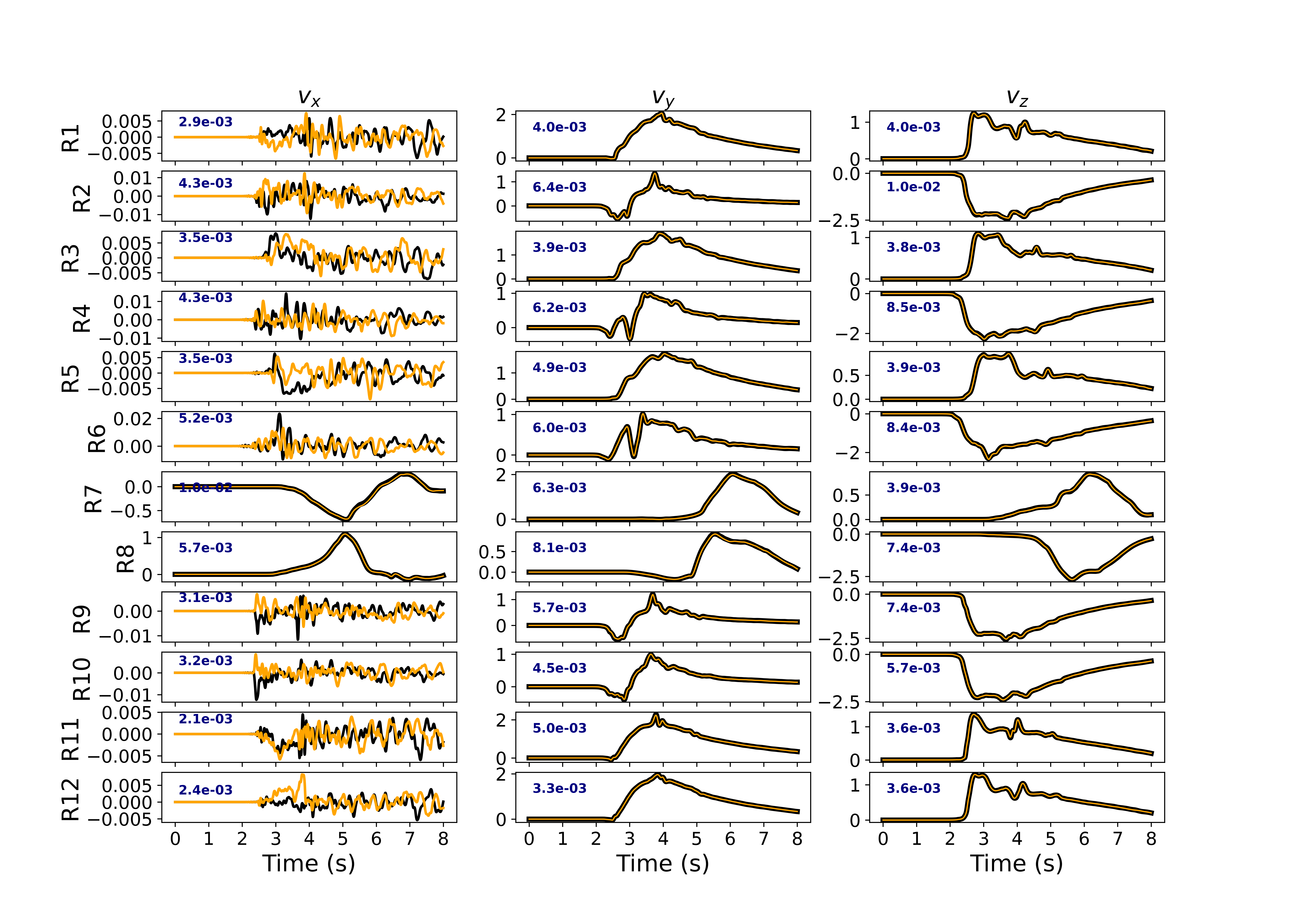}
     \caption{Measured components of velocity at each of the 12 receivers for $\theta = 70$. Output from the exactly generated mesh is in black, while output from the morphed mesh is in orange.}
     \label{fig:receiver_dip_70}
\end{figure}

\begin{figure}
     \centering
     \includegraphics[width=\textwidth]{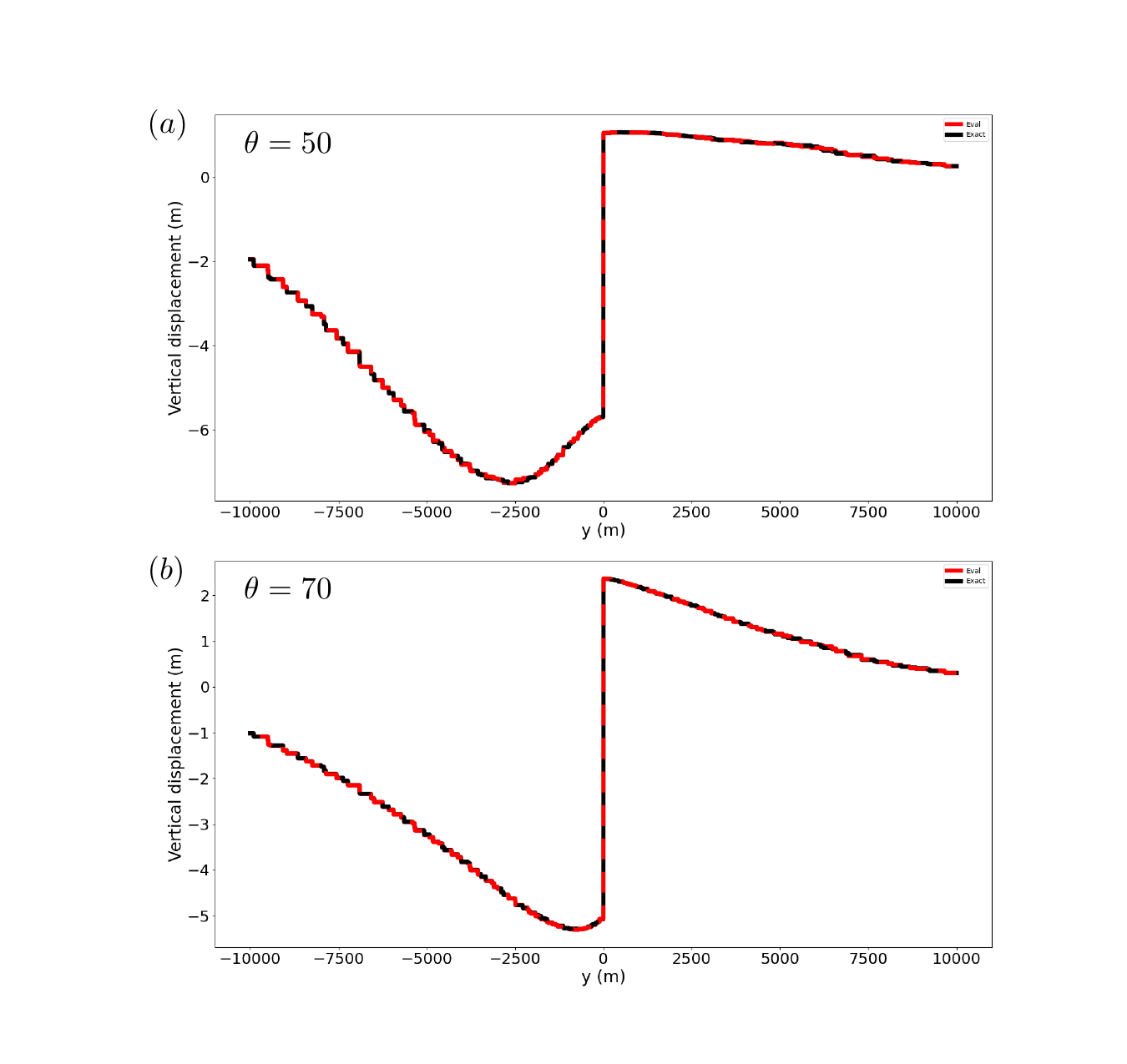}
     \caption{Uplift and subsidence profile along line A-B (shown in Figure~\ref{fig:surface_with_receivers}), for (a) $\theta = 50$ and (b) $\theta = 70$. }
     \label{fig:uplift_profiles}
\end{figure}

\begin{figure}
     \centering
     \includegraphics[width=\textwidth]{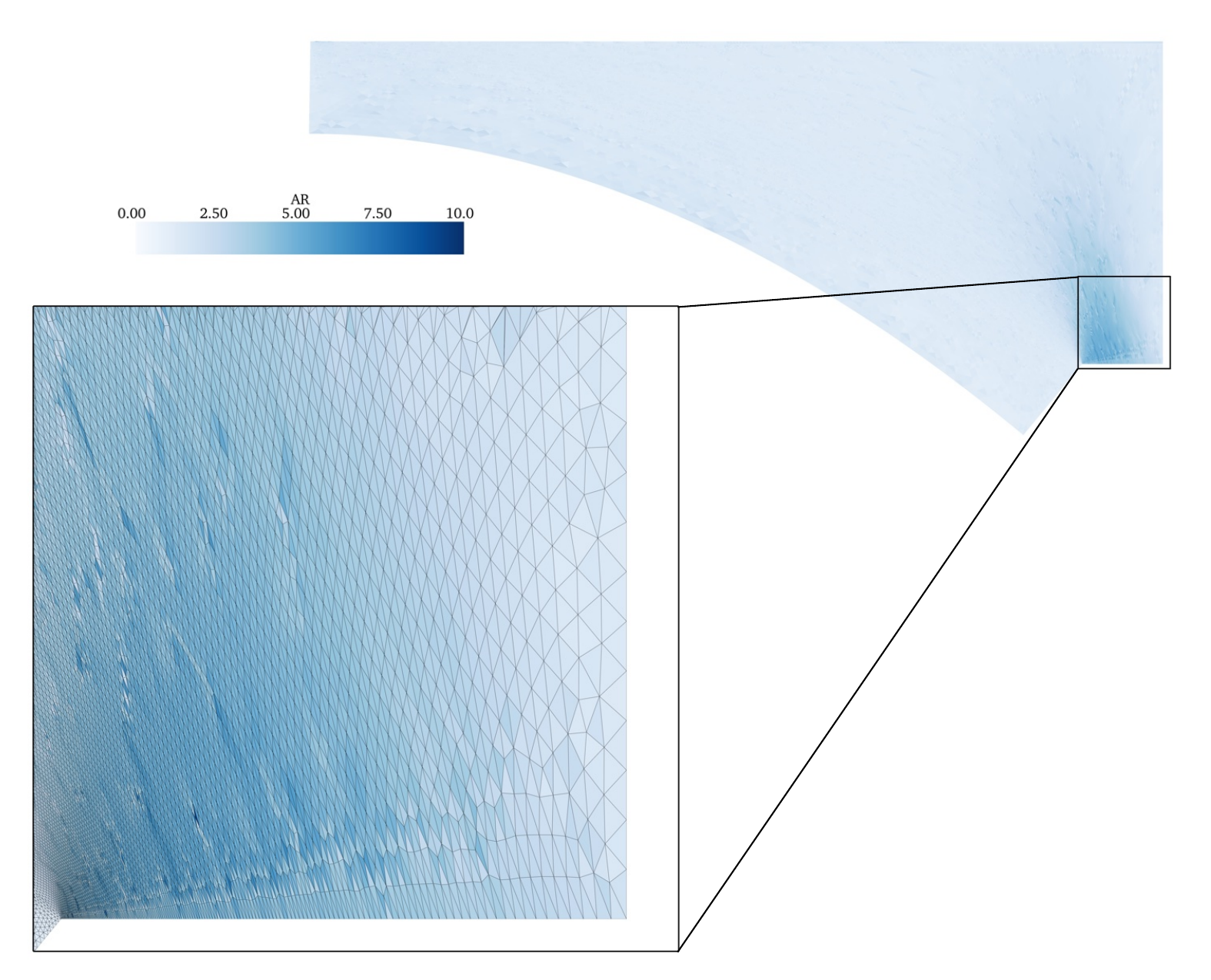}
     \caption{The morphed mesh with  $\alpha = 0.5 \times 10^{-3}$, colored by the Aspect Ratio (AR) of each cell. The colormap is saturated at AR = 10 for better visualization, though the max AR is 51.}
     \label{fig:mesh_quality_spatial_alpha_low}
\end{figure}

\begin{figure}
     \centering
     \includegraphics[width=\textwidth]{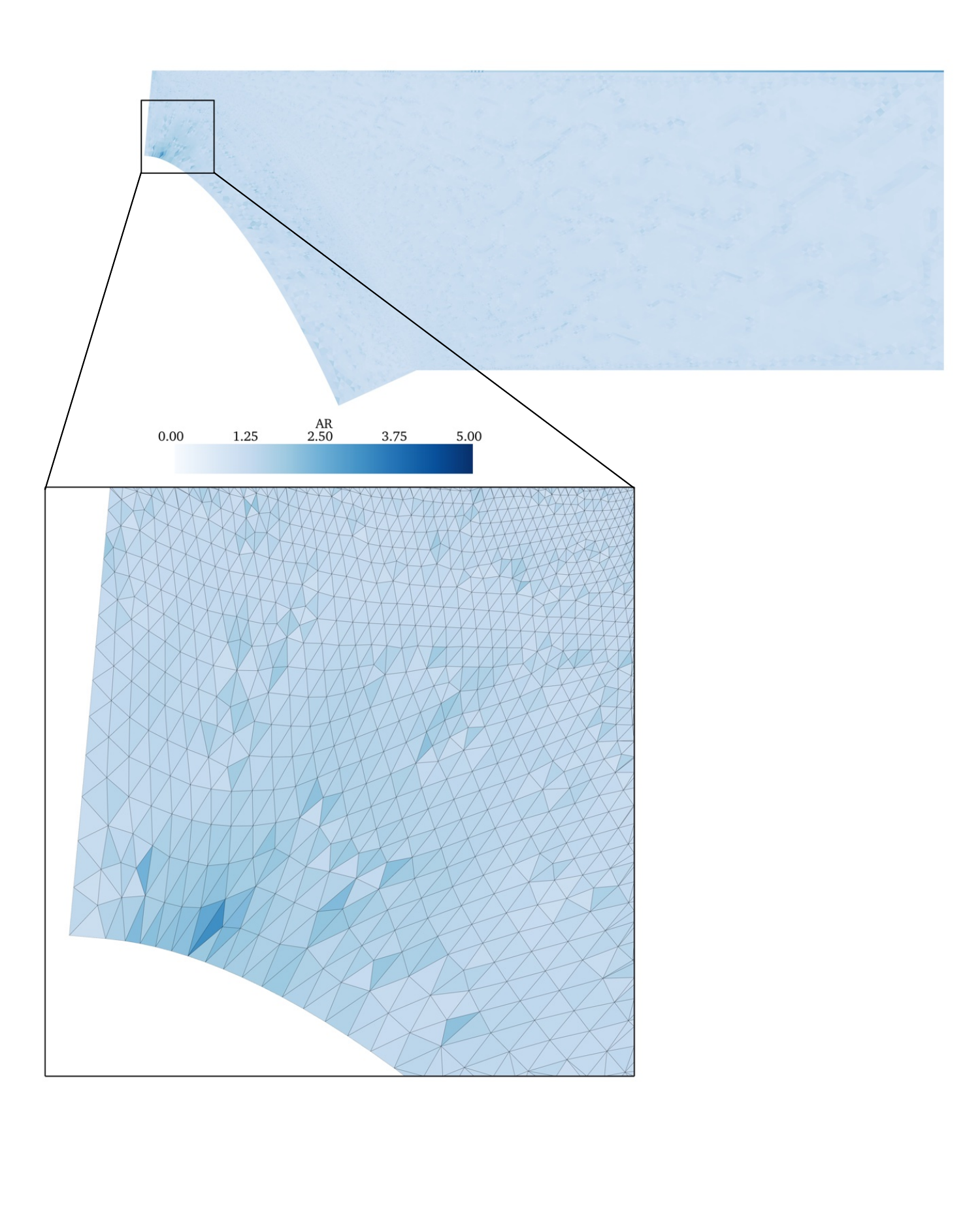}
     \caption{The morphed mesh with  $\alpha = 3.5 \times 10^{-3}$, colored by the Aspect Ratio (AR) of each cell. The colormap is saturated at AR = 5 for better visualization, though the max AR is 13.}
     \label{fig:mesh_quality_spatial_alpha_high}
\end{figure}

\begin{figure}
     \centering
     \includegraphics[width=\textwidth]{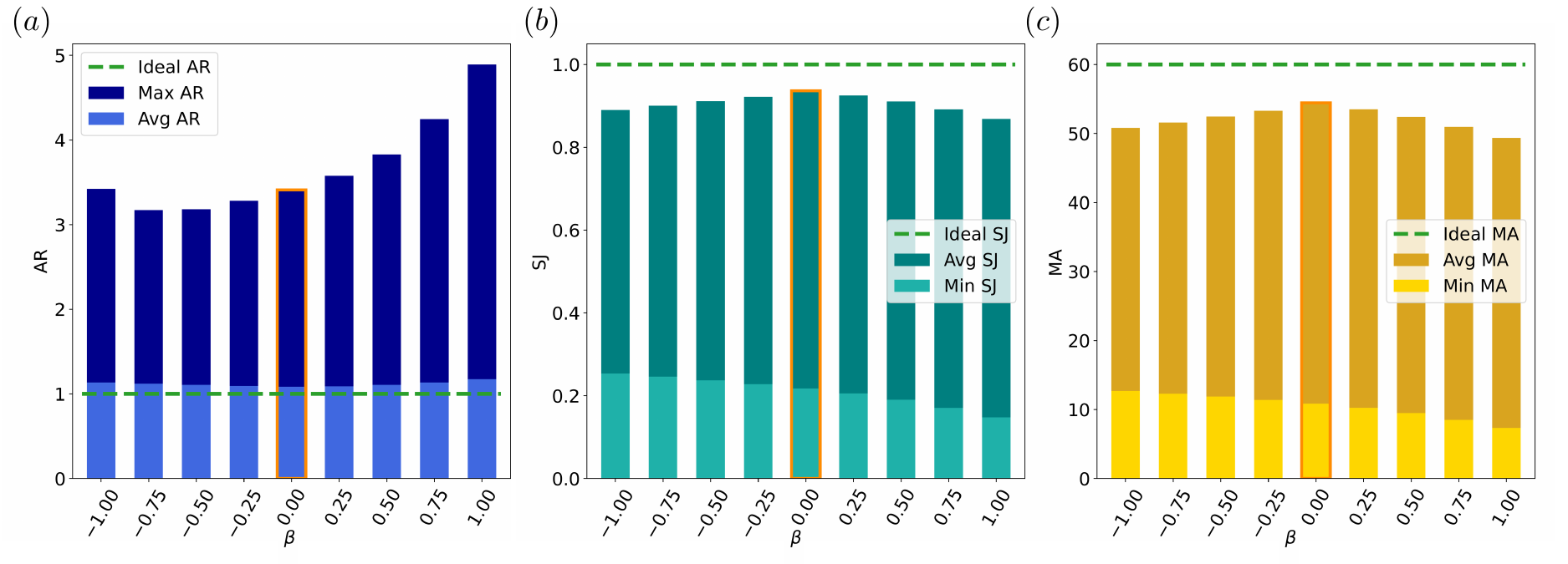}
     \caption{Bar plots of the mesh quality as $\beta$ varies, measured by (a) the Aspect Ratio (AR), (b) the Scaled Jacobian (SJ), and (c) the Minimum Angle (MA) metrics. The ideal values for each metric is shown by the dotted green line, and the mesh quality for the reference (not morphed) mesh is outlined in orange.}
     \label{fig:mesh_qual_barplots_beta}
\end{figure}

\begin{figure}
     \centering
     \includegraphics[width=\textwidth]{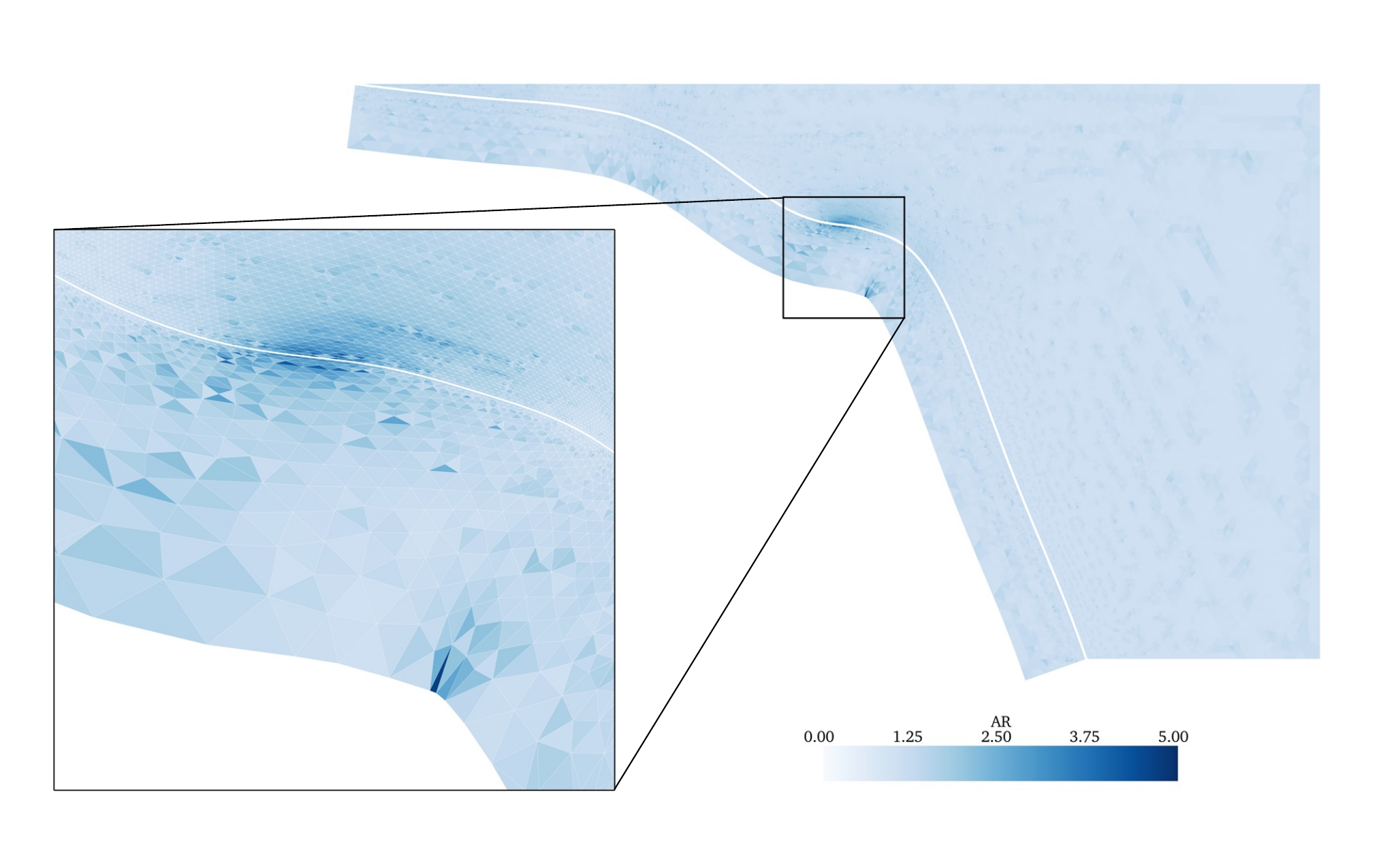}
     \caption{The morphed mesh with  $\beta = 1.0 $, colored by the Aspect Ratio of each cell. The colormap is saturated at AR = 5 for better visualization, though the max AR is 26.}
     \label{fig:mesh_quality_spatial_beta}
\end{figure}

\newpage

\ 

\

\newpage

\bibliography{references}

\begin{thebibliography}{}

\bibitem [\protect \citeauthoryear {%
Aki%
}{%
Aki%
}{%
{\protect \APACyear {1989}}%
}]{%
aki_1989}
\APACinsertmetastar {%
aki_1989}%
\begin{APACrefauthors}%
Aki, K.%
\end{APACrefauthors}%
\unskip\
\newblock
\APACrefYearMonthDay{1989}{}{}.
\newblock
{\BBOQ}\APACrefatitle {Geometric features of a fault zone related to the
  nucleation and termination of an earthquake rupture} {Geometric features of a
  fault zone related to the nucleation and termination of an earthquake
  rupture}.{\BBCQ}
\newblock
\APACjournalVolNumPages{US Geological Survey Open File Repository,
  89}{315}{}{1--9}.
\PrintBackRefs{\CurrentBib}

\bibitem [\protect \citeauthoryear {%
Alexa%
}{%
Alexa%
}{%
{\protect \APACyear {2002}}%
}]{%
alexa_2002}
\APACinsertmetastar {%
alexa_2002}%
\begin{APACrefauthors}%
Alexa, M.%
\end{APACrefauthors}%
\unskip\
\newblock
\APACrefYearMonthDay{2002}{}{}.
\newblock
{\BBOQ}\APACrefatitle {Recent Advances in Mesh Morphing} {Recent advances in
  mesh morphing}.{\BBCQ}
\newblock
\APACjournalVolNumPages{Computer Graphics Forum}{21}{2}{173-198}.
\newblock
\begin{APACrefDOI} \doi{https://doi.org/10.1111/1467-8659.00575}
  \end{APACrefDOI}
\PrintBackRefs{\CurrentBib}

\bibitem [\protect \citeauthoryear {%
Andrews%
}{%
Andrews%
}{%
{\protect \APACyear {1976}}%
}]{%
andrews_1976}
\APACinsertmetastar {%
andrews_1976}%
\begin{APACrefauthors}%
Andrews, D\BPBI J.%
\end{APACrefauthors}%
\unskip\
\newblock
\APACrefYearMonthDay{1976}{}{}.
\newblock
{\BBOQ}\APACrefatitle {Rupture velocity of plane strain shear cracks} {Rupture
  velocity of plane strain shear cracks}.{\BBCQ}
\newblock
\APACjournalVolNumPages{Journal of Geophysical Research
  (1896-1977)}{81}{32}{5679-5687}.
\newblock
\begin{APACrefDOI} \doi{https://doi.org/10.1029/JB081i032p05679}
  \end{APACrefDOI}
\PrintBackRefs{\CurrentBib}

\bibitem [\protect \citeauthoryear {%
Andrews%
}{%
Andrews%
}{%
{\protect \APACyear {2005}}%
}]{%
andrews_2005}
\APACinsertmetastar {%
andrews_2005}%
\begin{APACrefauthors}%
Andrews, D\BPBI J.%
\end{APACrefauthors}%
\unskip\
\newblock
\APACrefYearMonthDay{2005}{}{}.
\newblock
{\BBOQ}\APACrefatitle {Rupture dynamics with energy loss outside the slip zone}
  {Rupture dynamics with energy loss outside the slip zone}.{\BBCQ}
\newblock
\APACjournalVolNumPages{Journal of Geophysical Research: Solid
  Earth}{110}{B1}{}.
\newblock
\begin{APACrefDOI} \doi{https://doi.org/10.1029/2004JB003191} \end{APACrefDOI}
\PrintBackRefs{\CurrentBib}

\bibitem [\protect \citeauthoryear {%
Bai%
\ \BBA {} Ampuero%
}{%
Bai%
\ \BBA {} Ampuero%
}{%
{\protect \APACyear {2017}}%
}]{%
BaiAmpuero2017}
\APACinsertmetastar {%
BaiAmpuero2017}%
\begin{APACrefauthors}%
Bai, K.%
\BCBT {}\ \BBA {} Ampuero, J\BHBI P.%
\end{APACrefauthors}%
\unskip\
\newblock
\APACrefYearMonthDay{2017}{}{}.
\newblock
{\BBOQ}\APACrefatitle {Effect of seismogenic depth and background stress on
  physical limits of earthquake rupture across fault step overs} {Effect of
  seismogenic depth and background stress on physical limits of earthquake
  rupture across fault step overs}.{\BBCQ}
\newblock
\APACjournalVolNumPages{Journal of Geophysical Research: Solid
  Earth}{122}{12}{10--280}.
\newblock
\begin{APACrefDOI} \doi{https://doi.org/10.1002/2017JB014848} \end{APACrefDOI}
\PrintBackRefs{\CurrentBib}

\bibitem [\protect \citeauthoryear {%
Baker%
}{%
Baker%
}{%
{\protect \APACyear {2002}}%
}]{%
baker_2002}
\APACinsertmetastar {%
baker_2002}%
\begin{APACrefauthors}%
Baker, T\BPBI J.%
\end{APACrefauthors}%
\unskip\
\newblock
\APACrefYearMonthDay{2002}{}{}.
\newblock
{\BBOQ}\APACrefatitle {Mesh movement and metamorphosis} {Mesh movement and
  metamorphosis}.{\BBCQ}
\newblock
\APACjournalVolNumPages{Engineering with Computers}{18}{}{188--198}.
\newblock
\begin{APACrefDOI} \doi{https://doi.org/10.1007/s003660200017} \end{APACrefDOI}
\PrintBackRefs{\CurrentBib}

\bibitem [\protect \citeauthoryear {%
Ben-Zion%
\ \BBA {} Sammis%
}{%
Ben-Zion%
\ \BBA {} Sammis%
}{%
{\protect \APACyear {2003}}%
}]{%
BenZionSammis2003}
\APACinsertmetastar {%
BenZionSammis2003}%
\begin{APACrefauthors}%
Ben-Zion, Y.%
\BCBT {}\ \BBA {} Sammis, C\BPBI G.%
\end{APACrefauthors}%
\unskip\
\newblock
\APACrefYearMonthDay{2003}{}{}.
\newblock
{\BBOQ}\APACrefatitle {Characterization of fault zones} {Characterization of
  fault zones}.{\BBCQ}
\newblock
\APACjournalVolNumPages{Pure and Applied Geophysics}{160}{3}{677--715}.
\newblock
\begin{APACrefDOI} \doi{https://doi.org/10.1007/PL00012554} \end{APACrefDOI}
\PrintBackRefs{\CurrentBib}

\bibitem [\protect \citeauthoryear {%
Berkooz%
, Holmes%
\BCBL {}\ \BBA {} Lumley%
}{%
Berkooz%
\ \protect \BOthers {.}}{%
{\protect \APACyear {1993}}%
}]{%
berkooz_holmes_lumley_1993}
\APACinsertmetastar {%
berkooz_holmes_lumley_1993}%
\begin{APACrefauthors}%
Berkooz, G.%
, Holmes, P.%
\BCBL {}\ \BBA {} Lumley, J\BPBI L.%
\end{APACrefauthors}%
\unskip\
\newblock
\APACrefYearMonthDay{1993}{}{}.
\newblock
{\BBOQ}\APACrefatitle {The Proper Orthogonal Decomposition in the Analysis of
  Turbulent Flows} {The proper orthogonal decomposition in the analysis of
  turbulent flows}{\BBCQ}\ [Journal Article].
\newblock
\APACjournalVolNumPages{Annual Review of Fluid Mechanics}{25}{Volume 25,
  1993}{539-575}.
\newblock
\begin{APACrefDOI} \doi{https://doi.org/10.1146/annurev.fl.25.010193.002543}
  \end{APACrefDOI}
\PrintBackRefs{\CurrentBib}

\bibitem [\protect \citeauthoryear {%
Biancolini%
}{%
Biancolini%
}{%
{\protect \APACyear {2017}}%
}]{%
biancolini_2017}
\APACinsertmetastar {%
biancolini_2017}%
\begin{APACrefauthors}%
Biancolini, M\BPBI E.%
\end{APACrefauthors}%
\unskip\
\newblock
\APACrefYearMonthDay{2017}{}{}.
\newblock
{\BBOQ}\APACrefatitle {{FSI} Workflow Using Advanced {RBF} Mesh Morphing}
  {{FSI} workflow using advanced {RBF} mesh morphing}.{\BBCQ}
\newblock
\BIn{} \APACrefbtitle {{F}ast {R}adial {B}asis {F}unctions for {E}ngineering
  {A}pplications} {{F}ast {R}adial {B}asis {F}unctions for {E}ngineering
  {A}pplications}\ (\BPGS\ 225--256).
\newblock
\APACaddressPublisher{Cham}{Springer International Publishing}.
\newblock
\begin{APACrefDOI} \doi{https://doi.org/10.1007/978-3-319-75011-8_10}
  \end{APACrefDOI}
\PrintBackRefs{\CurrentBib}

\bibitem [\protect \citeauthoryear {%
Biemiller%
, Gabriel%
, May%
\BCBL {}\ \BBA {} Staisch%
}{%
Biemiller%
\ \protect \BOthers {.}}{%
{\protect \APACyear {2024}}%
}]{%
biemiller_et_al_2024}
\APACinsertmetastar {%
biemiller_et_al_2024}%
\begin{APACrefauthors}%
Biemiller, J.%
, Gabriel, A\BHBI A.%
, May, D\BPBI A.%
\BCBL {}\ \BBA {} Staisch, L.%
\end{APACrefauthors}%
\unskip\
\newblock
\APACrefYearMonthDay{2024}{}{}.
\newblock
{\BBOQ}\APACrefatitle {Subduction Zone Geometry Modulates the Megathrust
  Earthquake Cycle: Magnitude, Recurrence, and Variability} {Subduction zone
  geometry modulates the megathrust earthquake cycle: Magnitude, recurrence,
  and variability}.{\BBCQ}
\newblock
\APACjournalVolNumPages{Journal of Geophysical Research: Solid
  Earth}{129}{8}{e2024JB029191}.
\newblock
\begin{APACrefDOI} \doi{https://doi.org/10.1029/2024JB029191} \end{APACrefDOI}
\PrintBackRefs{\CurrentBib}

\bibitem [\protect \citeauthoryear {%
Bletery%
\ \protect \BOthers {.}}{%
Bletery%
\ \protect \BOthers {.}}{%
{\protect \APACyear {2016}}%
}]{%
bletery_et_al_2016}
\APACinsertmetastar {%
bletery_et_al_2016}%
\begin{APACrefauthors}%
Bletery, Q.%
, Thomas, A\BPBI M.%
, Rempel, A\BPBI W.%
, Karlstrom, L.%
, Sladen, A.%
\BCBL {}\ \BBA {} Barros, L\BPBI D.%
\end{APACrefauthors}%
\unskip\
\newblock
\APACrefYearMonthDay{2016}{}{}.
\newblock
{\BBOQ}\APACrefatitle {Mega-earthquakes rupture flat megathrusts}
  {Mega-earthquakes rupture flat megathrusts}.{\BBCQ}
\newblock
\APACjournalVolNumPages{Science}{354}{6315}{1027-1031}.
\newblock
\begin{APACrefDOI}
  \doi{https://www.science.org/doi/abs/10.1126/science.aag0482}
  \end{APACrefDOI}
\PrintBackRefs{\CurrentBib}

\bibitem [\protect \citeauthoryear {%
Botsch%
\ \BBA {} Kobbelt%
}{%
Botsch%
\ \BBA {} Kobbelt%
}{%
{\protect \APACyear {2005}}%
}]{%
botsch_kobbelt_2005}
\APACinsertmetastar {%
botsch_kobbelt_2005}%
\begin{APACrefauthors}%
Botsch, M.%
\BCBT {}\ \BBA {} Kobbelt, L.%
\end{APACrefauthors}%
\unskip\
\newblock
\APACrefYearMonthDay{2005}{}{}.
\newblock
{\BBOQ}\APACrefatitle {Real-Time Shape Editing using Radial Basis Functions}
  {Real-time shape editing using radial basis functions}.{\BBCQ}
\newblock
\APACjournalVolNumPages{Computer Graphics Forum}{24}{3}{611-621}.
\newblock
\begin{APACrefDOI} \doi{https://doi.org/10.1111/j.1467-8659.2005.00886.x}
  \end{APACrefDOI}
\PrintBackRefs{\CurrentBib}

\bibitem [\protect \citeauthoryear {%
Bouchon%
, Campillo%
\BCBL {}\ \BBA {} Cotton%
}{%
Bouchon%
\ \protect \BOthers {.}}{%
{\protect \APACyear {1998}}%
}]{%
bouchon_et_al_1998}
\APACinsertmetastar {%
bouchon_et_al_1998}%
\begin{APACrefauthors}%
Bouchon, M.%
, Campillo, M.%
\BCBL {}\ \BBA {} Cotton, F.%
\end{APACrefauthors}%
\unskip\
\newblock
\APACrefYearMonthDay{1998}{}{}.
\newblock
{\BBOQ}\APACrefatitle {Stress field associated with the rupture of the 1992
  {L}anders, {C}alifornia, earthquake and its implications concerning the fault
  strength at the onset of the earthquake} {Stress field associated with the
  rupture of the 1992 {L}anders, {C}alifornia, earthquake and its implications
  concerning the fault strength at the onset of the earthquake}.{\BBCQ}
\newblock
\APACjournalVolNumPages{Journal of Geophysical Research: Solid
  Earth}{103}{B9}{21091-21097}.
\newblock
\begin{APACrefDOI} \doi{https://doi.org/10.1029/98JB01982} \end{APACrefDOI}
\PrintBackRefs{\CurrentBib}

\bibitem [\protect \citeauthoryear {%
Breuer%
, Heinecke%
\BCBL {}\ \BBA {} Bader%
}{%
Breuer%
\ \protect \BOthers {.}}{%
{\protect \APACyear {2016}}%
}]{%
Breuer_2016}
\APACinsertmetastar {%
Breuer_2016}%
\begin{APACrefauthors}%
Breuer, A.%
, Heinecke, A.%
\BCBL {}\ \BBA {} Bader, M.%
\end{APACrefauthors}%
\unskip\
\newblock
\APACrefYearMonthDay{2016}{}{}.
\newblock
{\BBOQ}\APACrefatitle {Petascale Local Time Stepping for the {ADER-DG} Finite
  Element Method} {Petascale local time stepping for the {ADER-DG} finite
  element method}.{\BBCQ}
\newblock
\BIn{} \APACrefbtitle {{2016 IEEE International Parallel and Distributed
  Processing Symposium (IPDPS)}} {{2016 IEEE International Parallel and
  Distributed Processing Symposium (IPDPS)}}\ (\BPG~854-863).
\newblock
\begin{APACrefDOI} \doi{https://doi.org/10.1109/IPDPS.2016.109}
  \end{APACrefDOI}
\PrintBackRefs{\CurrentBib}

\bibitem [\protect \citeauthoryear {%
Brown%
\ \BBA {} Scholz%
}{%
Brown%
\ \BBA {} Scholz%
}{%
{\protect \APACyear {1985}}%
}]{%
Brown1985}
\APACinsertmetastar {%
Brown1985}%
\begin{APACrefauthors}%
Brown, S\BPBI R.%
\BCBT {}\ \BBA {} Scholz, C\BPBI H.%
\end{APACrefauthors}%
\unskip\
\newblock
\APACrefYearMonthDay{1985}{}{}.
\newblock
{\BBOQ}\APACrefatitle {Broad bandwidth study of the topography of natural rock
  surfaces} {Broad bandwidth study of the topography of natural rock
  surfaces}.{\BBCQ}
\newblock
\APACjournalVolNumPages{Journal of Geophysical Research: Solid
  Earth}{90}{B14}{12575--12582}.
\newblock
\begin{APACrefDOI} \doi{https://doi.org/10.1029/JB090iB14p12575}
  \end{APACrefDOI}
\PrintBackRefs{\CurrentBib}

\bibitem [\protect \citeauthoryear {%
Bui-Thanh%
, Damodaran%
\BCBL {}\ \BBA {} Willcox%
}{%
Bui-Thanh%
\ \protect \BOthers {.}}{%
{\protect \APACyear {2003}}%
}]{%
bui-thanh_2003}
\APACinsertmetastar {%
bui-thanh_2003}%
\begin{APACrefauthors}%
Bui-Thanh, T.%
, Damodaran, M.%
\BCBL {}\ \BBA {} Willcox, K.%
\end{APACrefauthors}%
\unskip\
\newblock
\APACrefYearMonthDay{2003}{}{}.
\newblock
{\BBOQ}\APACrefatitle {Proper Orthogonal Decomposition Extensions for
  Parametric Applications in Compressible Aerodynamics} {Proper orthogonal
  decomposition extensions for parametric applications in compressible
  aerodynamics}.{\BBCQ}
\newblock
\BIn{} \APACrefbtitle {21st {AIAA} {A}pplied {A}erodynamics {C}onference.}
  {21st {AIAA} {A}pplied {A}erodynamics {C}onference.}
\newblock
\begin{APACrefDOI} \doi{https://doi.org/10.2514/6.2003-4213} \end{APACrefDOI}
\PrintBackRefs{\CurrentBib}

\bibitem [\protect \citeauthoryear {%
Candela%
\ \protect \BOthers {.}}{%
Candela%
\ \protect \BOthers {.}}{%
{\protect \APACyear {2012}}%
}]{%
Candela2012}
\APACinsertmetastar {%
Candela2012}%
\begin{APACrefauthors}%
Candela, T.%
, Renard, F.%
, Klinger, Y.%
, Mair, K.%
, Schmittbuhl, J.%
\BCBL {}\ \BBA {} Brodsky, E\BPBI E.%
\end{APACrefauthors}%
\unskip\
\newblock
\APACrefYearMonthDay{2012}{}{}.
\newblock
{\BBOQ}\APACrefatitle {Roughness of fault surfaces over nine decades of length
  scales} {Roughness of fault surfaces over nine decades of length
  scales}.{\BBCQ}
\newblock
\APACjournalVolNumPages{Journal of Geophysical Research: Solid
  Earth}{117}{B8}{}.
\newblock
\begin{APACrefDOI} \doi{https://doi.org/10.1029/2011JB009041} \end{APACrefDOI}
\PrintBackRefs{\CurrentBib}

\bibitem [\protect \citeauthoryear {%
Currie%
, Wang%
, Hyndman%
\BCBL {}\ \BBA {} He%
}{%
Currie%
\ \protect \BOthers {.}}{%
{\protect \APACyear {2004}}%
}]{%
currie_et_al_2004}
\APACinsertmetastar {%
currie_et_al_2004}%
\begin{APACrefauthors}%
Currie, C.%
, Wang, K.%
, Hyndman, R\BPBI D.%
\BCBL {}\ \BBA {} He, J.%
\end{APACrefauthors}%
\unskip\
\newblock
\APACrefYearMonthDay{2004}{}{}.
\newblock
{\BBOQ}\APACrefatitle {The thermal effects of steady-state slab-driven mantle
  flow above a subducting plate: {T}he {C}ascadia subduction zone and backarc}
  {The thermal effects of steady-state slab-driven mantle flow above a
  subducting plate: {T}he {C}ascadia subduction zone and backarc}.{\BBCQ}
\newblock
\APACjournalVolNumPages{Earth and Planetary Science Letters}{223}{1}{35-48}.
\newblock
\begin{APACrefDOI} \doi{https://doi.org/10.1016/j.epsl.2004.04.020}
  \end{APACrefDOI}
\PrintBackRefs{\CurrentBib}

\bibitem [\protect \citeauthoryear {%
{de Boer}%
, {van der Schoot}%
\BCBL {}\ \BBA {} Bijl%
}{%
{de Boer}%
\ \protect \BOthers {.}}{%
{\protect \APACyear {2007}}%
}]{%
de_boer_2007}
\APACinsertmetastar {%
de_boer_2007}%
\begin{APACrefauthors}%
{de Boer}, A.%
, {van der Schoot}, M.%
\BCBL {}\ \BBA {} Bijl, H.%
\end{APACrefauthors}%
\unskip\
\newblock
\APACrefYearMonthDay{2007}{}{}.
\newblock
{\BBOQ}\APACrefatitle {Mesh deformation based on radial basis function
  interpolation} {Mesh deformation based on radial basis function
  interpolation}.{\BBCQ}
\newblock
\APACjournalVolNumPages{Computers \& Structures}{85}{11}{784-795}.
\newblock
\begin{APACrefDOI} \doi{https://doi.org/10.1016/j.compstruc.2007.01.013}
  \end{APACrefDOI}
\PrintBackRefs{\CurrentBib}

\bibitem [\protect \citeauthoryear {%
Demo%
, Tezzele%
, Mola%
\BCBL {}\ \BBA {} Rozza%
}{%
Demo%
\ \protect \BOthers {.}}{%
{\protect \APACyear {2021}}%
}]{%
demo_et_al_2021}
\APACinsertmetastar {%
demo_et_al_2021}%
\begin{APACrefauthors}%
Demo, N.%
, Tezzele, M.%
, Mola, A.%
\BCBL {}\ \BBA {} Rozza, G.%
\end{APACrefauthors}%
\unskip\
\newblock
\APACrefYearMonthDay{2021}{}{}.
\newblock
{\BBOQ}\APACrefatitle {Hull Shape Design Optimization with Parameter Space and
  Model Reductions, and Self-Learning Mesh Morphing} {Hull shape design
  optimization with parameter space and model reductions, and self-learning
  mesh morphing}.{\BBCQ}
\newblock
\APACjournalVolNumPages{Journal of Marine Science and Engineering}{9}{2}{}.
\newblock
\begin{APACrefDOI} \doi{https://doi.org/10.3390/jmse9020185} \end{APACrefDOI}
\PrintBackRefs{\CurrentBib}

\bibitem [\protect \citeauthoryear {%
Du%
, Liang%
, Lei%
, Xu%
\BCBL {}\ \BBA {} Xie%
}{%
Du%
\ \protect \BOthers {.}}{%
{\protect \APACyear {2023}}%
}]{%
du_et_al_2023}
\APACinsertmetastar {%
du_et_al_2023}%
\begin{APACrefauthors}%
Du, X.%
, Liang, J.%
, Lei, J.%
, Xu, J.%
\BCBL {}\ \BBA {} Xie, P.%
\end{APACrefauthors}%
\unskip\
\newblock
\APACrefYearMonthDay{2023}{}{}.
\newblock
{\BBOQ}\APACrefatitle {A radial-basis function mesh morphing and {B}ayesian
  optimization framework for vehicle crashworthiness design} {A radial-basis
  function mesh morphing and {B}ayesian optimization framework for vehicle
  crashworthiness design}.{\BBCQ}
\newblock
\APACjournalVolNumPages{Structural and Multidisciplinary
  Optimization}{66}{3}{64}.
\newblock
\begin{APACrefDOI} \doi{https://doi.org/10.1007/s00158-023-03496-x}
  \end{APACrefDOI}
\PrintBackRefs{\CurrentBib}

\bibitem [\protect \citeauthoryear {%
Dumbser%
, Käser%
\BCBL {}\ \BBA {} Toro%
}{%
Dumbser%
\ \protect \BOthers {.}}{%
{\protect \APACyear {2007}}%
}]{%
Dumbser_2007}
\APACinsertmetastar {%
Dumbser_2007}%
\begin{APACrefauthors}%
Dumbser, M.%
, Käser, M.%
\BCBL {}\ \BBA {} Toro, E\BPBI F.%
\end{APACrefauthors}%
\unskip\
\newblock
\APACrefYearMonthDay{2007}{}{}.
\newblock
{\BBOQ}\APACrefatitle {{An arbitrary high-order Discontinuous Galerkin method
  for elastic waves on unstructured meshes - V. Local time stepping and
  p-adaptivity}} {{An arbitrary high-order Discontinuous Galerkin method for
  elastic waves on unstructured meshes - V. Local time stepping and
  p-adaptivity}}.{\BBCQ}
\newblock
\APACjournalVolNumPages{Geophysical Journal International}{171}{2}{695-717}.
\newblock
\begin{APACrefDOI} \doi{https://doi.org/10.1111/j.1365-246X.2007.03427.x}
  \end{APACrefDOI}
\PrintBackRefs{\CurrentBib}

\bibitem [\protect \citeauthoryear {%
England%
\ \BBA {} May%
}{%
England%
\ \BBA {} May%
}{%
{\protect \APACyear {2021}}%
}]{%
england_may_2021}
\APACinsertmetastar {%
england_may_2021}%
\begin{APACrefauthors}%
England, P.%
\BCBT {}\ \BBA {} May, D.%
\end{APACrefauthors}%
\unskip\
\newblock
\APACrefYearMonthDay{2021}{}{}.
\newblock
{\BBOQ}\APACrefatitle {The Global Range of Temperatures on Convergent Plate
  Interfaces} {The global range of temperatures on convergent plate
  interfaces}.{\BBCQ}
\newblock
\APACjournalVolNumPages{Geochemistry, Geophysics,
  Geosystems}{22}{8}{e2021GC009849}.
\newblock
\begin{APACrefDOI} \doi{https://doi.org/10.1029/2021GC009849} \end{APACrefDOI}
\PrintBackRefs{\CurrentBib}

\bibitem [\protect \citeauthoryear {%
Fasshauer%
}{%
Fasshauer%
}{%
{\protect \APACyear {2007}}%
}]{%
fasshauer_2007}
\APACinsertmetastar {%
fasshauer_2007}%
\begin{APACrefauthors}%
Fasshauer, G\BPBI E.%
\end{APACrefauthors}%
\unskip\
\newblock
\APACrefYear{2007}.
\newblock
\APACrefbtitle {{Meshfree Approximation Methods with Matlab}} {{Meshfree
  Approximation Methods with Matlab}}.
\newblock
\APACaddressPublisher{}{WORLD SCIENTIFIC}.
\newblock
\begin{APACrefDOI} \doi{https://doi.org/10.1142/6437} \end{APACrefDOI}
\PrintBackRefs{\CurrentBib}

\bibitem [\protect \citeauthoryear {%
Gabriel%
, Garagash%
, Palgunadi%
\BCBL {}\ \BBA {} Mai%
}{%
Gabriel%
\ \protect \BOthers {.}}{%
{\protect \APACyear {2024}}%
}]{%
Gabriel2024}
\APACinsertmetastar {%
Gabriel2024}%
\begin{APACrefauthors}%
Gabriel, A\BHBI A.%
, Garagash, D\BPBI I.%
, Palgunadi, K\BPBI H.%
\BCBL {}\ \BBA {} Mai, P\BPBI M.%
\end{APACrefauthors}%
\unskip\
\newblock
\APACrefYearMonthDay{2024}{}{}.
\newblock
{\BBOQ}\APACrefatitle {Fault size–dependent fracture energy explains
  multiscale seismicity and cascading earthquakes} {Fault size–dependent
  fracture energy explains multiscale seismicity and cascading
  earthquakes}.{\BBCQ}
\newblock
\APACjournalVolNumPages{Science}{385}{6707}{eadj9587}.
\newblock
\begin{APACrefDOI} \doi{https://doi.org/10.1126/science.adj9587}
  \end{APACrefDOI}
\PrintBackRefs{\CurrentBib}

\bibitem [\protect \citeauthoryear {%
Gabriel%
\ \protect \BOthers {.}}{%
Gabriel%
\ \protect \BOthers {.}}{%
{\protect \APACyear {2025}}%
}]{%
seissol}
\APACinsertmetastar {%
seissol}%
\begin{APACrefauthors}%
Gabriel, A\BHBI A.%
, Kurapati, V.%
, Niu, Z.%
, Schliwa, N.%
, Schneller, D.%
, Ulrich, T.%
\BDBL {}Bader, M.%
\end{APACrefauthors}%
\unskip\
\newblock
\APACrefYearMonthDay{2025}{{\APACmonth{06}}}{}.
\newblock
\APACrefbtitle {SeisSol.} {Seissol.}
\newblock
\APACaddressPublisher{}{Zenodo}.
\newblock
\begin{APACrefURL} \url{https://doi.org/10.5281/zenodo.15685917}
  \end{APACrefURL}
\newblock
\begin{APACrefDOI} \doi{10.5281/zenodo.15685917} \end{APACrefDOI}
\PrintBackRefs{\CurrentBib}

\bibitem [\protect \citeauthoryear {%
Galis%
\ \protect \BOthers {.}}{%
Galis%
\ \protect \BOthers {.}}{%
{\protect \APACyear {2014}}%
}]{%
galis_et_al_2014}
\APACinsertmetastar {%
galis_et_al_2014}%
\begin{APACrefauthors}%
Galis, M.%
, Pelties, C.%
, Kristek, J.%
, Moczo, P.%
, Ampuero, J\BHBI P.%
\BCBL {}\ \BBA {} Mai, P\BPBI M.%
\end{APACrefauthors}%
\unskip\
\newblock
\APACrefYearMonthDay{2014}{}{}.
\newblock
{\BBOQ}\APACrefatitle {On the initiation of sustained slip-weakening ruptures
  by localized stresses} {On the initiation of sustained slip-weakening
  ruptures by localized stresses}.{\BBCQ}
\newblock
\APACjournalVolNumPages{Geophysical Journal International}{200}{2}{890-909}.
\newblock
\begin{APACrefDOI} \doi{https://doi.org/10.1093/gji/ggu436} \end{APACrefDOI}
\PrintBackRefs{\CurrentBib}

\bibitem [\protect \citeauthoryear {%
Geronzi%
\ \protect \BOthers {.}}{%
Geronzi%
\ \protect \BOthers {.}}{%
{\protect \APACyear {2020}}%
}]{%
geronzi_et_al_2020}
\APACinsertmetastar {%
geronzi_et_al_2020}%
\begin{APACrefauthors}%
Geronzi, L.%
, Gasparotti, E.%
, Capellini, K.%
, Cella, U.%
, Groth, C.%
, Porziani, S.%
\BDBL {}Biancolini, M\BPBI E.%
\end{APACrefauthors}%
\unskip\
\newblock
\APACrefYearMonthDay{2020}{}{}.
\newblock
{\BBOQ}\APACrefatitle {Advanced {R}adial {B}asis {F}unctions mesh morphing for
  high fidelity {F}luid-{S}tructure {I}nteraction with known movement of the
  walls: simulation of an aortic valve} {Advanced {R}adial {B}asis {F}unctions
  mesh morphing for high fidelity {F}luid-{S}tructure {I}nteraction with known
  movement of the walls: simulation of an aortic valve}.{\BBCQ}
\newblock
\BIn{} \APACrefbtitle {{Computational Science--ICCS 2020: 20th International
  Conference, Amsterdam, The Netherlands, June 3--5, 2020, Proceedings, Part VI
  20}} {{Computational Science--ICCS 2020: 20th International Conference,
  Amsterdam, The Netherlands, June 3--5, 2020, Proceedings, Part VI 20}}\
  (\BPGS\ 280--293).
\newblock
\begin{APACrefDOI} \doi{https://doi.org/10.1007/978-3-030-50433-5_22}
  \end{APACrefDOI}
\PrintBackRefs{\CurrentBib}

\bibitem [\protect \citeauthoryear {%
T.~Gerya%
}{%
T.~Gerya%
}{%
{\protect \APACyear {2022}}%
}]{%
gerya_2022}
\APACinsertmetastar {%
gerya_2022}%
\begin{APACrefauthors}%
Gerya, T.%
\end{APACrefauthors}%
\unskip\
\newblock
\APACrefYearMonthDay{2022}{}{}.
\newblock
{\BBOQ}\APACrefatitle {{Numerical modeling of subduction: State of the art and
  future directions}} {{Numerical modeling of subduction: State of the art and
  future directions}}.{\BBCQ}
\newblock
\APACjournalVolNumPages{Geosphere}{18}{2}{503--561}.
\PrintBackRefs{\CurrentBib}

\bibitem [\protect \citeauthoryear {%
T\BPBI V.~Gerya%
, Stöckhert%
\BCBL {}\ \BBA {} Perchuk%
}{%
T\BPBI V.~Gerya%
\ \protect \BOthers {.}}{%
{\protect \APACyear {2002}}%
}]{%
gerya_et_al_2002}
\APACinsertmetastar {%
gerya_et_al_2002}%
\begin{APACrefauthors}%
Gerya, T\BPBI V.%
, Stöckhert, B.%
\BCBL {}\ \BBA {} Perchuk, A\BPBI L.%
\end{APACrefauthors}%
\unskip\
\newblock
\APACrefYearMonthDay{2002}{}{}.
\newblock
{\BBOQ}\APACrefatitle {Exhumation of high-pressure metamorphic rocks in a
  subduction channel: {A} numerical simulation} {Exhumation of high-pressure
  metamorphic rocks in a subduction channel: {A} numerical simulation}.{\BBCQ}
\newblock
\APACjournalVolNumPages{Tectonics}{21}{6}{6-1-6-19}.
\newblock
\begin{APACrefDOI} \doi{https://doi.org/10.1029/2002TC001406} \end{APACrefDOI}
\PrintBackRefs{\CurrentBib}

\bibitem [\protect \citeauthoryear {%
Geuzaine%
\ \BBA {} Remacle%
}{%
Geuzaine%
\ \BBA {} Remacle%
}{%
{\protect \APACyear {2009}}%
}]{%
gmsh}
\APACinsertmetastar {%
gmsh}%
\begin{APACrefauthors}%
Geuzaine, C.%
\BCBT {}\ \BBA {} Remacle, J\BHBI F.%
\end{APACrefauthors}%
\unskip\
\newblock
\APACrefYearMonthDay{2009}{}{}.
\newblock
{\BBOQ}\APACrefatitle {Gmsh: a three-dimensional finite element mesh generator
  with built-in pre- and post-processing facilities.} {Gmsh: a
  three-dimensional finite element mesh generator with built-in pre- and
  post-processing facilities.}{\BBCQ}
\newblock
\APACjournalVolNumPages{International Journal for Numerical Methods in
  Engineering}{79(11)}{}{1309-1331}.
\newblock
\begin{APACrefDOI} \doi{https://doi.org/10.1002/nme.2579} \end{APACrefDOI}
\PrintBackRefs{\CurrentBib}

\bibitem [\protect \citeauthoryear {%
Harris%
\ \protect \BOthers {.}}{%
Harris%
\ \protect \BOthers {.}}{%
{\protect \APACyear {2018}}%
}]{%
harris_et_al_2018}
\APACinsertmetastar {%
harris_et_al_2018}%
\begin{APACrefauthors}%
Harris, R\BPBI A.%
, Barall, M.%
, Aagaard, B.%
, Ma, S.%
, Roten, D.%
, Olsen, K.%
\BDBL {}Dalguer, L.%
\end{APACrefauthors}%
\unskip\
\newblock
\APACrefYearMonthDay{2018}{}{}.
\newblock
{\BBOQ}\APACrefatitle {A Suite of Exercises for Verifying Dynamic Earthquake
  Rupture Codes} {A suite of exercises for verifying dynamic earthquake rupture
  codes}.{\BBCQ}
\newblock
\APACjournalVolNumPages{Seismological Research Letters}{89}{3}{1146-1162}.
\newblock
\begin{APACrefDOI} \doi{https://doi.org/10.1785/0220170222} \end{APACrefDOI}
\PrintBackRefs{\CurrentBib}

\bibitem [\protect \citeauthoryear {%
Hayek%
\ \protect \BOthers {.}}{%
Hayek%
\ \protect \BOthers {.}}{%
{\protect \APACyear {2024}}%
}]{%
hayek_et_al_2024}
\APACinsertmetastar {%
hayek_et_al_2024}%
\begin{APACrefauthors}%
Hayek, J\BPBI N.%
, Marchandon, M.%
, Li, D.%
, Pousse-Beltran, L.%
, Hollingsworth, J.%
, Li, T.%
\BCBL {}\ \BBA {} Gabriel, A\BHBI A.%
\end{APACrefauthors}%
\unskip\
\newblock
\APACrefYearMonthDay{2024}{}{}.
\newblock
{\BBOQ}\APACrefatitle {Non-Typical Supershear Rupture: Fault Heterogeneity and
  Segmentation Govern Unilateral Supershear and Cascading Multi-Fault Rupture
  in the 2021 \${M}\_{w}\$7.4 {Maduo} Earthquake} {Non-typical supershear
  rupture: Fault heterogeneity and segmentation govern unilateral supershear
  and cascading multi-fault rupture in the 2021 \${M}\_{w}\$7.4 {Maduo}
  earthquake}.{\BBCQ}
\newblock
\APACjournalVolNumPages{Geophysical Research Letters}{51}{20}{e2024GL110128}.
\newblock
\begin{APACrefDOI} \doi{https://doi.org/10.1029/2024GL110128} \end{APACrefDOI}
\PrintBackRefs{\CurrentBib}

\bibitem [\protect \citeauthoryear {%
Hayes%
\ \protect \BOthers {.}}{%
Hayes%
\ \protect \BOthers {.}}{%
{\protect \APACyear {2018}}%
}]{%
slab2}
\APACinsertmetastar {%
slab2}%
\begin{APACrefauthors}%
Hayes, G.%
, Moore, G.%
, Portner, D.%
, Hearne, M.%
, Flamme, H.%
, Furtney, M.%
\BCBL {}\ \BBA {} Smoczyk, G.%
\end{APACrefauthors}%
\unskip\
\newblock
\APACrefYearMonthDay{2018}{}{}.
\newblock
{\BBOQ}\APACrefatitle {Slab2, a comprehensive subduction zone geometry model}
  {Slab2, a comprehensive subduction zone geometry model}.{\BBCQ}
\newblock
\APACjournalVolNumPages{Science}{362}{}{eaat4723}.
\newblock
\begin{APACrefDOI} \doi{https://doi.org/10.1126/science.aat4723}
  \end{APACrefDOI}
\PrintBackRefs{\CurrentBib}

\bibitem [\protect \citeauthoryear {%
Heinecke%
\ \protect \BOthers {.}}{%
Heinecke%
\ \protect \BOthers {.}}{%
{\protect \APACyear {2014}}%
}]{%
heinecke_et_al_2014}
\APACinsertmetastar {%
heinecke_et_al_2014}%
\begin{APACrefauthors}%
Heinecke, A.%
, Breuer, A.%
, Rettenberger, S.%
, Bader, M.%
, Gabriel, A\BHBI A.%
, Pelties, C.%
\BDBL {}Dubey, P.%
\end{APACrefauthors}%
\unskip\
\newblock
\APACrefYearMonthDay{2014}{}{}.
\newblock
{\BBOQ}\APACrefatitle {Petascale High Order Dynamic Rupture Earthquake
  Simulations on Heterogeneous Supercomputers} {Petascale high order dynamic
  rupture earthquake simulations on heterogeneous supercomputers}.{\BBCQ}
\newblock
\BIn{} \APACrefbtitle {{SC '14: Proceedings of the International Conference for
  High Performance Computing, Networking, Storage and Analysis}} {{SC '14:
  Proceedings of the International Conference for High Performance Computing,
  Networking, Storage and Analysis}}\ (\BPG~3-14).
\newblock
\begin{APACrefDOI} \doi{https://doi.org/10.1109/SC.2014.6} \end{APACrefDOI}
\PrintBackRefs{\CurrentBib}

\bibitem [\protect \citeauthoryear {%
Herrera%
, Crempien%
, Cembrano%
\BCBL {}\ \BBA {} Moreno%
}{%
Herrera%
\ \protect \BOthers {.}}{%
{\protect \APACyear {2024}}%
}]{%
Herrera2024}
\APACinsertmetastar {%
Herrera2024}%
\begin{APACrefauthors}%
Herrera, M\BPBI T.%
, Crempien, J\BPBI G\BPBI F.%
, Cembrano, J.%
\BCBL {}\ \BBA {} Moreno, M.%
\end{APACrefauthors}%
\unskip\
\newblock
\APACrefYearMonthDay{2024}{}{}.
\newblock
{\BBOQ}\APACrefatitle {Seismic cycle controlled by subduction geometry: novel
  3-D quasi-dynamic model of Central Chile megathrust} {Seismic cycle
  controlled by subduction geometry: novel 3-d quasi-dynamic model of central
  chile megathrust}.{\BBCQ}
\newblock
\APACjournalVolNumPages{Geophysical Journal International}{237}{2}{772-787}.
\newblock
\begin{APACrefDOI} \doi{https://doi.org/10.1093/gji/ggae069} \end{APACrefDOI}
\PrintBackRefs{\CurrentBib}

\bibitem [\protect \citeauthoryear {%
Hobson%
\ \BBA {} May%
}{%
Hobson%
\ \BBA {} May%
}{%
{\protect \APACyear {2025}}%
}]{%
hobson_may_2025}
\APACinsertmetastar {%
hobson_may_2025}%
\begin{APACrefauthors}%
Hobson, G\BPBI M.%
\BCBT {}\ \BBA {} May, D\BPBI A.%
\end{APACrefauthors}%
\unskip\
\newblock
\APACrefYearMonthDay{2025}{}{}.
\newblock
{\BBOQ}\APACrefatitle {Sensitivity Analysis of the Thermal Structure Within
  Subduction Zones Using Reduced-Order Modeling} {Sensitivity analysis of the
  thermal structure within subduction zones using reduced-order
  modeling}.{\BBCQ}
\newblock
\APACjournalVolNumPages{Geochemistry, Geophysics,
  Geosystems}{26}{5}{e2024GC011937}.
\newblock
\begin{APACrefDOI} \doi{https://doi.org/10.1029/2024GC011937} \end{APACrefDOI}
\PrintBackRefs{\CurrentBib}

\bibitem [\protect \citeauthoryear {%
Hobson%
, May%
\BCBL {}\ \BBA {} Gabriel%
}{%
Hobson%
\ \protect \BOthers {.}}{%
{\protect \APACyear {2025}}%
}]{%
zenodo_repo}
\APACinsertmetastar {%
zenodo_repo}%
\begin{APACrefauthors}%
Hobson, G\BPBI M.%
, May, D\BPBI A.%
\BCBL {}\ \BBA {} Gabriel, A\BHBI A.%
\end{APACrefauthors}%
\unskip\
\newblock
\APACrefYearMonthDay{2025}{}{}.
\newblock
\APACrefbtitle {Quantifying the influence of fault geometry via mesh morphing
  with applications to earthquake dynamic rupture and thermal models of
  subduction.} {Quantifying the influence of fault geometry via mesh morphing
  with applications to earthquake dynamic rupture and thermal models of
  subduction.}
\newblock
\begin{APACrefDOI} \doi{https://doi.org/10.5281/zenodo.15610248}
  \end{APACrefDOI}
\PrintBackRefs{\CurrentBib}

\bibitem [\protect \citeauthoryear {%
Holmes%
, Lumley%
, Berkooz%
\BCBL {}\ \BBA {} Rowley%
}{%
Holmes%
\ \protect \BOthers {.}}{%
{\protect \APACyear {2012}}%
}]{%
lumley}
\APACinsertmetastar {%
lumley}%
\begin{APACrefauthors}%
Holmes, P.%
, Lumley, J\BPBI L.%
, Berkooz, G.%
\BCBL {}\ \BBA {} Rowley, C\BPBI W.%
\end{APACrefauthors}%
\unskip\
\newblock
\APACrefYear{2012}.
\newblock
\APACrefbtitle {Turbulence, Coherent Structures, Dynamical Systems and
  Symmetry} {Turbulence, coherent structures, dynamical systems and symmetry}.
\newblock
\APACaddressPublisher{}{Cambridge University Press, Cambridge}.
\newblock
\begin{APACrefDOI} \doi{https://doi.org/10.1017/CBO9780511622700}
  \end{APACrefDOI}
\PrintBackRefs{\CurrentBib}

\bibitem [\protect \citeauthoryear {%
Ida%
}{%
Ida%
}{%
{\protect \APACyear {1972}}%
}]{%
ida_1972}
\APACinsertmetastar {%
ida_1972}%
\begin{APACrefauthors}%
Ida, Y.%
\end{APACrefauthors}%
\unskip\
\newblock
\APACrefYearMonthDay{1972}{}{}.
\newblock
{\BBOQ}\APACrefatitle {Cohesive force across the tip of a longitudinal-shear
  crack and Griffith's specific surface energy} {Cohesive force across the tip
  of a longitudinal-shear crack and griffith's specific surface energy}.{\BBCQ}
\newblock
\APACjournalVolNumPages{Journal of Geophysical Research
  (1896-1977)}{77}{20}{3796-3805}.
\newblock
\begin{APACrefDOI} \doi{https://doi.org/10.1029/JB077i020p03796}
  \end{APACrefDOI}
\PrintBackRefs{\CurrentBib}

\bibitem [\protect \citeauthoryear {%
Jakobsson%
\ \BBA {} Amoignon%
}{%
Jakobsson%
\ \BBA {} Amoignon%
}{%
{\protect \APACyear {2007}}%
}]{%
jakobsson_amoignon_2007}
\APACinsertmetastar {%
jakobsson_amoignon_2007}%
\begin{APACrefauthors}%
Jakobsson, S.%
\BCBT {}\ \BBA {} Amoignon, O.%
\end{APACrefauthors}%
\unskip\
\newblock
\APACrefYearMonthDay{2007}{}{}.
\newblock
{\BBOQ}\APACrefatitle {Mesh deformation using radial basis functions for
  gradient-based aerodynamic shape optimization} {Mesh deformation using radial
  basis functions for gradient-based aerodynamic shape optimization}.{\BBCQ}
\newblock
\APACjournalVolNumPages{Computers \& Fluids}{36}{6}{1119-1136}.
\newblock
\begin{APACrefDOI} \doi{https://doi.org/10.1016/j.compfluid.2006.11.002}
  \end{APACrefDOI}
\PrintBackRefs{\CurrentBib}

\bibitem [\protect \citeauthoryear {%
King%
\ \BBA {} Nábělek%
}{%
King%
\ \BBA {} Nábělek%
}{%
{\protect \APACyear {1985}}%
}]{%
king_nabelek_1985}
\APACinsertmetastar {%
king_nabelek_1985}%
\begin{APACrefauthors}%
King, G.%
\BCBT {}\ \BBA {} Nábělek, J.%
\end{APACrefauthors}%
\unskip\
\newblock
\APACrefYearMonthDay{1985}{}{}.
\newblock
{\BBOQ}\APACrefatitle {Role of Fault Bends in the Initiation and Termination of
  Earthquake Rupture} {Role of fault bends in the initiation and termination of
  earthquake rupture}.{\BBCQ}
\newblock
\APACjournalVolNumPages{Science}{228}{4702}{984-987}.
\newblock
\begin{APACrefDOI} \doi{https://doi.org/10.1126/science.228.4702.984}
  \end{APACrefDOI}
\PrintBackRefs{\CurrentBib}

\bibitem [\protect \citeauthoryear {%
Kneller%
\ \BBA {} van Keken%
}{%
Kneller%
\ \BBA {} van Keken%
}{%
{\protect \APACyear {2008}}%
}]{%
kneller2008effect}
\APACinsertmetastar {%
kneller2008effect}%
\begin{APACrefauthors}%
Kneller, E\BPBI A.%
\BCBT {}\ \BBA {} van Keken, P\BPBI E.%
\end{APACrefauthors}%
\unskip\
\newblock
\APACrefYearMonthDay{2008}{}{}.
\newblock
{\BBOQ}\APACrefatitle {Effect of three-dimensional slab geometry on deformation
  in the mantle wedge: Implications for shear wave anisotropy} {Effect of
  three-dimensional slab geometry on deformation in the mantle wedge:
  Implications for shear wave anisotropy}.{\BBCQ}
\newblock
\APACjournalVolNumPages{Geochemistry, Geophysics, Geosystems}{9}{1}{}.
\PrintBackRefs{\CurrentBib}

\bibitem [\protect \citeauthoryear {%
Knupp%
}{%
Knupp%
}{%
{\protect \APACyear {2000}}%
}]{%
knupp_2000}
\APACinsertmetastar {%
knupp_2000}%
\begin{APACrefauthors}%
Knupp, P\BPBI M.%
\end{APACrefauthors}%
\unskip\
\newblock
\APACrefYearMonthDay{2000}{}{}.
\newblock
{\BBOQ}\APACrefatitle {Achieving finite element mesh quality via optimization
  of the {Jacobian} matrix norm and associated quantities. {Part I}—a
  framework for surface mesh optimization} {Achieving finite element mesh
  quality via optimization of the {Jacobian} matrix norm and associated
  quantities. {Part I}—a framework for surface mesh optimization}.{\BBCQ}
\newblock
\APACjournalVolNumPages{International Journal for Numerical Methods in
  Engineering}{48}{3}{401-420}.
\newblock
\begin{APACrefDOI}
  \doi{https://doi.org/10.1002/(SICI)1097-0207(20000530)48:3<401::AID-NME880>3.0.CO;2-D}
  \end{APACrefDOI}
\PrintBackRefs{\CurrentBib}

\bibitem [\protect \citeauthoryear {%
Kohn%
, Castro%
, Kerswell%
, Ranero%
\BCBL {}\ \BBA {} Spear%
}{%
Kohn%
\ \protect \BOthers {.}}{%
{\protect \APACyear {2018}}%
}]{%
kohn2018shear}
\APACinsertmetastar {%
kohn2018shear}%
\begin{APACrefauthors}%
Kohn, M\BPBI J.%
, Castro, A\BPBI E.%
, Kerswell, B\BPBI C.%
, Ranero, C\BPBI R.%
\BCBL {}\ \BBA {} Spear, F\BPBI S.%
\end{APACrefauthors}%
\unskip\
\newblock
\APACrefYearMonthDay{2018}{}{}.
\newblock
{\BBOQ}\APACrefatitle {Shear heating reconciles thermal models with the
  metamorphic rock record of subduction} {Shear heating reconciles thermal
  models with the metamorphic rock record of subduction}.{\BBCQ}
\newblock
\APACjournalVolNumPages{Proceedings of the National Academy of
  Sciences}{115}{46}{11706--11711}.
\PrintBackRefs{\CurrentBib}

\bibitem [\protect \citeauthoryear {%
Käser%
\ \BBA {} Dumbser%
}{%
Käser%
\ \BBA {} Dumbser%
}{%
{\protect \APACyear {2006}}%
}]{%
kaser_dumbser_2006}
\APACinsertmetastar {%
kaser_dumbser_2006}%
\begin{APACrefauthors}%
Käser, M.%
\BCBT {}\ \BBA {} Dumbser, M.%
\end{APACrefauthors}%
\unskip\
\newblock
\APACrefYearMonthDay{2006}{}{}.
\newblock
{\BBOQ}\APACrefatitle {An arbitrary high-order discontinuous {G}alerkin method
  for elastic waves on unstructured meshes — {I. The} two-dimensional
  isotropic case with external source terms} {An arbitrary high-order
  discontinuous {G}alerkin method for elastic waves on unstructured meshes —
  {I. The} two-dimensional isotropic case with external source terms}.{\BBCQ}
\newblock
\APACjournalVolNumPages{Geophysical Journal International}{166}{2}{855-877}.
\newblock
\begin{APACrefDOI} \doi{https://doi.org/10.1111/j.1365-246X.2006.03051.x}
  \end{APACrefDOI}
\PrintBackRefs{\CurrentBib}

\bibitem [\protect \citeauthoryear {%
Lallemand%
, Peyret%
, van Rijsingen%
, Arcay%
\BCBL {}\ \BBA {} Heuret%
}{%
Lallemand%
\ \protect \BOthers {.}}{%
{\protect \APACyear {2018}}%
}]{%
lallemand_et_al_2018}
\APACinsertmetastar {%
lallemand_et_al_2018}%
\begin{APACrefauthors}%
Lallemand, S.%
, Peyret, M.%
, van Rijsingen, E.%
, Arcay, D.%
\BCBL {}\ \BBA {} Heuret, A.%
\end{APACrefauthors}%
\unskip\
\newblock
\APACrefYearMonthDay{2018}{}{}.
\newblock
{\BBOQ}\APACrefatitle {Roughness Characteristics of Oceanic Seafloor Prior to
  Subduction in Relation to the Seismogenic Potential of Subduction Zones}
  {Roughness characteristics of oceanic seafloor prior to subduction in
  relation to the seismogenic potential of subduction zones}.{\BBCQ}
\newblock
\APACjournalVolNumPages{Geochemistry, Geophysics,
  Geosystems}{19}{7}{2121-2146}.
\newblock
\begin{APACrefDOI} \doi{https://doi.org/10.1029/2018GC007434} \end{APACrefDOI}
\PrintBackRefs{\CurrentBib}

\bibitem [\protect \citeauthoryear {%
Lauer%
\ \BBA {} Saffer%
}{%
Lauer%
\ \BBA {} Saffer%
}{%
{\protect \APACyear {2012}}%
}]{%
Lauer_Saffer_2012}
\APACinsertmetastar {%
Lauer_Saffer_2012}%
\begin{APACrefauthors}%
Lauer, R\BPBI M.%
\BCBT {}\ \BBA {} Saffer, D\BPBI M.%
\end{APACrefauthors}%
\unskip\
\newblock
\APACrefYearMonthDay{2012}{}{}.
\newblock
{\BBOQ}\APACrefatitle {Fluid budgets of subduction zone forearcs: The
  contribution of splay faults} {Fluid budgets of subduction zone forearcs: The
  contribution of splay faults}.{\BBCQ}
\newblock
\APACjournalVolNumPages{Geophysical Research Letters}{39}{13}{}.
\newblock
\begin{APACrefDOI} \doi{https://doi.org/10.1029/2012GL052182} \end{APACrefDOI}
\PrintBackRefs{\CurrentBib}

\bibitem [\protect \citeauthoryear {%
Lee%
\ \protect \BOthers {.}}{%
Lee%
\ \protect \BOthers {.}}{%
{\protect \APACyear {2025}}%
}]{%
Lee2025}
\APACinsertmetastar {%
Lee2025}%
\begin{APACrefauthors}%
Lee, M\BPBI K.%
, Carbotte, S\BPBI M.%
, Han, S.%
, Shuck, B.%
, Gurun, P.%
, Boston, B.%
\BDBL {}Spinelli, G.%
\end{APACrefauthors}%
\unskip\
\newblock
\APACrefYearMonthDay{2025}{}{}.
\newblock
{\BBOQ}\APACrefatitle {{Anomalous sediment consolidation and alteration from
  buried incoming plate seamounts along the Cascadia margin}} {{Anomalous
  sediment consolidation and alteration from buried incoming plate seamounts
  along the Cascadia margin}}.{\BBCQ}
\newblock
\APACjournalVolNumPages{Geochemistry, Geophysics,
  Geosystems}{26}{2}{e2024GC011949}.
\newblock
\begin{APACrefDOI} \doi{https://doi.org/10.1029/2024GC011949} \end{APACrefDOI}
\PrintBackRefs{\CurrentBib}

\bibitem [\protect \citeauthoryear {%
Logg%
, Mardal%
\BCBL {}\ \BBA {} Wells%
}{%
Logg%
\ \protect \BOthers {.}}{%
{\protect \APACyear {2012}}%
}]{%
fenics}
\APACinsertmetastar {%
fenics}%
\begin{APACrefauthors}%
Logg, A.%
, Mardal, K\BHBI A.%
\BCBL {}\ \BBA {} Wells, G.%
\end{APACrefauthors}%
\unskip\
\newblock
\APACrefYear{2012}.
\newblock
\APACrefbtitle {{Automated Solution of Differential Equations by the Finite
  Element Method: The FEniCS Book}} {{Automated Solution of Differential
  Equations by the Finite Element Method: The FEniCS Book}}.
\newblock
\APACaddressPublisher{}{Springer, Berlin, Heidelberg}.
\newblock
\begin{APACrefDOI} \doi{https://doi.org/10.1007/978-3-642-23099-8}
  \end{APACrefDOI}
\PrintBackRefs{\CurrentBib}

\bibitem [\protect \citeauthoryear {%
Ly%
\ \BBA {} Tran%
}{%
Ly%
\ \BBA {} Tran%
}{%
{\protect \APACyear {2001}}%
}]{%
ly_tran_2001}
\APACinsertmetastar {%
ly_tran_2001}%
\begin{APACrefauthors}%
Ly, H\BPBI V.%
\BCBT {}\ \BBA {} Tran, H\BPBI T.%
\end{APACrefauthors}%
\unskip\
\newblock
\APACrefYearMonthDay{2001}{}{}.
\newblock
{\BBOQ}\APACrefatitle {Modeling and control of physical processes using proper
  orthogonal decomposition} {Modeling and control of physical processes using
  proper orthogonal decomposition}.{\BBCQ}
\newblock
\APACjournalVolNumPages{Mathematical and Computer Modelling}{33}{1}{223-236}.
\newblock
\begin{APACrefDOI} \doi{https://doi.org/10.1016/S0895-7177(00)00240-5}
  \end{APACrefDOI}
\PrintBackRefs{\CurrentBib}

\bibitem [\protect \citeauthoryear {%
Maunder%
, van Hunen%
, Bouilhol%
\BCBL {}\ \BBA {} Magni%
}{%
Maunder%
\ \protect \BOthers {.}}{%
{\protect \APACyear {2019}}%
}]{%
maunder_et_al_2019}
\APACinsertmetastar {%
maunder_et_al_2019}%
\begin{APACrefauthors}%
Maunder, B.%
, van Hunen, J.%
, Bouilhol, P.%
\BCBL {}\ \BBA {} Magni, V.%
\end{APACrefauthors}%
\unskip\
\newblock
\APACrefYearMonthDay{2019}{}{}.
\newblock
{\BBOQ}\APACrefatitle {Modeling Slab Temperature: A Reevaluation of the Thermal
  Parameter} {Modeling slab temperature: A reevaluation of the thermal
  parameter}.{\BBCQ}
\newblock
\APACjournalVolNumPages{Geochemistry, Geophysics, Geosystems}{20}{2}{673-687}.
\newblock
\begin{APACrefDOI} \doi{https://doi.org/10.1029/2018GC007641} \end{APACrefDOI}
\PrintBackRefs{\CurrentBib}

\bibitem [\protect \citeauthoryear {%
Michler%
}{%
Michler%
}{%
{\protect \APACyear {2011}}%
}]{%
michler_2011}
\APACinsertmetastar {%
michler_2011}%
\begin{APACrefauthors}%
Michler, A\BPBI K.%
\end{APACrefauthors}%
\unskip\
\newblock
\APACrefYearMonthDay{2011}{}{}.
\newblock
{\BBOQ}\APACrefatitle {Aircraft control surface deflection using {RBF}-based
  mesh deformation} {Aircraft control surface deflection using {RBF}-based mesh
  deformation}.{\BBCQ}
\newblock
\APACjournalVolNumPages{International Journal for Numerical Methods in
  Engineering}{88}{10}{986-1007}.
\newblock
\begin{APACrefDOI} \doi{https://doi.org/10.1002/nme.3208} \end{APACrefDOI}
\PrintBackRefs{\CurrentBib}

\bibitem [\protect \citeauthoryear {%
Moreno%
\ \protect \BOthers {.}}{%
Moreno%
\ \protect \BOthers {.}}{%
{\protect \APACyear {2012}}%
}]{%
moreno_et_al_2012}
\APACinsertmetastar {%
moreno_et_al_2012}%
\begin{APACrefauthors}%
Moreno, M.%
, Melnick, D.%
, Rosenau, M.%
, Baez, J.%
, Klotz, J.%
, Oncken, O.%
\BDBL {}Hase, H.%
\end{APACrefauthors}%
\unskip\
\newblock
\APACrefYearMonthDay{2012}{}{}.
\newblock
{\BBOQ}\APACrefatitle {{Toward understanding tectonic control on the Mw 8.8
  2010 Maule Chile earthquake}} {{Toward understanding tectonic control on the
  Mw 8.8 2010 Maule Chile earthquake}}.{\BBCQ}
\newblock
\APACjournalVolNumPages{Earth and Planetary Science
  Letters}{321-322}{}{152-165}.
\newblock
\begin{APACrefDOI} \doi{https://doi.org/10.1016/j.epsl.2012.01.006}
  \end{APACrefDOI}
\PrintBackRefs{\CurrentBib}

\bibitem [\protect \citeauthoryear {%
Muldashev%
\ \BBA {} Sobolev%
}{%
Muldashev%
\ \BBA {} Sobolev%
}{%
{\protect \APACyear {2020}}%
}]{%
muldashev_sobolev_2020}
\APACinsertmetastar {%
muldashev_sobolev_2020}%
\begin{APACrefauthors}%
Muldashev, I\BPBI A.%
\BCBT {}\ \BBA {} Sobolev, S\BPBI V.%
\end{APACrefauthors}%
\unskip\
\newblock
\APACrefYearMonthDay{2020}{}{}.
\newblock
{\BBOQ}\APACrefatitle {What Controls Maximum Magnitudes of Giant Subduction
  Earthquakes?} {What controls maximum magnitudes of giant subduction
  earthquakes?}{\BBCQ}
\newblock
\APACjournalVolNumPages{Geochemistry, Geophysics, Geosystems}{21}{9}{}.
\newblock
\begin{APACrefDOI} \doi{https://doi.org/10.1029/2020gc009145} \end{APACrefDOI}
\PrintBackRefs{\CurrentBib}

\bibitem [\protect \citeauthoryear {%
My-Ha%
, Lim%
, Khoo%
\BCBL {}\ \BBA {} Willcox%
}{%
My-Ha%
\ \protect \BOthers {.}}{%
{\protect \APACyear {2007}}%
}]{%
my-ha_2007}
\APACinsertmetastar {%
my-ha_2007}%
\begin{APACrefauthors}%
My-Ha, D.%
, Lim, K.%
, Khoo, B.%
\BCBL {}\ \BBA {} Willcox, K.%
\end{APACrefauthors}%
\unskip\
\newblock
\APACrefYearMonthDay{2007}{}{}.
\newblock
{\BBOQ}\APACrefatitle {Real-time optimization using proper orthogonal
  decomposition: {F}ree surface shape prediction due to underwater bubble
  dynamics} {Real-time optimization using proper orthogonal decomposition:
  {F}ree surface shape prediction due to underwater bubble dynamics}.{\BBCQ}
\newblock
\APACjournalVolNumPages{Computers \& Fluids}{36}{3}{499-512}.
\newblock
\begin{APACrefDOI} \doi{https://doi.org/10.1016/j.compfluid.2006.01.016}
  \end{APACrefDOI}
\PrintBackRefs{\CurrentBib}

\bibitem [\protect \citeauthoryear {%
Okubo%
\ \BBA {} Aki%
}{%
Okubo%
\ \BBA {} Aki%
}{%
{\protect \APACyear {1987}}%
}]{%
okubo_aki_1987}
\APACinsertmetastar {%
okubo_aki_1987}%
\begin{APACrefauthors}%
Okubo, P\BPBI G.%
\BCBT {}\ \BBA {} Aki, K.%
\end{APACrefauthors}%
\unskip\
\newblock
\APACrefYearMonthDay{1987}{}{}.
\newblock
{\BBOQ}\APACrefatitle {Fractal geometry in the {San Andreas Fault System}}
  {Fractal geometry in the {San Andreas Fault System}}.{\BBCQ}
\newblock
\APACjournalVolNumPages{Journal of Geophysical Research: Solid
  Earth}{92}{B1}{345-355}.
\newblock
\begin{APACrefDOI} \doi{https://doi.org/10.1029/JB092iB01p00345}
  \end{APACrefDOI}
\PrintBackRefs{\CurrentBib}

\bibitem [\protect \citeauthoryear {%
Palmer%
, Rice%
\BCBL {}\ \BBA {} Hill%
}{%
Palmer%
\ \protect \BOthers {.}}{%
{\protect \APACyear {1973}}%
}]{%
palmer_et_al_1973}
\APACinsertmetastar {%
palmer_et_al_1973}%
\begin{APACrefauthors}%
Palmer, A\BPBI C.%
, Rice, J\BPBI R.%
\BCBL {}\ \BBA {} Hill, R.%
\end{APACrefauthors}%
\unskip\
\newblock
\APACrefYearMonthDay{1973}{}{}.
\newblock
{\BBOQ}\APACrefatitle {The growth of slip surfaces in the progressive failure
  of over-consolidated clay} {The growth of slip surfaces in the progressive
  failure of over-consolidated clay}.{\BBCQ}
\newblock
\APACjournalVolNumPages{Proceedings of the Royal Society of London. A.
  Mathematical and Physical Sciences}{332}{1591}{527-548}.
\newblock
\begin{APACrefDOI} \doi{10.1098/rspa.1973.0040} \end{APACrefDOI}
\PrintBackRefs{\CurrentBib}

\bibitem [\protect \citeauthoryear {%
Peacock%
}{%
Peacock%
}{%
{\protect \APACyear {2020}}%
}]{%
peacock_2020}
\APACinsertmetastar {%
peacock_2020}%
\begin{APACrefauthors}%
Peacock, S\BPBI M.%
\end{APACrefauthors}%
\unskip\
\newblock
\APACrefYearMonthDay{2020}{}{}.
\newblock
{\BBOQ}\APACrefatitle {{Advances in the thermal and petrologic modeling of
  subduction zones}} {{Advances in the thermal and petrologic modeling of
  subduction zones}}.{\BBCQ}
\newblock
\APACjournalVolNumPages{Geosphere}{16}{4}{936-952}.
\newblock
\begin{APACrefURL} \url{https://doi.org/10.1130/GES02213.1} \end{APACrefURL}
\newblock
\begin{APACrefDOI} \doi{10.1130/GES02213.1} \end{APACrefDOI}
\PrintBackRefs{\CurrentBib}

\bibitem [\protect \citeauthoryear {%
Pelties%
, Gabriel%
\BCBL {}\ \BBA {} Ampuero%
}{%
Pelties%
\ \protect \BOthers {.}}{%
{\protect \APACyear {2014}}%
}]{%
pelties_gabriel_ampuero_2014}
\APACinsertmetastar {%
pelties_gabriel_ampuero_2014}%
\begin{APACrefauthors}%
Pelties, C.%
, Gabriel, A\BHBI A.%
\BCBL {}\ \BBA {} Ampuero, J\BHBI P.%
\end{APACrefauthors}%
\unskip\
\newblock
\APACrefYearMonthDay{2014}{}{}.
\newblock
{\BBOQ}\APACrefatitle {Verification of an {ADER-DG} method for complex dynamic
  rupture problems} {Verification of an {ADER-DG} method for complex dynamic
  rupture problems}.{\BBCQ}
\newblock
\APACjournalVolNumPages{Geoscientific Model Development}{7}{3}{847--866}.
\newblock
\begin{APACrefDOI} \doi{https://doi.org/10.5194/gmd-7-847-2014}
  \end{APACrefDOI}
\PrintBackRefs{\CurrentBib}

\bibitem [\protect \citeauthoryear {%
Plesch%
\ \protect \BOthers {.}}{%
Plesch%
\ \protect \BOthers {.}}{%
{\protect \APACyear {2007}}%
}]{%
Plesch2007}
\APACinsertmetastar {%
Plesch2007}%
\begin{APACrefauthors}%
Plesch, A.%
, Shaw, J\BPBI H.%
, Benson, C.%
, Bryant, W\BPBI A.%
, Carena, S.%
, Cooke, M.%
\BDBL {}others%
\end{APACrefauthors}%
\unskip\
\newblock
\APACrefYearMonthDay{2007}{}{}.
\newblock
{\BBOQ}\APACrefatitle {{Community fault model (CFM) for southern California}}
  {{Community fault model (CFM) for southern California}}.{\BBCQ}
\newblock
\APACjournalVolNumPages{Bulletin of the Seismological Society of
  America}{97}{6}{1793--1802}.
\newblock
\begin{APACrefDOI} \doi{https://doi.org/10.1785/0120050211} \end{APACrefDOI}
\PrintBackRefs{\CurrentBib}

\bibitem [\protect \citeauthoryear {%
Plescia%
\ \BBA {} Hayes%
}{%
Plescia%
\ \BBA {} Hayes%
}{%
{\protect \APACyear {2020}}%
}]{%
plescia_hayes_2020}
\APACinsertmetastar {%
plescia_hayes_2020}%
\begin{APACrefauthors}%
Plescia, S\BPBI M.%
\BCBT {}\ \BBA {} Hayes, G\BPBI P.%
\end{APACrefauthors}%
\unskip\
\newblock
\APACrefYearMonthDay{2020}{}{}.
\newblock
{\BBOQ}\APACrefatitle {Geometric controls on megathrust earthquakes} {Geometric
  controls on megathrust earthquakes}.{\BBCQ}
\newblock
\APACjournalVolNumPages{Geophysical Journal International}{222}{2}{1270-1282}.
\newblock
\begin{APACrefDOI} \doi{https://doi.org/10.1093/gji/ggaa254} \end{APACrefDOI}
\PrintBackRefs{\CurrentBib}

\bibitem [\protect \citeauthoryear {%
Ragon%
, Sladen%
\BCBL {}\ \BBA {} Simons%
}{%
Ragon%
\ \protect \BOthers {.}}{%
{\protect \APACyear {2018}}%
}]{%
ragon_2018}
\APACinsertmetastar {%
ragon_2018}%
\begin{APACrefauthors}%
Ragon, T.%
, Sladen, A.%
\BCBL {}\ \BBA {} Simons, M.%
\end{APACrefauthors}%
\unskip\
\newblock
\APACrefYearMonthDay{2018}{}{}.
\newblock
{\BBOQ}\APACrefatitle {{Accounting for uncertain fault geometry in earthquake
  source inversions – I: theory and simplified application}} {{Accounting for
  uncertain fault geometry in earthquake source inversions – I: theory and
  simplified application}}.{\BBCQ}
\newblock
\APACjournalVolNumPages{Geophysical Journal International}{214}{2}{1174-1190}.
\newblock
\begin{APACrefDOI} \doi{https://doi.org/10.1093/gji/ggy187} \end{APACrefDOI}
\PrintBackRefs{\CurrentBib}

\bibitem [\protect \citeauthoryear {%
Ragon%
, Sladen%
\BCBL {}\ \BBA {} Simons%
}{%
Ragon%
\ \protect \BOthers {.}}{%
{\protect \APACyear {2019}}%
}]{%
ragon_et_al_2019}
\APACinsertmetastar {%
ragon_et_al_2019}%
\begin{APACrefauthors}%
Ragon, T.%
, Sladen, A.%
\BCBL {}\ \BBA {} Simons, M.%
\end{APACrefauthors}%
\unskip\
\newblock
\APACrefYearMonthDay{2019}{}{}.
\newblock
{\BBOQ}\APACrefatitle {{Accounting for uncertain fault geometry in earthquake
  source inversions – II: application to the Mw 6.2 Amatrice earthquake,
  central Italy}} {{Accounting for uncertain fault geometry in earthquake
  source inversions – II: application to the Mw 6.2 Amatrice earthquake,
  central Italy}}.{\BBCQ}
\newblock
\APACjournalVolNumPages{Geophysical Journal International}{218}{1}{689-707}.
\newblock
\begin{APACrefDOI} \doi{https://doi.org/10.1093/gji/ggz180} \end{APACrefDOI}
\PrintBackRefs{\CurrentBib}

\bibitem [\protect \citeauthoryear {%
Ramos%
, Thakur%
, Huang%
, Harris%
\BCBL {}\ \BBA {} Ryan%
}{%
Ramos%
\ \protect \BOthers {.}}{%
{\protect \APACyear {2022}}%
}]{%
ramos_et_al_2022}
\APACinsertmetastar {%
ramos_et_al_2022}%
\begin{APACrefauthors}%
Ramos, M\BPBI D.%
, Thakur, P.%
, Huang, Y.%
, Harris, R\BPBI A.%
\BCBL {}\ \BBA {} Ryan, K\BPBI J.%
\end{APACrefauthors}%
\unskip\
\newblock
\APACrefYearMonthDay{2022}{04}{}.
\newblock
{\BBOQ}\APACrefatitle {Working with Dynamic Earthquake Rupture Models: A
  Practical Guide} {Working with dynamic earthquake rupture models: A practical
  guide}.{\BBCQ}
\newblock
\APACjournalVolNumPages{Seismological Research Letters}{93}{4}{2096-2110}.
\newblock
\begin{APACrefDOI} \doi{https://doi.org/10.1785/0220220022} \end{APACrefDOI}
\PrintBackRefs{\CurrentBib}

\bibitem [\protect \citeauthoryear {%
Rekoske%
, Gabriel%
\BCBL {}\ \BBA {} May%
}{%
Rekoske%
\ \protect \BOthers {.}}{%
{\protect \APACyear {2023}}%
}]{%
rekoske_2023}
\APACinsertmetastar {%
rekoske_2023}%
\begin{APACrefauthors}%
Rekoske, J\BPBI M.%
, Gabriel, A\BHBI A.%
\BCBL {}\ \BBA {} May, D\BPBI A.%
\end{APACrefauthors}%
\unskip\
\newblock
\APACrefYearMonthDay{2023}{}{}.
\newblock
{\BBOQ}\APACrefatitle {Instantaneous Physics-Based Ground Motion Maps Using
  Reduced-Order Modeling} {Instantaneous physics-based ground motion maps using
  reduced-order modeling}.{\BBCQ}
\newblock
\APACjournalVolNumPages{Journal of Geophysical Research: Solid
  Earth}{128}{8}{e2023JB026975}.
\newblock
\begin{APACrefDOI} \doi{https://doi.org/10.1029/2023JB026975} \end{APACrefDOI}
\PrintBackRefs{\CurrentBib}

\bibitem [\protect \citeauthoryear {%
Rekoske%
, May%
\BCBL {}\ \BBA {} Gabriel%
}{%
Rekoske%
\ \protect \BOthers {.}}{%
{\protect \APACyear {2025}}%
}]{%
rekoske_et_al_2025}
\APACinsertmetastar {%
rekoske_et_al_2025}%
\begin{APACrefauthors}%
Rekoske, J\BPBI M.%
, May, D\BPBI A.%
\BCBL {}\ \BBA {} Gabriel, A\BHBI A.%
\end{APACrefauthors}%
\unskip\
\newblock
\APACrefYearMonthDay{2025}{}{}.
\newblock
{\BBOQ}\APACrefatitle {Reduced-order modelling for complex three-dimensional
  seismic wave propagation} {Reduced-order modelling for complex
  three-dimensional seismic wave propagation}.{\BBCQ}
\newblock
\APACjournalVolNumPages{Geophysical Journal International}{241}{1}{526-548}.
\newblock
\begin{APACrefDOI} \doi{https://doi.org/10.1093/gji/ggaf049} \end{APACrefDOI}
\PrintBackRefs{\CurrentBib}

\bibitem [\protect \citeauthoryear {%
Ross%
\ \protect \BOthers {.}}{%
Ross%
\ \protect \BOthers {.}}{%
{\protect \APACyear {2019}}%
}]{%
Ross2019}
\APACinsertmetastar {%
Ross2019}%
\begin{APACrefauthors}%
Ross, Z\BPBI E.%
, Idini, B.%
, Jia, Z.%
, Stephenson, O\BPBI L.%
, Zhong, M.%
, Wang, X.%
\BDBL {}others%
\end{APACrefauthors}%
\unskip\
\newblock
\APACrefYearMonthDay{2019}{}{}.
\newblock
{\BBOQ}\APACrefatitle {{Hierarchical interlocked orthogonal faulting in the
  2019 Ridgecrest earthquake sequence}} {{Hierarchical interlocked orthogonal
  faulting in the 2019 Ridgecrest earthquake sequence}}.{\BBCQ}
\newblock
\APACjournalVolNumPages{Science}{366}{6463}{346--351}.
\newblock
\begin{APACrefDOI} \doi{https://doi.org/10.1126/science.aaz0109}
  \end{APACrefDOI}
\PrintBackRefs{\CurrentBib}

\bibitem [\protect \citeauthoryear {%
Schellart%
\ \BBA {} Rawlinson%
}{%
Schellart%
\ \BBA {} Rawlinson%
}{%
{\protect \APACyear {2013}}%
}]{%
schellart_rawlinson_2013}
\APACinsertmetastar {%
schellart_rawlinson_2013}%
\begin{APACrefauthors}%
Schellart, W.%
\BCBT {}\ \BBA {} Rawlinson, N.%
\end{APACrefauthors}%
\unskip\
\newblock
\APACrefYearMonthDay{2013}{}{}.
\newblock
{\BBOQ}\APACrefatitle {Global correlations between maximum magnitudes of
  subduction zone interface thrust earthquakes and physical parameters of
  subduction zones} {Global correlations between maximum magnitudes of
  subduction zone interface thrust earthquakes and physical parameters of
  subduction zones}.{\BBCQ}
\newblock
\APACjournalVolNumPages{Physics of the Earth and Planetary
  Interiors}{225}{}{41-67}.
\newblock
\begin{APACrefDOI} \doi{https://doi.org/10.1016/j.pepi.2013.10.001}
  \end{APACrefDOI}
\PrintBackRefs{\CurrentBib}

\bibitem [\protect \citeauthoryear {%
Sederberg%
\ \BBA {} Parry%
}{%
Sederberg%
\ \BBA {} Parry%
}{%
{\protect \APACyear {1986}}%
}]{%
sederberg_parry_1986}
\APACinsertmetastar {%
sederberg_parry_1986}%
\begin{APACrefauthors}%
Sederberg, T\BPBI W.%
\BCBT {}\ \BBA {} Parry, S\BPBI R.%
\end{APACrefauthors}%
\unskip\
\newblock
\APACrefYearMonthDay{1986}{}{}.
\newblock
{\BBOQ}\APACrefatitle {Free-form deformation of solid geometric models}
  {Free-form deformation of solid geometric models}.{\BBCQ}
\newblock
\APACjournalVolNumPages{SIGGRAPH Computer Graphics}{20}{4}{151–160}.
\newblock
\begin{APACrefDOI} \doi{https://doi.org/10.1145/15886.15903} \end{APACrefDOI}
\PrintBackRefs{\CurrentBib}

\bibitem [\protect \citeauthoryear {%
Segall%
\ \BBA {} Pollard%
}{%
Segall%
\ \BBA {} Pollard%
}{%
{\protect \APACyear {1980}}%
}]{%
segall_pollard_1980}
\APACinsertmetastar {%
segall_pollard_1980}%
\begin{APACrefauthors}%
Segall, P.%
\BCBT {}\ \BBA {} Pollard, D\BPBI D.%
\end{APACrefauthors}%
\unskip\
\newblock
\APACrefYearMonthDay{1980}{}{}.
\newblock
{\BBOQ}\APACrefatitle {Mechanics of discontinuous faults} {Mechanics of
  discontinuous faults}.{\BBCQ}
\newblock
\APACjournalVolNumPages{Journal of Geophysical Research: Solid
  Earth}{85}{B8}{4337-4350}.
\newblock
\begin{APACrefDOI} \doi{https://doi.org/10.1029/JB085iB08p04337}
  \end{APACrefDOI}
\PrintBackRefs{\CurrentBib}

\bibitem [\protect \citeauthoryear {%
Shrivastava%
\ \protect \BOthers {.}}{%
Shrivastava%
\ \protect \BOthers {.}}{%
{\protect \APACyear {2019}}%
}]{%
shrivastava_et_al_2019}
\APACinsertmetastar {%
shrivastava_et_al_2019}%
\begin{APACrefauthors}%
Shrivastava, M\BPBI N.%
, González, G.%
, Moreno, M.%
, Soto, H.%
, Schurr, B.%
, Salazar, P.%
\BCBL {}\ \BBA {} Báez, J\BPBI C.%
\end{APACrefauthors}%
\unskip\
\newblock
\APACrefYearMonthDay{2019}{}{}.
\newblock
{\BBOQ}\APACrefatitle {Earthquake segmentation in northern {Chile} correlates
  with curved plate geometry} {Earthquake segmentation in northern {Chile}
  correlates with curved plate geometry}.{\BBCQ}
\newblock
\APACjournalVolNumPages{Scientific Reports}{9}{1}{4403}.
\newblock
\begin{APACrefDOI} \doi{https://doi.org/10.1038/s41598-019-40282-6}
  \end{APACrefDOI}
\PrintBackRefs{\CurrentBib}

\bibitem [\protect \citeauthoryear {%
Sieger%
, Menzel%
\BCBL {}\ \BBA {} Botsch%
}{%
Sieger%
\ \protect \BOthers {.}}{%
{\protect \APACyear {2014}}%
}]{%
sieger_et_al_2014}
\APACinsertmetastar {%
sieger_et_al_2014}%
\begin{APACrefauthors}%
Sieger, D.%
, Menzel, S.%
\BCBL {}\ \BBA {} Botsch, M.%
\end{APACrefauthors}%
\unskip\
\newblock
\APACrefYearMonthDay{2014}{}{}.
\newblock
{\BBOQ}\APACrefatitle {{RBF} morphing techniques for simulation-based design
  optimization} {{RBF} morphing techniques for simulation-based design
  optimization}.{\BBCQ}
\newblock
\APACjournalVolNumPages{Engineering with Computers}{30}{}{161--174}.
\newblock
\begin{APACrefDOI} \doi{https://doi.org/10.1007/s00366-013-0330-1}
  \end{APACrefDOI}
\PrintBackRefs{\CurrentBib}

\bibitem [\protect \citeauthoryear {%
Sirovich%
}{%
Sirovich%
}{%
{\protect \APACyear {1987}}%
}]{%
sirovich}
\APACinsertmetastar {%
sirovich}%
\begin{APACrefauthors}%
Sirovich, L.%
\end{APACrefauthors}%
\unskip\
\newblock
\APACrefYearMonthDay{1987}{}{}.
\newblock
{\BBOQ}\APACrefatitle {Turbulence and the Dynamics of Coherent Structures,
  Parts {I–III}} {Turbulence and the dynamics of coherent structures, parts
  {I–III}}.{\BBCQ}
\newblock
\APACjournalVolNumPages{Quarterly of Applied Mathematics}{45(3)}{}{561–571}.
\newblock
\begin{APACrefDOI} \doi{http://www.jstor.org/stable/43637457} \end{APACrefDOI}
\PrintBackRefs{\CurrentBib}

\bibitem [\protect \citeauthoryear {%
Staten%
, Owen%
, Shontz%
, Salinger%
\BCBL {}\ \BBA {} Coffey%
}{%
Staten%
\ \protect \BOthers {.}}{%
{\protect \APACyear {2012}}%
}]{%
staten_et_al_2012}
\APACinsertmetastar {%
staten_et_al_2012}%
\begin{APACrefauthors}%
Staten, M\BPBI L.%
, Owen, S\BPBI J.%
, Shontz, S\BPBI M.%
, Salinger, A\BPBI G.%
\BCBL {}\ \BBA {} Coffey, T\BPBI S.%
\end{APACrefauthors}%
\unskip\
\newblock
\APACrefYearMonthDay{2012}{}{}.
\newblock
{\BBOQ}\APACrefatitle {A Comparison of Mesh Morphing Methods for {3D} Shape
  Optimization} {A comparison of mesh morphing methods for {3D} shape
  optimization}.{\BBCQ}
\newblock
\BIn{} W\BPBI R.~Quadros\ (\BED), \APACrefbtitle {{Proceedings of the 20th
  International Meshing Roundtable}} {{Proceedings of the 20th International
  Meshing Roundtable}}\ (\BPGS\ 293--311).
\newblock
\APACaddressPublisher{Berlin, Heidelberg}{Springer Berlin Heidelberg}.
\newblock
\begin{APACrefDOI} \doi{https://doi.org/10.1007/978-3-642-24734-7_16}
  \end{APACrefDOI}
\PrintBackRefs{\CurrentBib}

\bibitem [\protect \citeauthoryear {%
Stein%
\ \BBA {} Bird%
}{%
Stein%
\ \BBA {} Bird%
}{%
{\protect \APACyear {2024}}%
}]{%
Stein2024}
\APACinsertmetastar {%
Stein2024}%
\begin{APACrefauthors}%
Stein, R\BPBI S.%
\BCBT {}\ \BBA {} Bird, P.%
\end{APACrefauthors}%
\unskip\
\newblock
\APACrefYearMonthDay{2024}{}{}.
\newblock
{\BBOQ}\APACrefatitle {Why do great continental transform earthquakes nucleate
  on branch faults?} {Why do great continental transform earthquakes nucleate
  on branch faults?}{\BBCQ}
\newblock
\APACjournalVolNumPages{Seismological Research Letters}{95}{6}{3406--3415}.
\newblock
\begin{APACrefDOI} \doi{https://doi.org/10.1785/0220240175} \end{APACrefDOI}
\PrintBackRefs{\CurrentBib}

\bibitem [\protect \citeauthoryear {%
Syracuse%
, van Keken%
\BCBL {}\ \BBA {} Abers%
}{%
Syracuse%
\ \protect \BOthers {.}}{%
{\protect \APACyear {2010}}%
}]{%
syracuse_2010}
\APACinsertmetastar {%
syracuse_2010}%
\begin{APACrefauthors}%
Syracuse, E\BPBI M.%
, van Keken, P\BPBI E.%
\BCBL {}\ \BBA {} Abers, G\BPBI A.%
\end{APACrefauthors}%
\unskip\
\newblock
\APACrefYearMonthDay{2010}{}{}.
\newblock
{\BBOQ}\APACrefatitle {The global range of subduction zone thermal models} {The
  global range of subduction zone thermal models}.{\BBCQ}
\newblock
\APACjournalVolNumPages{Physics of the earth and planetary
  interiors}{183}{1-2}{73-90}.
\newblock
\begin{APACrefDOI} \doi{https://doi.org/10.1016/j.pepi.2010.02.004}
  \end{APACrefDOI}
\PrintBackRefs{\CurrentBib}

\bibitem [\protect \citeauthoryear {%
Taira%
\ \protect \BOthers {.}}{%
Taira%
\ \protect \BOthers {.}}{%
{\protect \APACyear {2017}}%
}]{%
taira_et_al_2017}
\APACinsertmetastar {%
taira_et_al_2017}%
\begin{APACrefauthors}%
Taira, K.%
, Brunton, S\BPBI L.%
, Dawson, S\BPBI T\BPBI M.%
, Rowley, C\BPBI W.%
, Colonius, T.%
, McKeon, B\BPBI J.%
\BDBL {}Ukeiley, L\BPBI S.%
\end{APACrefauthors}%
\unskip\
\newblock
\APACrefYearMonthDay{2017}{}{}.
\newblock
{\BBOQ}\APACrefatitle {Modal Analysis of Fluid Flows: {A}n Overview} {Modal
  analysis of fluid flows: {A}n overview}.{\BBCQ}
\newblock
\APACjournalVolNumPages{AIAA Journal}{55}{12}{4013-4041}.
\PrintBackRefs{\CurrentBib}

\bibitem [\protect \citeauthoryear {%
Thoutireddy%
\ \BBA {} Ortiz%
}{%
Thoutireddy%
\ \BBA {} Ortiz%
}{%
{\protect \APACyear {2004}}%
}]{%
thoutireddy_ortiz_2004}
\APACinsertmetastar {%
thoutireddy_ortiz_2004}%
\begin{APACrefauthors}%
Thoutireddy, P.%
\BCBT {}\ \BBA {} Ortiz, M.%
\end{APACrefauthors}%
\unskip\
\newblock
\APACrefYearMonthDay{2004}{}{}.
\newblock
{\BBOQ}\APACrefatitle {A variational r-adaption and shape-optimization method
  for finite-deformation elasticity} {A variational r-adaption and
  shape-optimization method for finite-deformation elasticity}.{\BBCQ}
\newblock
\APACjournalVolNumPages{International Journal for Numerical Methods in
  Engineering}{61}{1}{1--21}.
\newblock
\begin{APACrefDOI} \doi{https://doi.org/10.1002/nme.1052} \end{APACrefDOI}
\PrintBackRefs{\CurrentBib}

\bibitem [\protect \citeauthoryear {%
Uphoff%
\ \protect \BOthers {.}}{%
Uphoff%
\ \protect \BOthers {.}}{%
{\protect \APACyear {2017}}%
}]{%
uphoff_et_al_2017}
\APACinsertmetastar {%
uphoff_et_al_2017}%
\begin{APACrefauthors}%
Uphoff, C.%
, Rettenberger, S.%
, Bader, M.%
, Madden, E\BPBI H.%
, Ulrich, T.%
, Wollherr, S.%
\BCBL {}\ \BBA {} Gabriel, A\BHBI A.%
\end{APACrefauthors}%
\unskip\
\newblock
\APACrefYearMonthDay{2017}{}{}.
\newblock
{\BBOQ}\APACrefatitle {Extreme scale multi-physics simulations of the
  tsunamigenic 2004 sumatra megathrust earthquake} {Extreme scale multi-physics
  simulations of the tsunamigenic 2004 sumatra megathrust earthquake}.{\BBCQ}
\newblock
\BIn{} \APACrefbtitle {{Proceedings of the International Conference for High
  Performance Computing, Networking, Storage and Analysis}} {{Proceedings of
  the International Conference for High Performance Computing, Networking,
  Storage and Analysis}}\ (\BPGS\ 1--16).
\newblock
\begin{APACrefDOI} \doi{https://doi.org/10.1145/3126908.3126948}
  \end{APACrefDOI}
\PrintBackRefs{\CurrentBib}

\bibitem [\protect \citeauthoryear {%
{van Keken}%
\ \protect \BOthers {.}}{%
{van Keken}%
\ \protect \BOthers {.}}{%
{\protect \APACyear {2008}}%
}]{%
community_benchmark_van_keken_2008}
\APACinsertmetastar {%
community_benchmark_van_keken_2008}%
\begin{APACrefauthors}%
{van Keken}, P\BPBI E.%
, Currie, C.%
, King, S\BPBI D.%
, Behn, M\BPBI D.%
, Cagnioncle, A.%
, He, J.%
\BDBL {}Wang, K.%
\end{APACrefauthors}%
\unskip\
\newblock
\APACrefYearMonthDay{2008}{}{}.
\newblock
{\BBOQ}\APACrefatitle {A community benchmark for subduction zone modeling} {A
  community benchmark for subduction zone modeling}.{\BBCQ}
\newblock
\APACjournalVolNumPages{Physics of the Earth and Planetary
  Interiors}{171}{1}{187-197}.
\newblock
\begin{APACrefDOI} \doi{https://doi.org/10.1016/j.pepi.2008.04.015}
  \end{APACrefDOI}
\PrintBackRefs{\CurrentBib}

\bibitem [\protect \citeauthoryear {%
van Keken%
, Kiefer%
\BCBL {}\ \BBA {} Peacock%
}{%
van Keken%
\ \protect \BOthers {.}}{%
{\protect \APACyear {2002}}%
}]{%
van_keken_kiefer_peacock_2002}
\APACinsertmetastar {%
van_keken_kiefer_peacock_2002}%
\begin{APACrefauthors}%
van Keken, P\BPBI E.%
, Kiefer, B.%
\BCBL {}\ \BBA {} Peacock, S\BPBI M.%
\end{APACrefauthors}%
\unskip\
\newblock
\APACrefYearMonthDay{2002}{}{}.
\newblock
{\BBOQ}\APACrefatitle {High-resolution models of subduction zones: Implications
  for mineral dehydration reactions and the transport of water into the deep
  mantle} {High-resolution models of subduction zones: Implications for mineral
  dehydration reactions and the transport of water into the deep
  mantle}.{\BBCQ}
\newblock
\APACjournalVolNumPages{Geochemistry Geophysics Geosystems}{3(10)}{}{}.
\newblock
\begin{APACrefDOI} \doi{https://doi.org/10.1029/2001GC000256} \end{APACrefDOI}
\PrintBackRefs{\CurrentBib}

\bibitem [\protect \citeauthoryear {%
van Rijsingen%
\ \protect \BOthers {.}}{%
van Rijsingen%
\ \protect \BOthers {.}}{%
{\protect \APACyear {2018}}%
}]{%
van_rijsingen_et_al_2018}
\APACinsertmetastar {%
van_rijsingen_et_al_2018}%
\begin{APACrefauthors}%
van Rijsingen, E.%
, Lallemand, S.%
, Peyret, M.%
, Arcay, D.%
, Heuret, A.%
, Funiciello, F.%
\BCBL {}\ \BBA {} Corbi, F.%
\end{APACrefauthors}%
\unskip\
\newblock
\APACrefYearMonthDay{2018}{}{}.
\newblock
{\BBOQ}\APACrefatitle {How Subduction Interface Roughness Influences the
  Occurrence of Large Interplate Earthquakes} {How subduction interface
  roughness influences the occurrence of large interplate earthquakes}.{\BBCQ}
\newblock
\APACjournalVolNumPages{Geochemistry, Geophysics,
  Geosystems}{19}{8}{2342-2370}.
\newblock
\begin{APACrefDOI} \doi{https://doi.org/10.1029/2018GC007618} \end{APACrefDOI}
\PrintBackRefs{\CurrentBib}

\bibitem [\protect \citeauthoryear {%
Wada%
\ \BBA {} Wang%
}{%
Wada%
\ \BBA {} Wang%
}{%
{\protect \APACyear {2009}}%
}]{%
wada_wang_2009}
\APACinsertmetastar {%
wada_wang_2009}%
\begin{APACrefauthors}%
Wada, I.%
\BCBT {}\ \BBA {} Wang, K.%
\end{APACrefauthors}%
\unskip\
\newblock
\APACrefYearMonthDay{2009}{}{}.
\newblock
{\BBOQ}\APACrefatitle {Common depth of slab-mantle decoupling: {R}econciling
  diversity and uniformity of subduction zones} {Common depth of slab-mantle
  decoupling: {R}econciling diversity and uniformity of subduction
  zones}.{\BBCQ}
\newblock
\APACjournalVolNumPages{Geochemistry Geophysics Geosystems}{10}{}{}.
\newblock
\begin{APACrefDOI} \doi{https://doi.org/10.1029/2009GC002570} \end{APACrefDOI}
\PrintBackRefs{\CurrentBib}

\bibitem [\protect \citeauthoryear {%
Wada%
, Wang%
, He%
\BCBL {}\ \BBA {} Hyndman%
}{%
Wada%
\ \protect \BOthers {.}}{%
{\protect \APACyear {2008}}%
}]{%
wada_et_al_2008}
\APACinsertmetastar {%
wada_et_al_2008}%
\begin{APACrefauthors}%
Wada, I.%
, Wang, K.%
, He, J.%
\BCBL {}\ \BBA {} Hyndman, R\BPBI D.%
\end{APACrefauthors}%
\unskip\
\newblock
\APACrefYearMonthDay{2008}{}{}.
\newblock
{\BBOQ}\APACrefatitle {Weakening of the subduction interface and its effects on
  surface heat flow, slab dehydration, and mantle wedge serpentinization}
  {Weakening of the subduction interface and its effects on surface heat flow,
  slab dehydration, and mantle wedge serpentinization}.{\BBCQ}
\newblock
\APACjournalVolNumPages{Journal of Geophysical Research: Solid
  Earth}{113}{B4}{}.
\newblock
\begin{APACrefDOI} \doi{https://doi.org/10.1029/2007JB005190} \end{APACrefDOI}
\PrintBackRefs{\CurrentBib}

\bibitem [\protect \citeauthoryear {%
Walton%
, Hassan%
\BCBL {}\ \BBA {} Morgan%
}{%
Walton%
\ \protect \BOthers {.}}{%
{\protect \APACyear {2013}}%
}]{%
walton}
\APACinsertmetastar {%
walton}%
\begin{APACrefauthors}%
Walton, S.%
, Hassan, O.%
\BCBL {}\ \BBA {} Morgan, K.%
\end{APACrefauthors}%
\unskip\
\newblock
\APACrefYearMonthDay{2013}{}{}.
\newblock
{\BBOQ}\APACrefatitle {Reduced order modelling for unsteady fluid flow using
  proper orthogonal decomposition and radial basis functions} {Reduced order
  modelling for unsteady fluid flow using proper orthogonal decomposition and
  radial basis functions}.{\BBCQ}
\newblock
\APACjournalVolNumPages{Applied Mathematical Modelling}{37}{}{8930-8945}.
\newblock
\begin{APACrefDOI} \doi{http://dx.doi.org/10.1016/j.apm.2013.04.025}
  \end{APACrefDOI}
\PrintBackRefs{\CurrentBib}

\bibitem [\protect \citeauthoryear {%
Wesnousky%
}{%
Wesnousky%
}{%
{\protect \APACyear {2006}}%
}]{%
wesnousky_2006}
\APACinsertmetastar {%
wesnousky_2006}%
\begin{APACrefauthors}%
Wesnousky, S\BPBI G.%
\end{APACrefauthors}%
\unskip\
\newblock
\APACrefYearMonthDay{2006}{}{}.
\newblock
{\BBOQ}\APACrefatitle {Predicting the endpoints of earthquake ruptures}
  {Predicting the endpoints of earthquake ruptures}.{\BBCQ}
\newblock
\APACjournalVolNumPages{Nature}{444}{7117}{358--360}.
\newblock
\begin{APACrefDOI} \doi{https://doi.org/10.1038/nature05275} \end{APACrefDOI}
\PrintBackRefs{\CurrentBib}

\bibitem [\protect \citeauthoryear {%
Wesnousky%
}{%
Wesnousky%
}{%
{\protect \APACyear {2008}}%
}]{%
wesnousky_2008}
\APACinsertmetastar {%
wesnousky_2008}%
\begin{APACrefauthors}%
Wesnousky, S\BPBI G.%
\end{APACrefauthors}%
\unskip\
\newblock
\APACrefYearMonthDay{2008}{}{}.
\newblock
{\BBOQ}\APACrefatitle {Displacement and Geometrical Characteristics of
  Earthquake Surface Ruptures: Issues and Implications for Seismic-Hazard
  Analysis and the Process of Earthquake Rupture} {Displacement and geometrical
  characteristics of earthquake surface ruptures: Issues and implications for
  seismic-hazard analysis and the process of earthquake rupture}.{\BBCQ}
\newblock
\APACjournalVolNumPages{Bulletin of the Seismological Society of
  America}{98}{4}{1609-1632}.
\newblock
\begin{APACrefDOI} \doi{https://doi.org/10.1785/0120070111} \end{APACrefDOI}
\PrintBackRefs{\CurrentBib}

\bibitem [\protect \citeauthoryear {%
Wirth%
, Sahakian%
, Wallace%
\BCBL {}\ \BBA {} Melnick%
}{%
Wirth%
\ \protect \BOthers {.}}{%
{\protect \APACyear {2022}}%
}]{%
wirth_et_al_2022}
\APACinsertmetastar {%
wirth_et_al_2022}%
\begin{APACrefauthors}%
Wirth, E\BPBI A.%
, Sahakian, V\BPBI J.%
, Wallace, L\BPBI M.%
\BCBL {}\ \BBA {} Melnick, D.%
\end{APACrefauthors}%
\unskip\
\newblock
\APACrefYearMonthDay{2022}{}{}.
\newblock
{\BBOQ}\APACrefatitle {The occurrence and hazards of great subduction zone
  earthquakes} {The occurrence and hazards of great subduction zone
  earthquakes}.{\BBCQ}
\newblock
\APACjournalVolNumPages{Nature Reviews Earth \& Environment}{3}{2}{125--140}.
\newblock
\begin{APACrefDOI} \doi{https://doi.org/10.1038/s43017-021-00245-w}
  \end{APACrefDOI}
\PrintBackRefs{\CurrentBib}

\bibitem [\protect \citeauthoryear {%
Wollherr%
, Gabriel%
\BCBL {}\ \BBA {} Uphoff%
}{%
Wollherr%
\ \protect \BOthers {.}}{%
{\protect \APACyear {2018}}%
}]{%
wollherr_et_al_2018}
\APACinsertmetastar {%
wollherr_et_al_2018}%
\begin{APACrefauthors}%
Wollherr, S.%
, Gabriel, A\BHBI A.%
\BCBL {}\ \BBA {} Uphoff, C.%
\end{APACrefauthors}%
\unskip\
\newblock
\APACrefYearMonthDay{2018}{}{}.
\newblock
{\BBOQ}\APACrefatitle {Off-fault plasticity in three-dimensional dynamic
  rupture simulations using a modal Discontinuous Galerkin method on
  unstructured meshes: implementation, verification and application} {Off-fault
  plasticity in three-dimensional dynamic rupture simulations using a modal
  discontinuous galerkin method on unstructured meshes: implementation,
  verification and application}.{\BBCQ}
\newblock
\APACjournalVolNumPages{Geophysical Journal International}{214}{3}{1556-1584}.
\newblock
\begin{APACrefDOI} \doi{10.1093/gji/ggy213} \end{APACrefDOI}
\PrintBackRefs{\CurrentBib}

\bibitem [\protect \citeauthoryear {%
Zhang%
, Slemmons%
\BCBL {}\ \BBA {} Mao%
}{%
Zhang%
\ \protect \BOthers {.}}{%
{\protect \APACyear {1991}}%
}]{%
zhang_et_al_1991}
\APACinsertmetastar {%
zhang_et_al_1991}%
\begin{APACrefauthors}%
Zhang, P.%
, Slemmons, D.%
\BCBL {}\ \BBA {} Mao, F.%
\end{APACrefauthors}%
\unskip\
\newblock
\APACrefYearMonthDay{1991}{}{}.
\newblock
{\BBOQ}\APACrefatitle {{Geometric pattern, rupture termination and fault
  segmentation of the Dixie Valley—Pleasant Valley active normal fault
  system, Nevada, U.S.A.}} {{Geometric pattern, rupture termination and fault
  segmentation of the Dixie Valley—Pleasant Valley active normal fault
  system, Nevada, U.S.A.}}{\BBCQ}
\newblock
\APACjournalVolNumPages{Journal of Structural Geology}{13}{2}{165-176}.
\newblock
\begin{APACrefDOI} \doi{https://doi.org/10.1016/0191-8141(91)90064-P}
  \end{APACrefDOI}
\PrintBackRefs{\CurrentBib}

\end{thebibliography}

\end{document}